%% file: ms.tex
\documentclass[12pt,preprint]{aastex}
\newcommand{\Msol}{M$_{\odot}$}
\newcommand{\hk}{ }
\defcitealias{Luhman10}{L10}
\defcitealias{Rebull10}{R10}

\begin{document}

\title{Young Stellar Groups and Their Most Massive Stars}
\author{Helen Kirk\altaffilmark{1} \& Philip C. Myers\altaffilmark{1}}
\altaffiltext{1}{Radio and Geoastronomy Division, Harvard Smithsonian
	Center for Astrophysics, MS-42, Cambridge, MA, 02138, USA;
	hkirk@cfa.harvard.edu}
%\slugcomment{\today}

\begin{abstract}
 
We analyze the masses and spatial distributions of fourteen
young stellar groups in
Taurus, Lupus3, ChaI, and IC348.
These nearby groups, which typically contain 20 to 40 members,
have membership catalogs complete to $\sim$0.02~\Msol, and 
are sufficiently young that
their locations should be similar to where they formed.  
These groups show five properties seen in
clusters having many more stars and much greater surface density
of stars: (1) a broad range of masses, (2) a concentration of the most 
massive star towards the centre of the group, (3) an association of the 
most massive
star with a high surface density of lower-mass stars, (4) a correlation
of the mass of the most massive star with the total mass of the group,
and (5) the distribution of a large fraction of the mass in
a small fraction of the stars.
\end{abstract}

\section{INTRODUCTION}
Most stars are believed to form in clusters
\citep{Lada03}, and clusters also harbour most, if not all, massive stars
\citep[e.g,][]{Zinnecker07}.  The nearest star-forming regions
within a few hundred pc, however,
have few clusters according to most definitions \citep{Reipurth08}.
There, low mass stars form
primarily in an ``isolated'' mode, and high mass stars are absent.  
%In comparison to the large body of research into both isolated and clustered
%star formation

There has been relatively little work investigating
the transition between these two regimes.  Several authors have
considered this transition from the perspective of a continuum
of clustering, rather than two discrete states.
\citet{Elmegreen08}, for example, theoretically examined which 
physical conditions
within a turbulent molecular complex lead to the formation of
bound clusters versus more dispersed stellar systems.  Varied
physical conditions could 
account for the substantial variations in properties of
young massive star-forming regions seen in nearby galaxies 
\citep{MaizApel01}, as well as the presence or absence of clusters
within portions of the local Gould Belt structure \citep{Elias09}.  
Within the nearest few hundred parsecs, 
\citet{Bressert10} recently demonstrated that 
there is no evidence for a preferred length-scale
for clustering of YSOs detected with {\it Spitzer}.

Despite this evidence that there may not be a sharp boundary
between clustered and isolated modes of star-formation, there are
distinctions between the two at least in terms of processes that
will influence later evolution (discussed in Section~5.1).
In such a manner, the transition between the regime where
processes associated with clusters are important, and the regime
where they are not 
is expected to
occur for stellar groupings whose most massive member is between roughly
2 and 15~M$_{\odot}$ \citep{Testi99}.  
In this transition range, do stellar groups exhibit properties typical
of clusters (e.g., mass segregation and high surface density of sources),
or are their properties more similar to those of isolated star formation?

\citet[][and references therein]{Testi99} observed 44 Herbig Ae/Be stars
and their surroundings in order to study this transition.  They focussed
on cluster richness indicators (number and surface density of cluster
members) around each star targeted.  They found that both measures
tend to decrease from early to late B spectral types, and then show
relatively little trend through the A spectral types, although there is
large scatter at all spectral types.  A significant fraction of the late-Be
and Ae stars show little evidence for belonging to a group or cluster.
\citet{Massi03} find a similar result in the Vela C and D clouds, 
using the bolometric luminosity, $L_{bol}$, as a proxy for 
mass in these young star-forming
regions.  They found all
IRAS sources with L$_{bol}$ $\gtrsim 10^3 L_{\odot}$ and some 
lower $L_{bol}$ sources show evidence of clustering.

In this paper, we examine this transition further, focussing
on the properties of the most massive member or members of each group.  
With sensitive near-IR data from {\it Spitzer},
in combination with large ground-based spectral surveys, catalogs
of YSOs with excellent completeness are now available for several
nearby star-forming regions.  These catalogs allow us to analyze the
properties of stellar groupings in the transition range in ways not
possible for the more distant regions studied in \citet{Testi99} and
\citet{Massi03}, since the nearby regions considered here
have deeper and more uniform completeness levels.

Our main conclusions are that within the groups we identify,
similar to clusters, the members have a broad range in masses,
with a significant fraction of the mass being found within the
few most massive members, consistent with an IMF-like distribution.
Within each group, the most massive member is centrally-located, and
is in or near a region of enhanced source density.
We find a correlation between the mass of the most massive member 
and the total mass of the group similar to that found in larger clusters. 

In Section 2, we describe the YSO catalogs we use in our analysis,
with further detail in Appendices~A and B, while our procedure for
identifying groups of YSOs is discussed in Section~3.
We analyze the properties of the groups identified in Section~4,
focussing on mass segregation and clustering, then discuss the
implications in Section~5, and conclude in Section~6.  Our
procedure for estimating the YSO masses and our sources of
uncertainty are examined in Appendices~C and D.

\section{DATA}
To analyze the properties of young stellar groups, we
require the stars to be nearby (within $\sim$300~pc) to prevent
source confusion.  We require the stars to be younger than several
Myr, so that their natal groups have not had time for significant
dynamical evolution.  The stars must also be old enough (roughly
class I/II or higher) so that accurate spectral types can be
determined, and hence reasonable mass estimates made.
Finally, the census of stars must be complete to better than
90\% to masses below the brown dwarf limit (0.08~\Msol) 
in order for us to
apply our analysis.  Applying
these criteria, there are four nearby star-forming regions for
which suitable YSO catalogs exist -- Taurus, ChaI, Lupus 3 and IC348,
all of which include members out to late M (or even L0) spectral types.
This sample differs from other samples of nearby star-forming regions
such as those listed in \citet{Evans09}, \citet{Gutermuth09}, 
and \citet{Myers09a} largely due to our requirement of spectral 
classifications.  Most of the groups in this sample correspond
to previously identified groups, while ChaI 2 and IC348 1 correspond 
to small clusters, as discussed in Section~3.
%[HK note not to be incl in paper -- 
%the above statement from skimming through all entries
%in the Star Formation Handbook.  The only other nearby, young regions
%either have very embedded stellar populations (NGC 1333, Coronet cluster in
%CrA, Serpens), or too few optical / IR spectral types (cluster in
%Ophiuchus near L1688, only 171 of 316 likely candidate members have
%some sort of spectral measurement).]

The YSO catalogs we analyze do not include the most deeply embedded objects.  
Most class 0 and
some class I sources are likely missing from our catalogs,
as discussed further in Appendix~A.5.
These are estimated to comprise
fewer than $\sim$ 7\% of the members of any group 
considered, based on the
distribution of classes in \citet{Evans09}. Since our analysis
focusses on the relationship between the most massive member of each
group and the group as a whole, this incompleteness is unlikely to
affect our results. 
%For my full notes on completeness, see this directory, completeness_notes.txt

The YSO catalogs for the four nearby star-forming regions
are given in Tables~\ref{tab_Taurus} through \ref{tab_IC348}
{\it (available online only)},
which list the position, common name(s), spectral type, estimated
mass (see Section~\ref{sec_mass_est} and Appendix~C), 
and the group each source 
is associated with in our analysis (see Section~\ref{sec_mst}).
Appendix~\ref{app_catalogs} and B describe the catalogs in more detail.

\subsection{Mass Estimation}
\label{sec_mass_est}
For each of the YSOs, we estimate the mass based on the spectral
type.  We assume a constant age of 1~Myr for all YSOs and
follow the procedure outlined in \citet{Luhman03},
using a combination of models from \citet{PS99},
\citet{Baraffe98}, and \citet{Chabrier00} to estimate the
masses.  The assumption of a constant, 1~Myr age, as well as the
exact stellar models chosen, leads to uncertainties in the
masses of order 50\%.
This is discussed in more detail in Appendix~\ref{app_mass_est}.

\subsection{Additional Data -- ONC1}
\label{sec_onc1}
In several parts of the following analysis, we compare the results for 
the four nearby star-forming regions with
the large cluster encompassing the Trapezium in the Orion Nebula Cluster,
to represent properties typical to young clusters.  
To make this comparison, we use the ONC1 dataset from \citet{Hillenbrand97},
adopting the masses and positions given there.  We include all sources
listed as having a 70\% or higher probability of membership.
This list contains 721 sources, of which we identify 410 of these as 
belonging to the Trapezium cluster (see Section~3 for our method of 
identifying groups and clusters).
Of these 410 Trapezium cluster sources, 26 do not have estimated masses in
\citet{Hillenbrand97}; we assign each of these the median mass 
in the cluster (0.23~\Msol).

\section{IDENTIFICATION OF GROUPS}
\label{sec_mst}
Within each of the four regions in our dataset, there are clearly small
groupings of YSOs. 
In order to define these groups, we use the minimal spanning tree (MST)
algorithm \citep{Barrow85}, following the procedure of \citet{Gutermuth09}.
Many methods can and have been used to identify groups and clusters
in various studies; \citet{Bressert10} find that clustering of YSOs in
nearby molecular clouds tends to be best described by a continuum, rather
than a single discrete value separating clustered versus non-clustered
regions.
We adopt the MST to define stellar groups, as it 
can be applied to multiple
regions in an easily reproducible manner, does not require unique
thresholds to be set {\it a priori}, and is independent of the
distance to the region; other cluster-identification schemes are
discussed in Appendix~D.  
Our analysis is relatively insensitive to the continuum of 
clustering -- inclusion of more or fewer YSOs in each group does not
to have a major effect on our results, as discussed in Appendix~D.  

Minimal spanning trees are structures where all points in a region are
connected via the minimum distance between them (i.e., nearest neighbours);
each of these connections is a `branch'.
The MST structure mimics how the eye naturally connect points;
most constellations are connected via their MST, 
for example.  

\begin{figure}[h]
\plottwo{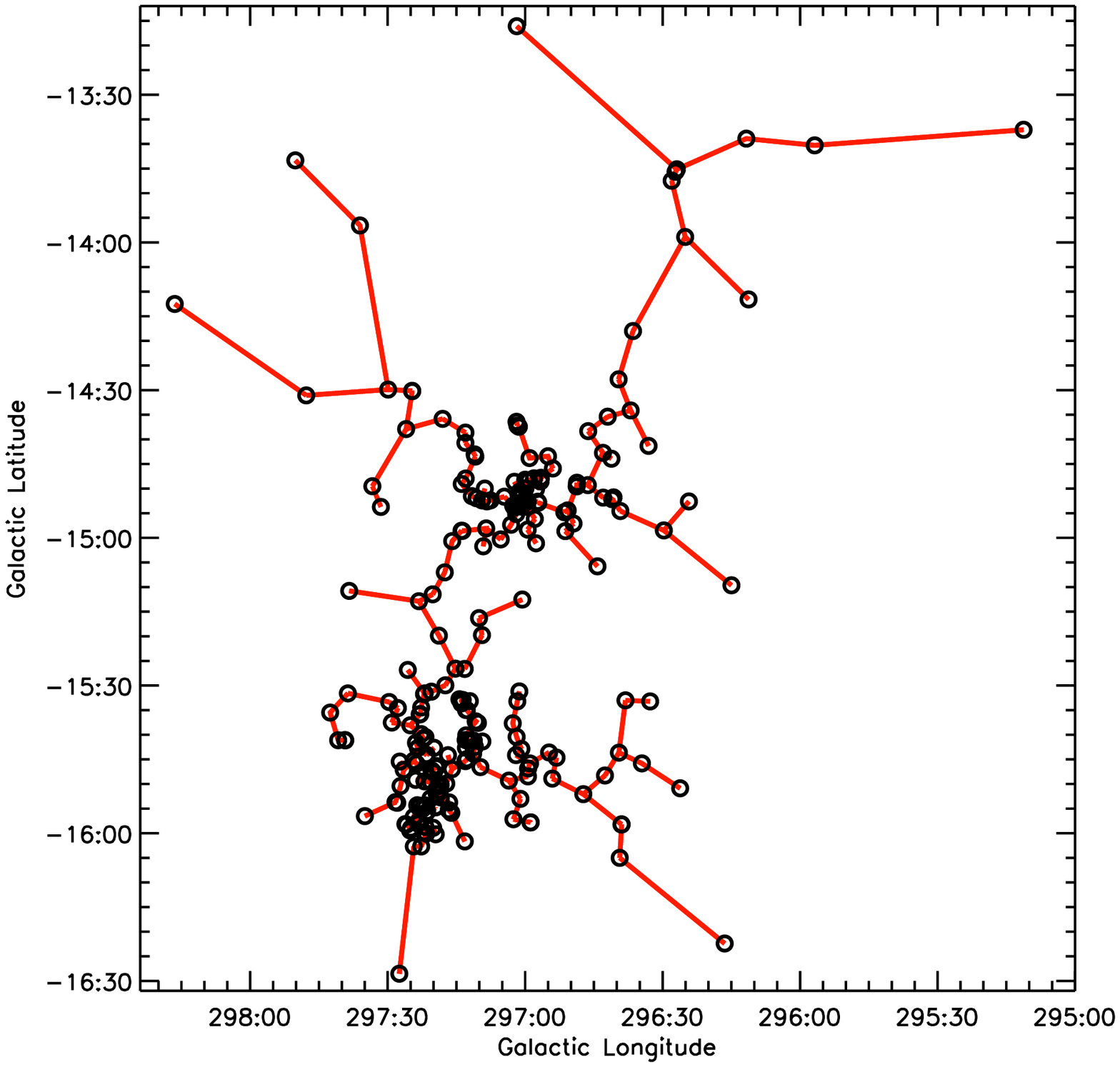}
	{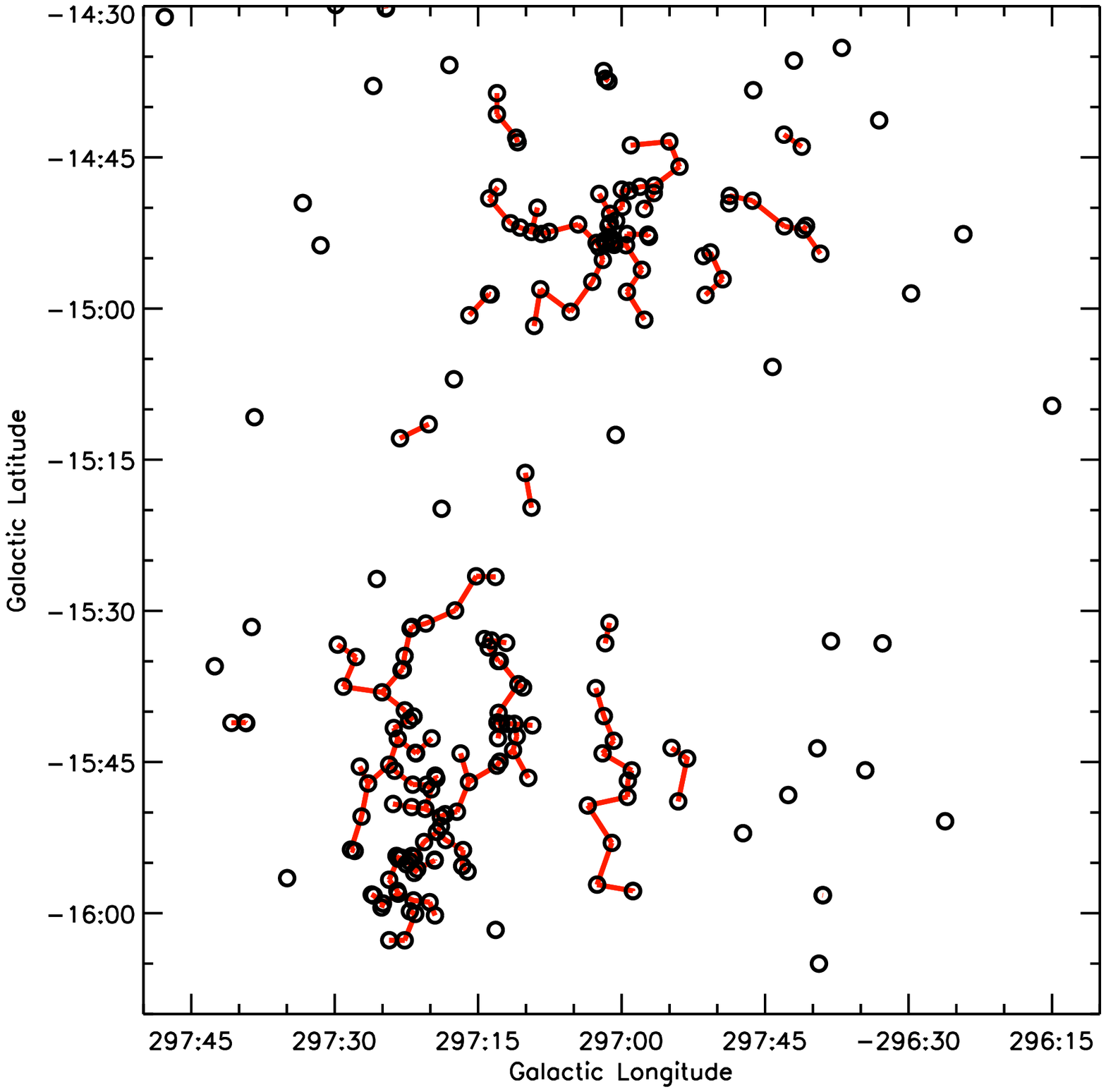}
\caption{
	An overview of the YSOs identified in ChaI.
	The black circles denote the locations of all of the YSOs
	in the region,
	while the red lines indicate the MST structure.  The left 
	panel shows the entire original MST structure, whereas
	the right panel shows the MST structure after branches
	longer than the critical length have been removed,
	zooming in on on the clustered region of the
	left panel.  The region is plotted in galactic projection
	for easier comparison with Figure~\ref{fig_MST_gallery3}.
	See Section~3 for more details.
	}
\label{fig_Cha_overview}
\end{figure}

The left panel of Figure~\ref{fig_Cha_overview} shows
the MST structure in the ChaI region.
Groups are apparent within a region by eye as having smaller
separations between members than typical in the region as a whole.
Within the MST structure, groups can be separated as having ``small''
branch lengths between all members, i.e., less than some cutoff
branch length.
Although this length could be defined in an absolute sense,
\citet{Gutermuth09} find it more effective to determine a critical length
based on the distribution of branch lengths within a given region.
This has the advantage of being insensitive to the uncertainty
in distance to the region, relying only on the clustering properties
of the sources within the region.  This method is discussed
in more detail in Appendix~D.1. 

%As illustrated in 
%Figure~\ref{fig_MST_cutoff}, the cumulative distribution of MST branch lengths
%within a region tends to be well-described by a steep linear rise
%at small branch lengths, followed by a turn-over and a shallow
%linear rise at large branch lengths.  Figure~\ref{fig_MST_cutoff} 
%shows the
%data for ChaI; the other three regions show a similar
%trend.  Following \citet{Gutermuth09}, linear fits
%to the small and large branch lengths are made, and the {\hk critical}
%branch length is defined as the length at which the two best-fit lines
%intersect.  
The right panel of Figure~\ref{fig_Cha_overview} 
shows the resulting groups in ChaI when branches with lengths larger
than the critical length are removed. 
After this `pruning' of the MST, isolated groupings of YSOs remain.
These groupings range from very small numbers of sources (e.g., isolated pairs)
to larger groups.  

For our analysis, we consider only groups which
have more than ten members.
This cutoff value is somewhat arbitrary, and is based on a visual
examination of the groupings identified in Taurus.  This 
examination indicates that groupings of
very small numbers of sources would not allow group properties to be
examined meaningfully, while a minimum group size of
twenty or thirty members would exclude some visually-striking smaller groups.

Using this MST procedure, we identify fourteen groups with more than
ten members in our dataset -- eight in Taurus, three in ChaI, two
in IC348, and one in Lupus3.  
The four star-forming regions are all much more 
`clustered' than expected from
a random distribution -- as a test, we ran our MST algorithm
on a set of N$_{YSO}$ randomly distributed points within each
region (where N$_{YSO}$ is the number of YSOs within each
region).  When the same critical length scale is adopted as
is measured for the observed YSOs, {\it no} groupings of more
than ten members are found for any of the four regions. 

Using the MST procedure, we also identify five groups in 
the \citet{Hillenbrand97}
ONC1 dataset, representing the main ONC1 cluster (with 410 members) and
four additional small groups (typically around 15 members).  We compare 
only the large ONC1 cluster with our groups, 
as this best represents the properties of a cluster.
The smaller groups identified in ONC1 may be
affected by incompleteness; the 70\% probability of 
membership criterion we applied  
may exclude some bona fide members, but this is unlikely to
affect our measures in the main ONC1 cluster.

\subsection{Groups and Their Environments}
The groups we identify in Taurus, Lupus3, ChaI, and IC348 are shown in
Figures~\ref{fig_MST_gallery1} through \ref{fig_MST_gallery4},
and their properties are summarized in Table~\ref{tab_groups}.
In the figures, the group members are indicated by blue circles
(with the radius scaling linearly with mass), the red lines denote
the MST structure, and the thin black circles denote nearby
YSOs that do not belong to the group.
The greyscale images in the background indicate the present-day
distribution of extinction
based on stellar reddening.  Displayed are extinction
data from \citet{Froebrich07} for Taurus, \citet{Dobashi05} for
ChaI, a combination of \citet{Teixeira05} and \citet{Rowles09}
for Lupus3, and \citet{Rowles09} for IC348.

As can be seen in Figures~\ref{fig_MST_gallery1} through 
\ref{fig_MST_gallery4}, the groups often lie near, but
not on, regions of high extinction, suggesting the YSOs have 
accreted and/or blown away a similar distribution of gas from their
immediate environs.  In most cases, the mass of gas required to create
the present-day stellar masses corresponds to a uniform
gas and dust having A$_V \simeq 3$~mag, assuming a formation efficiency
of 30\%.  This extinction due to smoothed-out stellar mass
is approximately the difference
between the maximum nearby extinction and the mean value of the
extinction within the group.  

The main group in IC348 is an
exception, requiring roughly ten times more material for the
formation of the YSOs, which is substantially higher than the
present-day nearby extinction.  The Lupus3 group also shows a much
larger difference in extinction than do most of the other groups, 
since the high-resolution extinction map from \citet{Teixeira05} 
allows much larger peaks in extinction to be measured.  At comparable
resolutions to the other regions, however, the difference in extinction
is similar to that in the other regions. 
At the same time, the total amount of material remaining in the area spanned
by each group is usually insufficient, or barely sufficient, to form a 
group with mass equal to the mass in the group we observe.  

In other
words, the immediate areas in which we see the YSOs in the groups 
appear largely
or entirely finished with the star-formation process, but there are
still reservoirs of material nearby which are capable of forming a significant
number of new stars.
%A more careful follow-up analysis
%would be required in order to make a quantitative comparison between
%the present day material and that required to create the YSOs.

\subsection{Comparison with Previously Identified Groups}
In Taurus, there is good correspondence between the groups
we identify and those previously identified using other methods.
The \citet{Gomez93} Groups I through VI correspond to our
groups 1, 2, 7, 5, 6, and 4 (B209, L1495E, L1527, L1529, L1536, and L1551)
respectively, with our remaining
two groups (3 / B213 and 8 / L1517) corresponding to un-named contours of higher
stellar density in their Figure~8 (right of their Group III and
the upper left of the figure respectively).  Group I in \citet{Jones79}
corresponds roughly to our Groups 1 and 2 (B209 \& L1495E),
their Group IIa corresponds to our Group 5 (L1529), their surrounding
Group II encompasses our Groups 6 and 7 (L1536 \& L1527), 
and their Group III is our Group 4 (L1551).  Our Group 3 (B213)
lies between their Group II and III with too few stars in their 
catalog to be classified as a group,
and our Group 8 (L1517) lies beyond their catalog.

The main group in IC348 appears most similar to a small cluster,
and has been studied in that vein by a variety of authors --
see \citet{Herbst08} and references therein.  The two largest groupings
of stars in ChaI have also been viewed in a similar manner -- see,
for example, \citet{Luhman07}.

\begin{figure}[hp]
\begin{tabular}{cc}
\includegraphics[height=8cm]{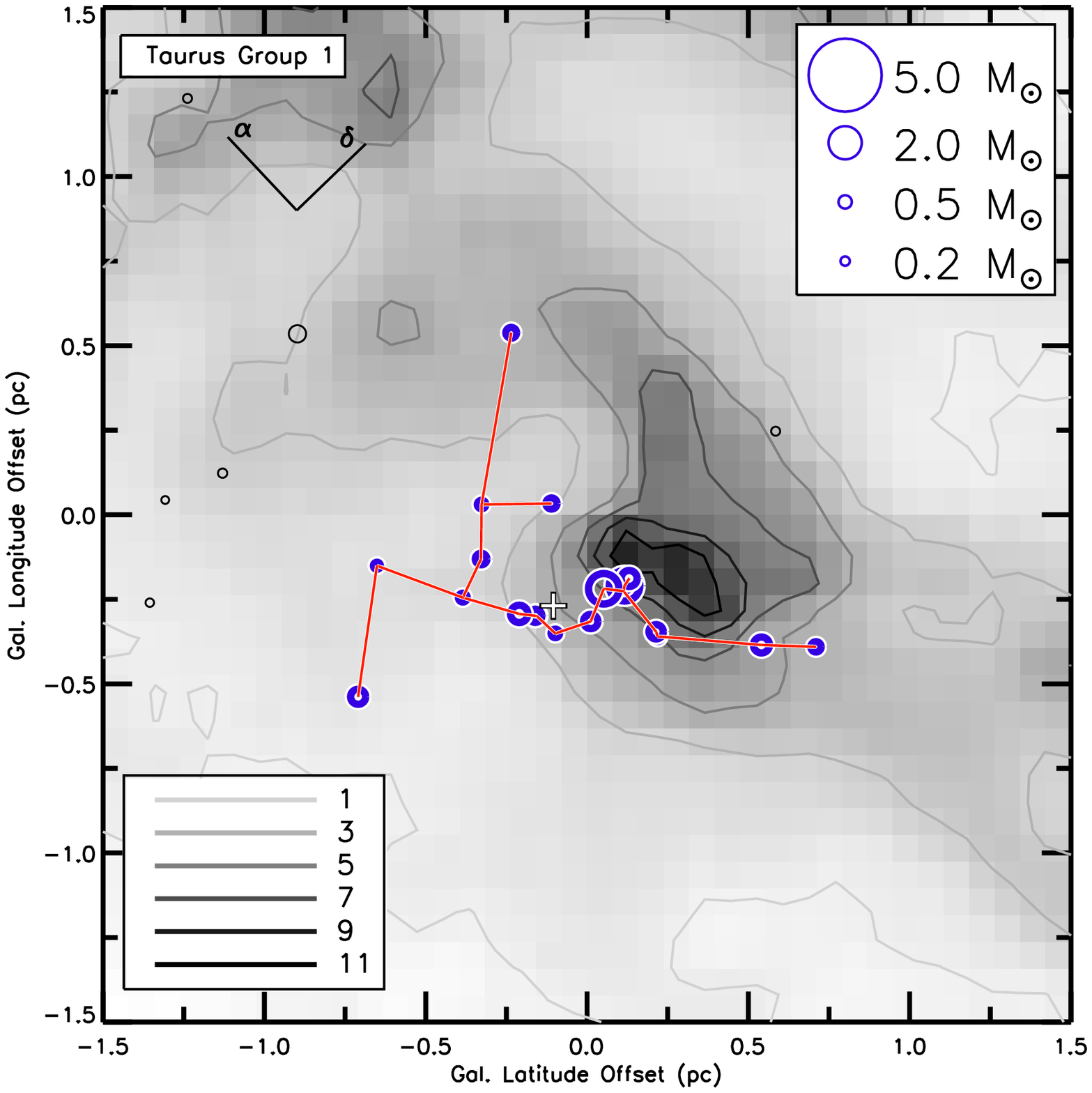} &
\includegraphics[height=8cm]{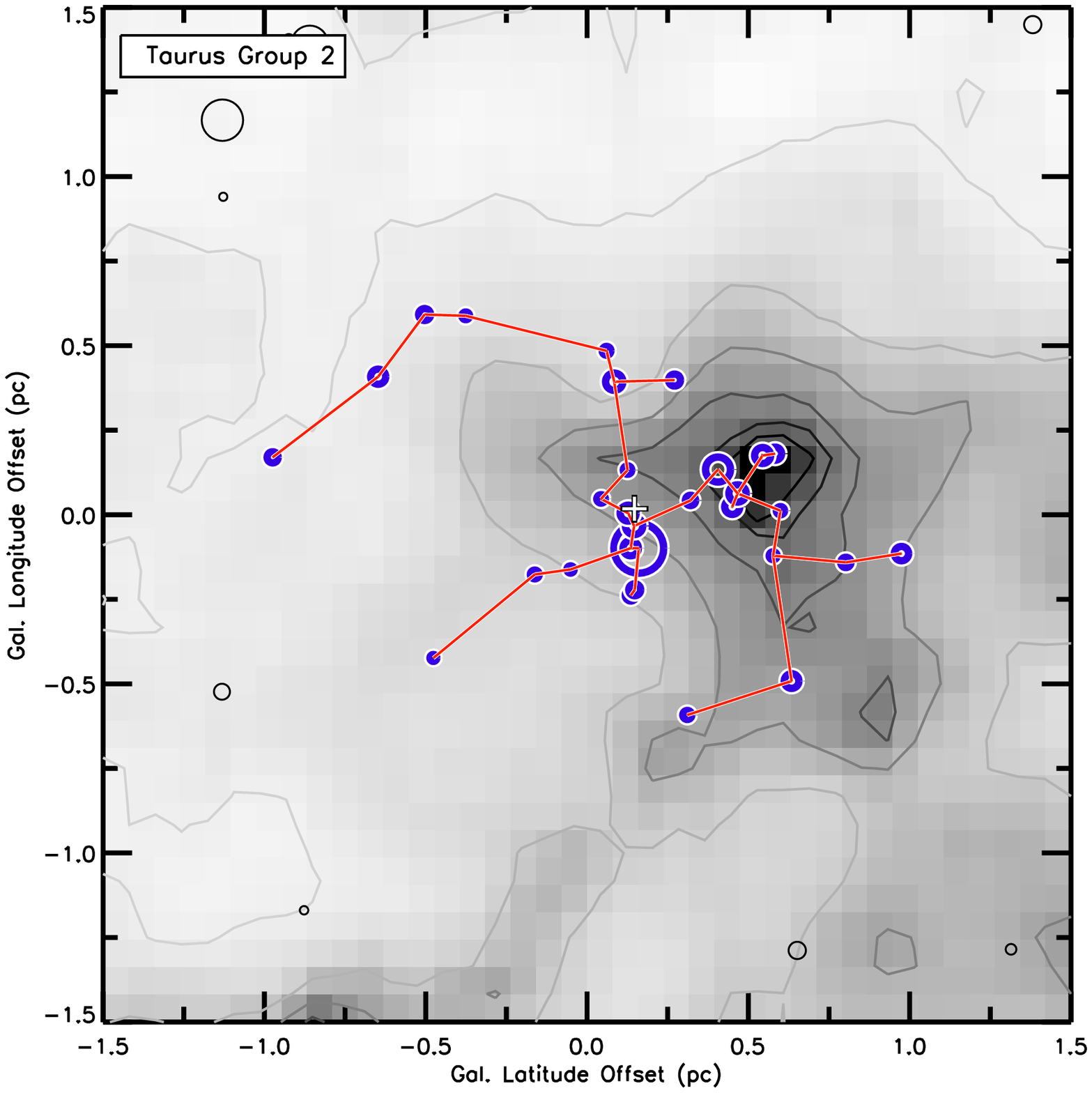} \\
\includegraphics[height=8cm]{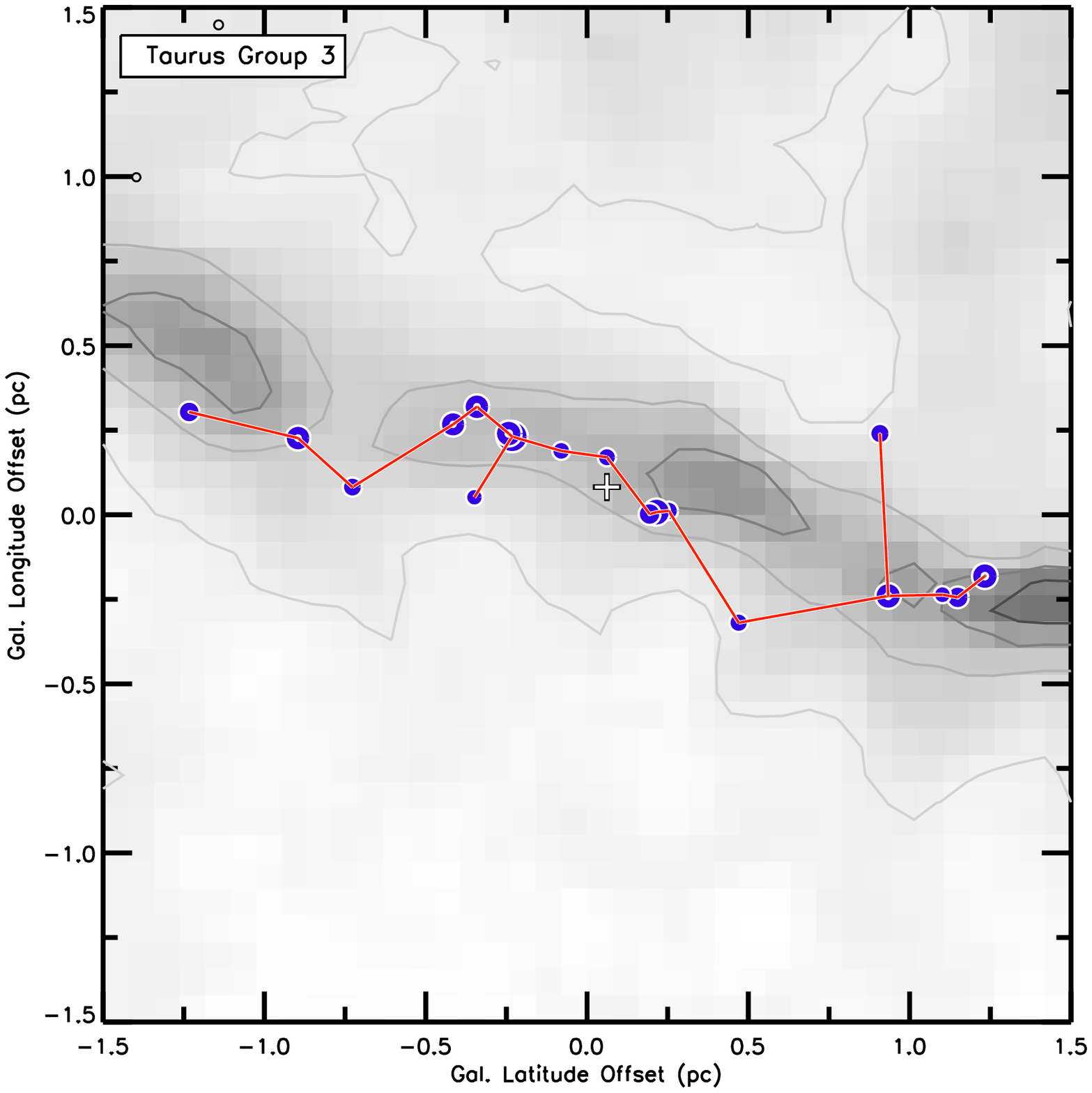} &
\includegraphics[height=8cm]{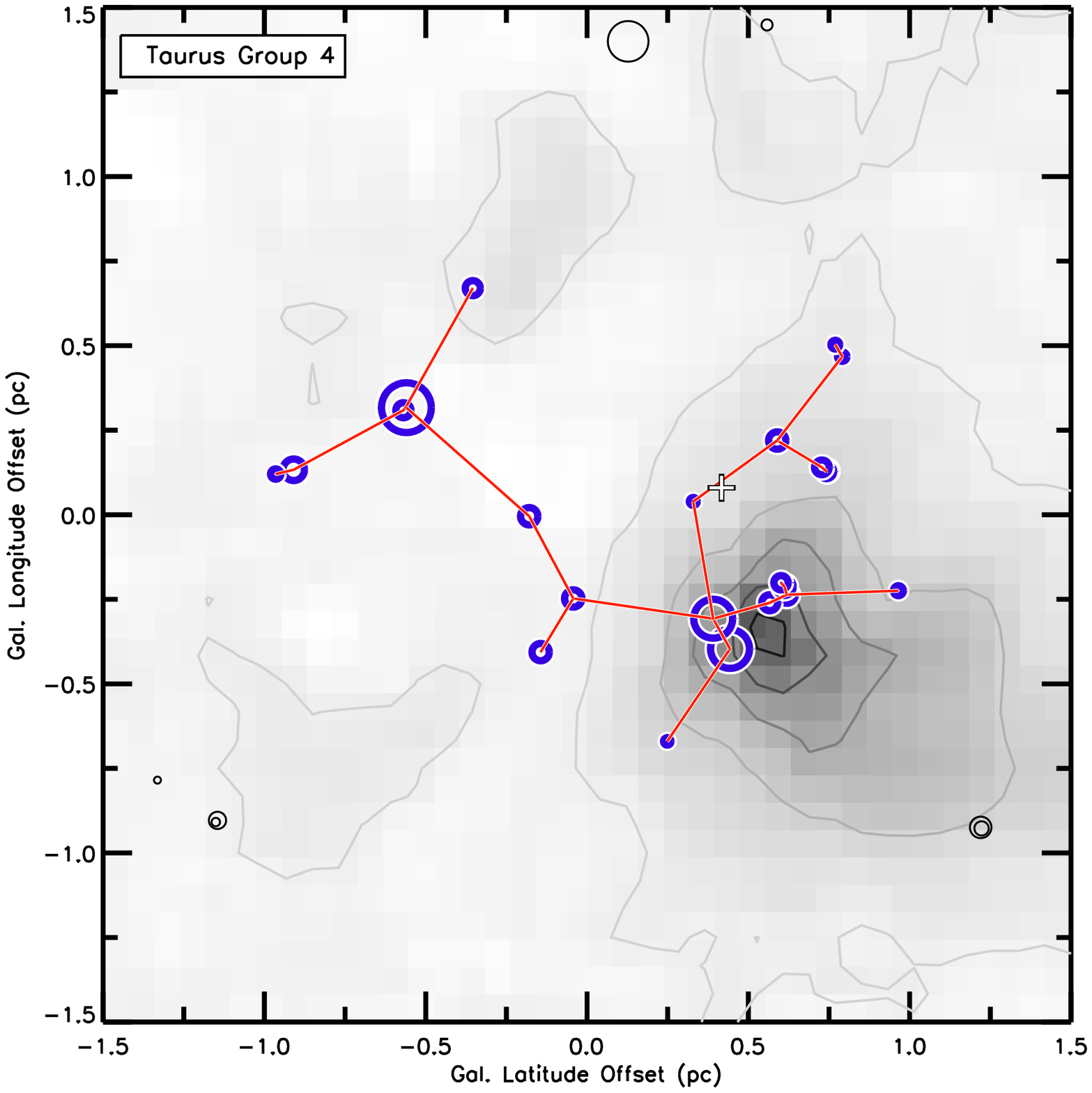} \\
\end{tabular}
\caption{ 
	The groups identified using the MST technique in each of the four
        regions.  Groups in Taurus 
        are shown in this figure; the other groups are in the following
        figures.  Blue circles indicate the YSOs within each group, with
        the circle size scaling linearly with the estimated mass (see
	first panel for scaling used).
        Red lines indicate the MST branches in the group.  Non-group
        YSOs in the vicinity are shown in black.
	The greyscale and contours in the background show the 
	extinction measured.  The greyscale ranges from A$_V$ of 15~mag
	(black) to 0 (white), with contours drawn at 1 to 11~mag
	(see scale bar in first panel).
	The orientation of the figures is in galactic co-ordinates
	to match the native projection of the extinction maps; the
	first panel shows the direction of increasing RA and dec.
	All figures are centred on the group's mid-point position.
	The median group member position given in Table~\ref{tab_groups} 
	is indicated by the white plus.}
\label{fig_MST_gallery1}
\end{figure}

\begin{figure}[hp]
\begin{tabular}{cc}
\includegraphics[height=8cm]{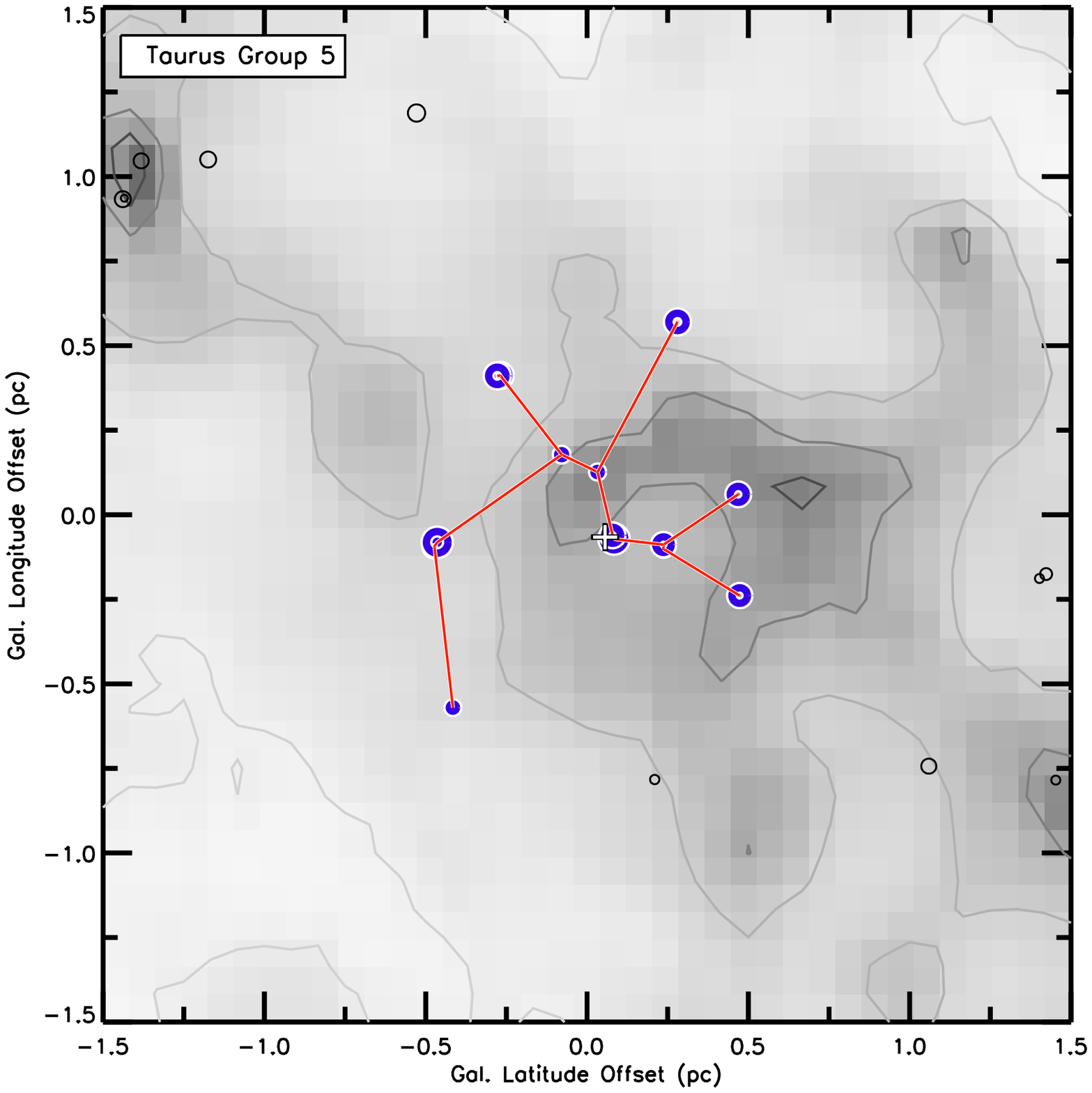} &
\includegraphics[height=8cm]{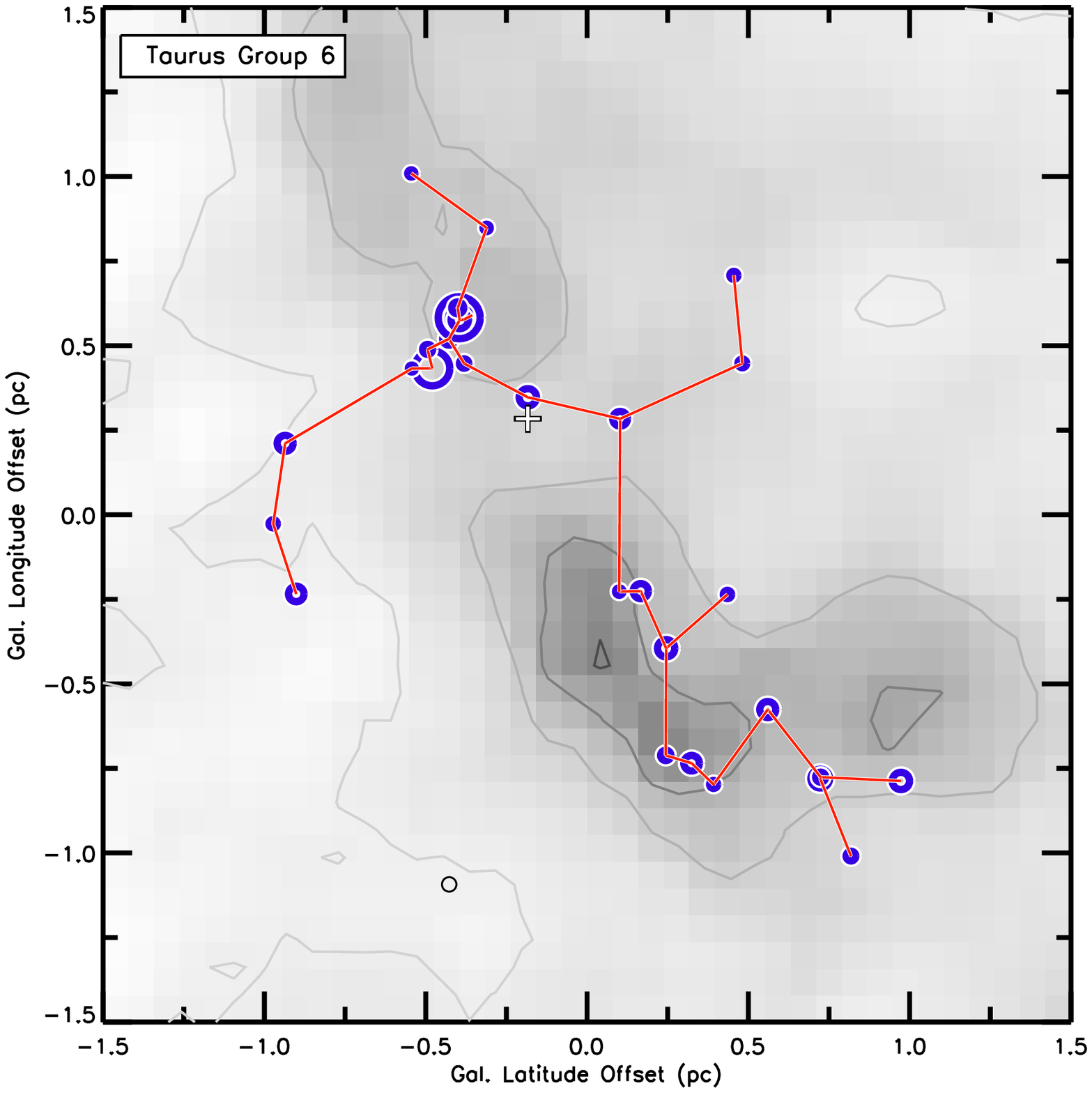} \\
\includegraphics[height=8cm]{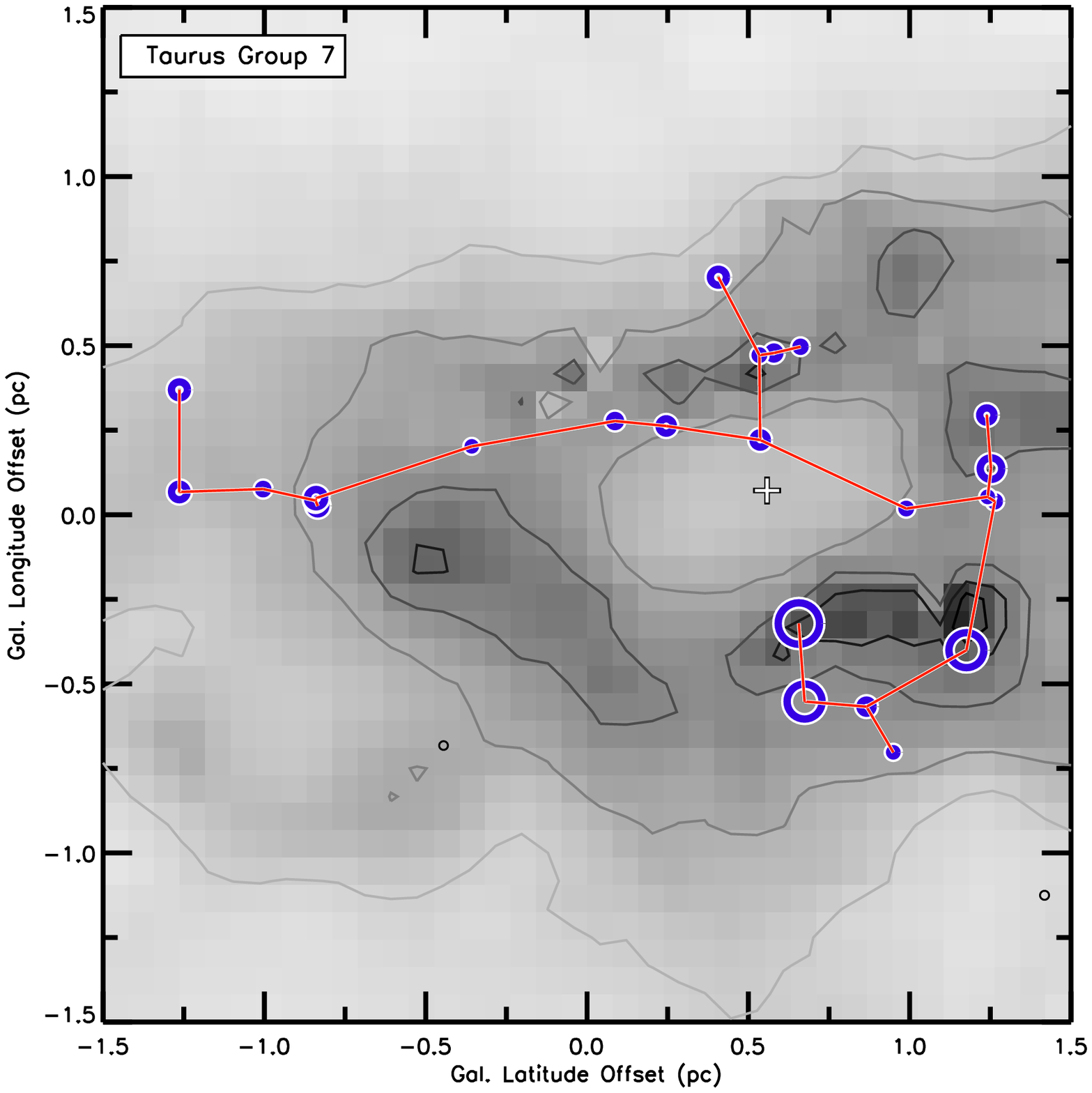} &
\includegraphics[height=8cm]{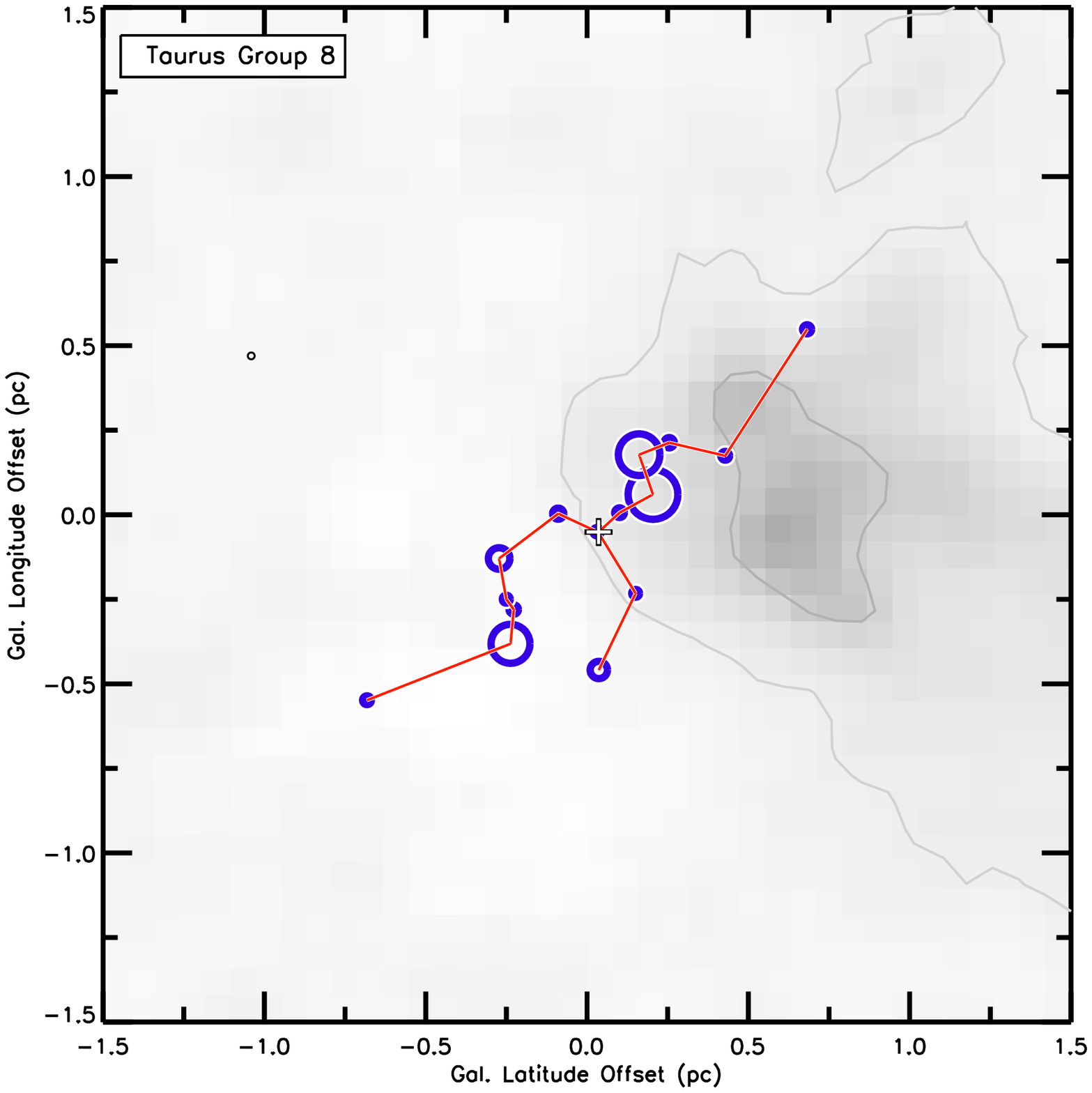} \\
\end{tabular}
\caption{More groups identified in Taurus.  See Figure~\ref{fig_MST_gallery1}
	caption for details.}
\label{fig_MST_gallery2}
\end{figure}

\begin{figure}[hp]
\begin{tabular}{cc}
\includegraphics[height=8cm]{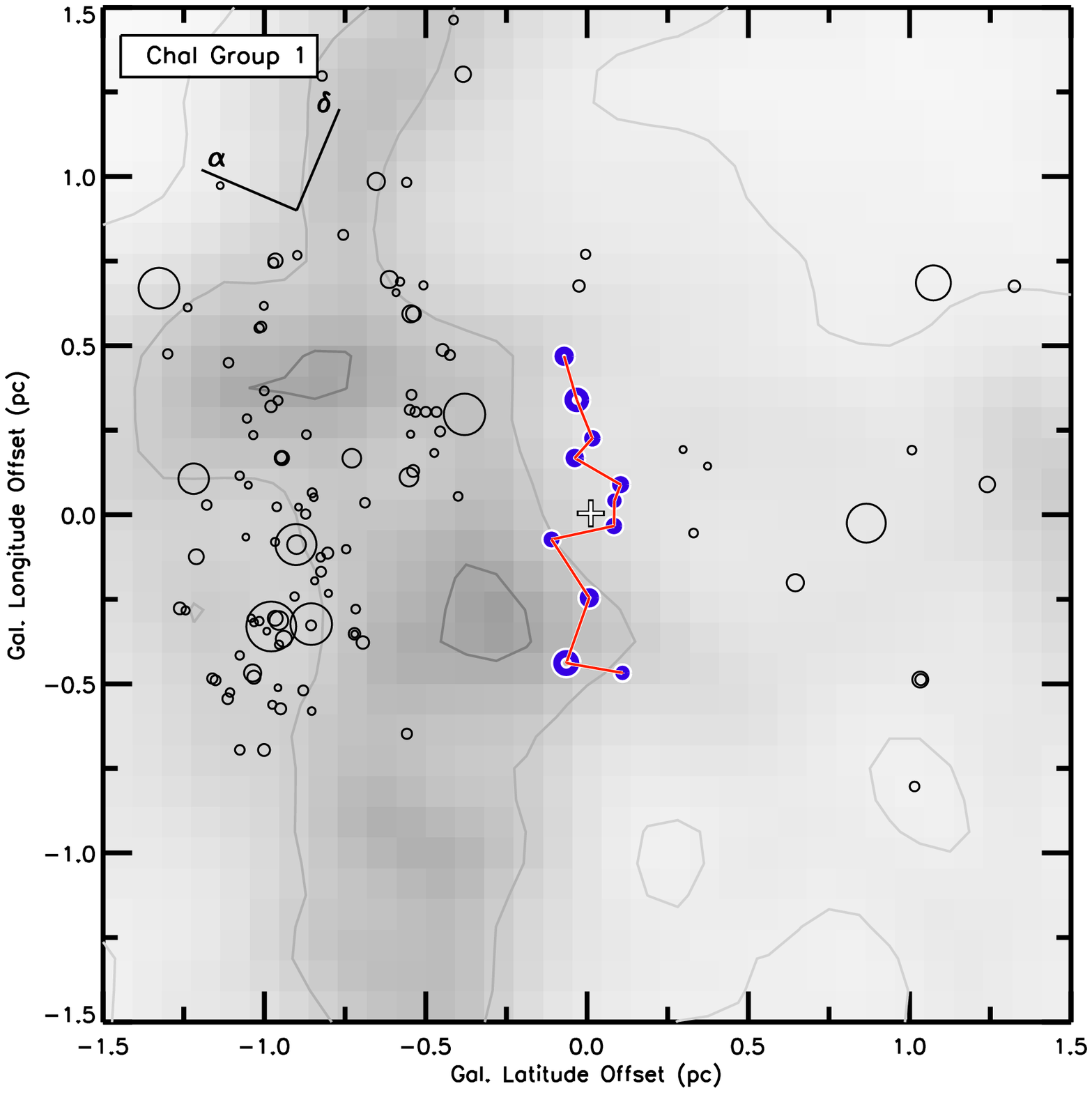} &
\includegraphics[height=8cm]{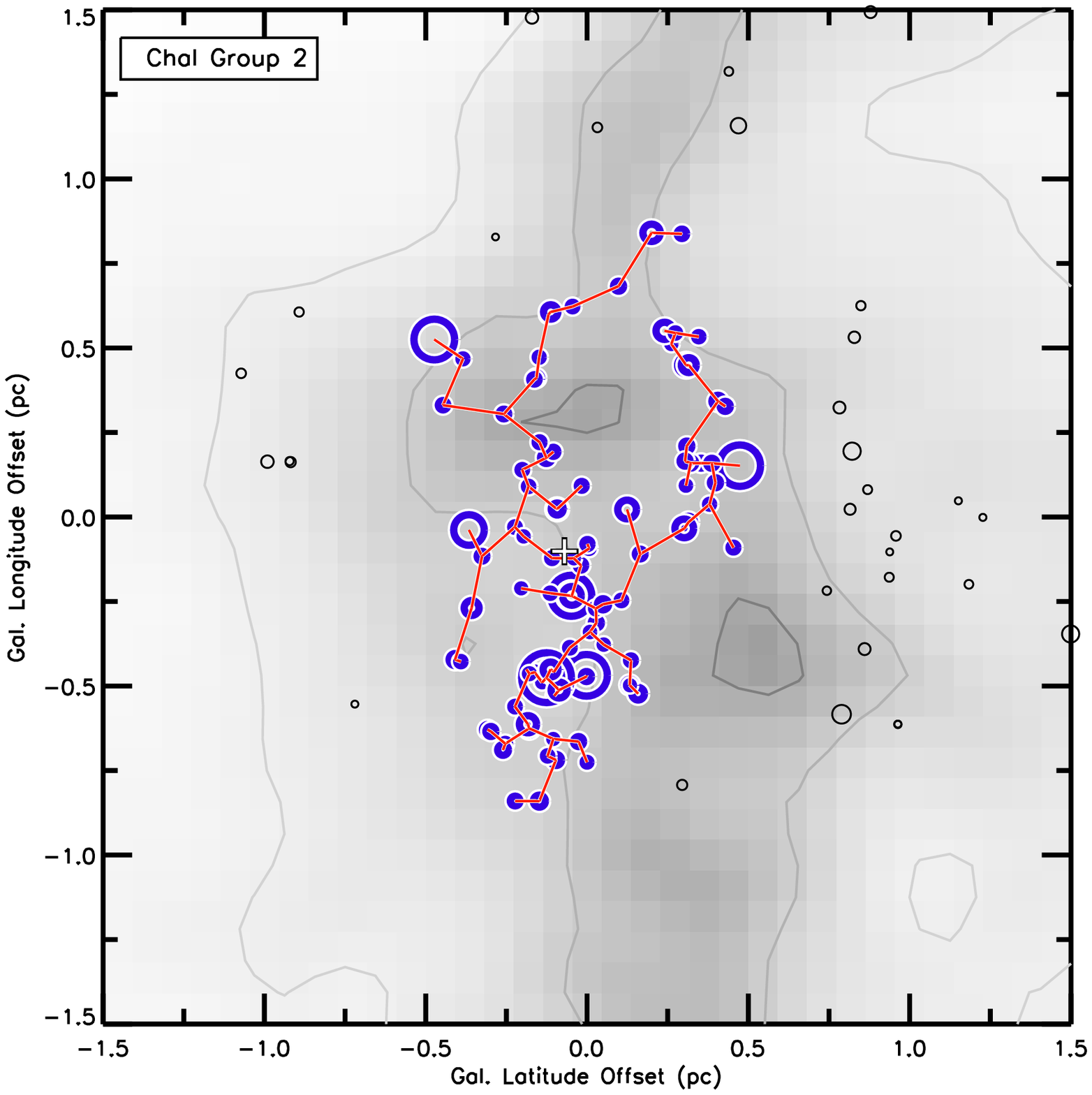} \\
\includegraphics[height=8cm]{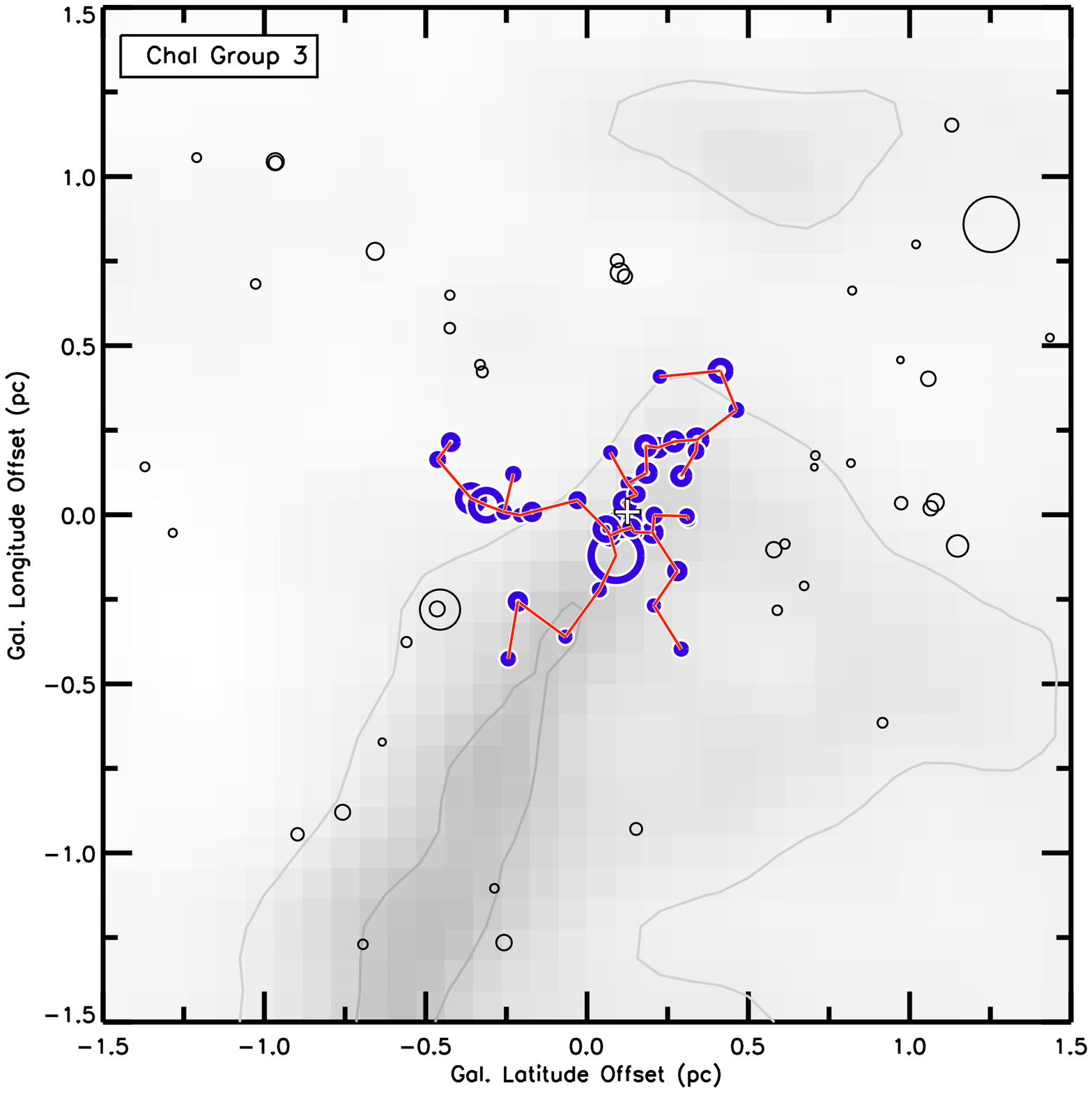} &
\includegraphics[height=8cm]{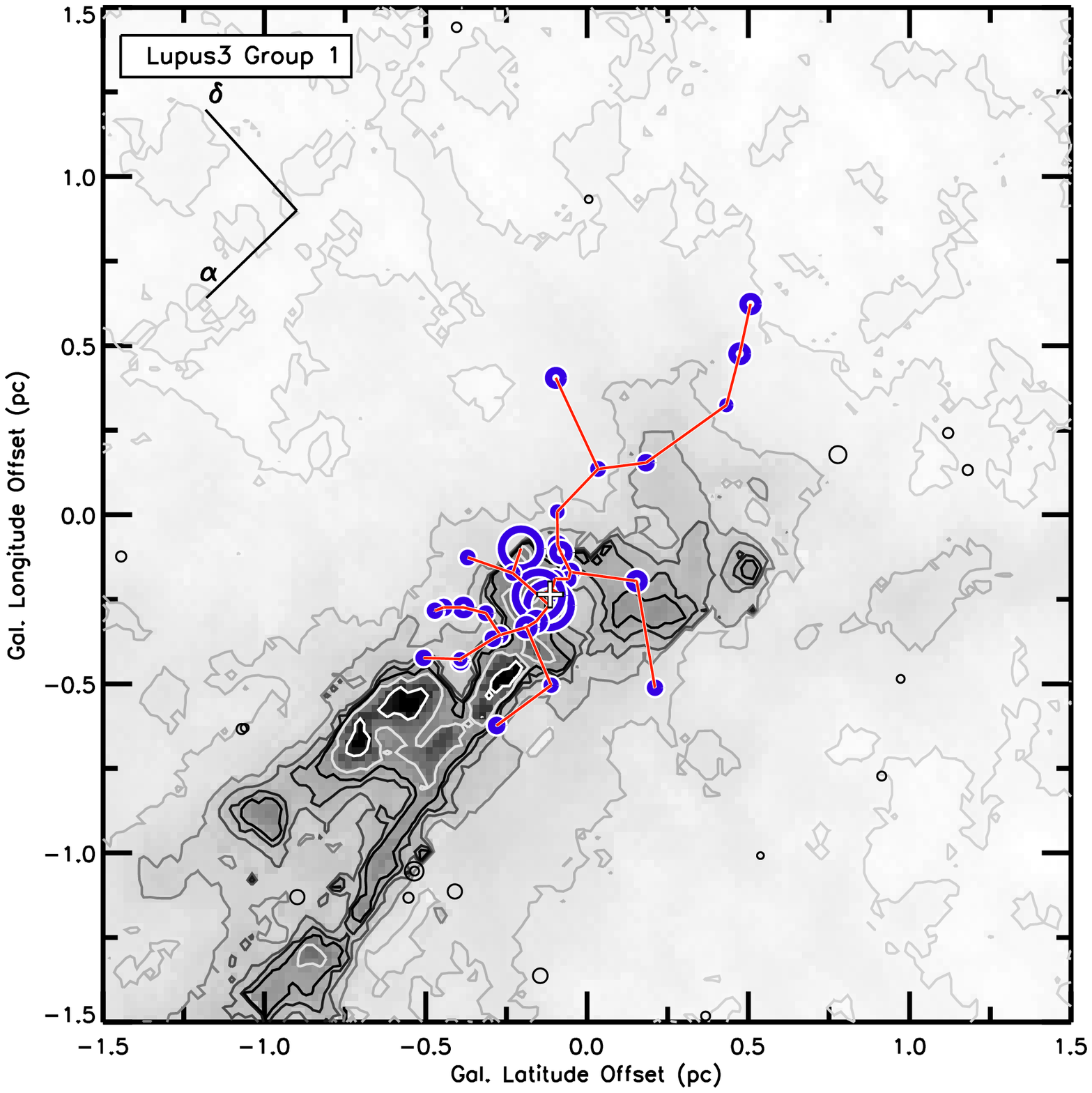} \\
\end{tabular}
\caption{Groups identified in ChaI (first three panels) and Lupus3 (final
	panel).  Directions of increasing RA and dec are shown for the
	first group in each region.  See Figure~\ref{fig_MST_gallery1} 
	caption for more details.  For clarity, the greyscale in
	the Lupus3 panel extends a factor of two higher in extinction
	than the other panels.  Additional contours at 15 and 25
	magnitudes are overlaid in light grey and white.}
\label{fig_MST_gallery3}
\end{figure}

\begin{figure}[h!]
\begin{tabular}{cc}
\includegraphics[height=8cm]{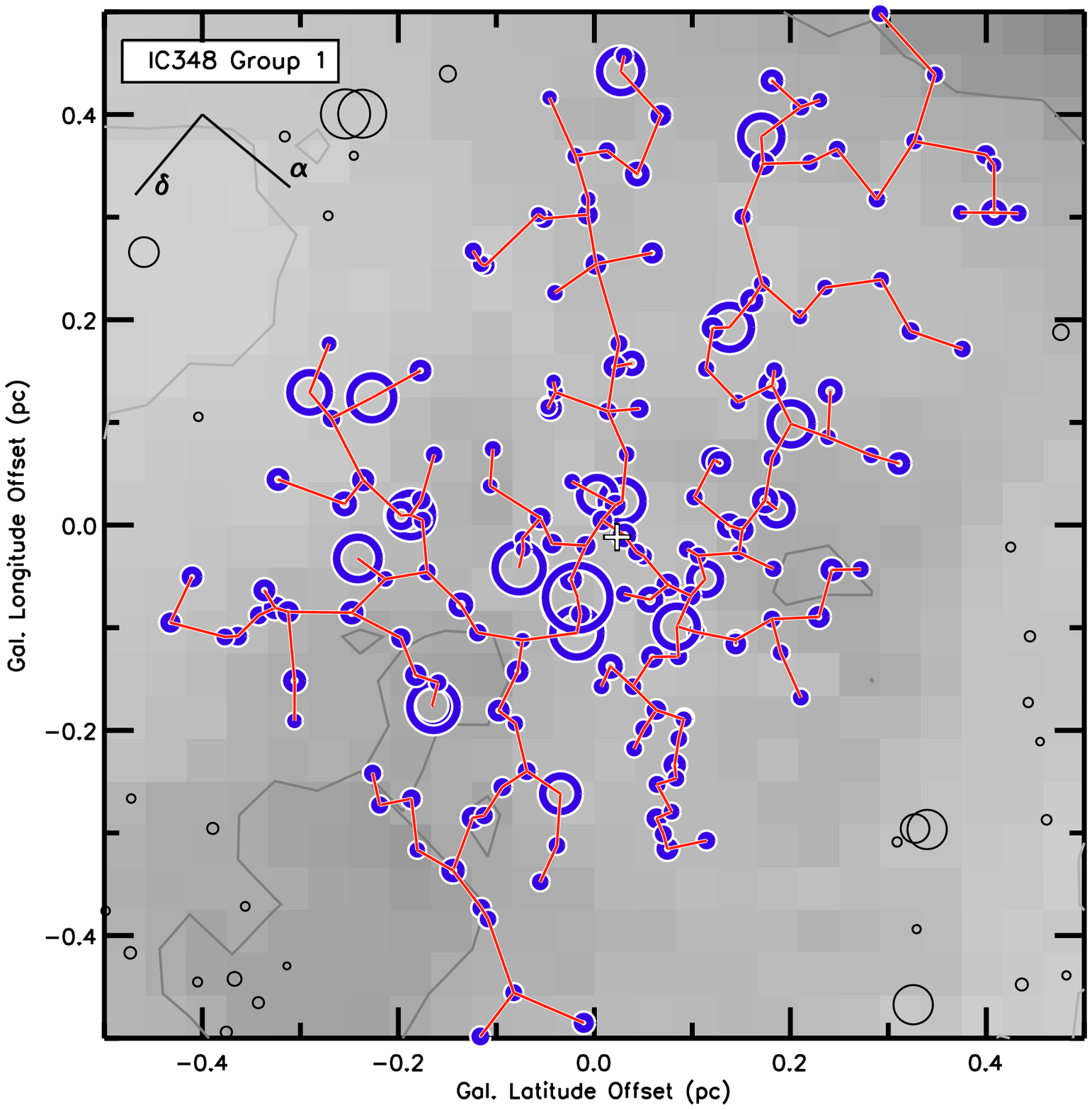} &
\includegraphics[height=8cm]{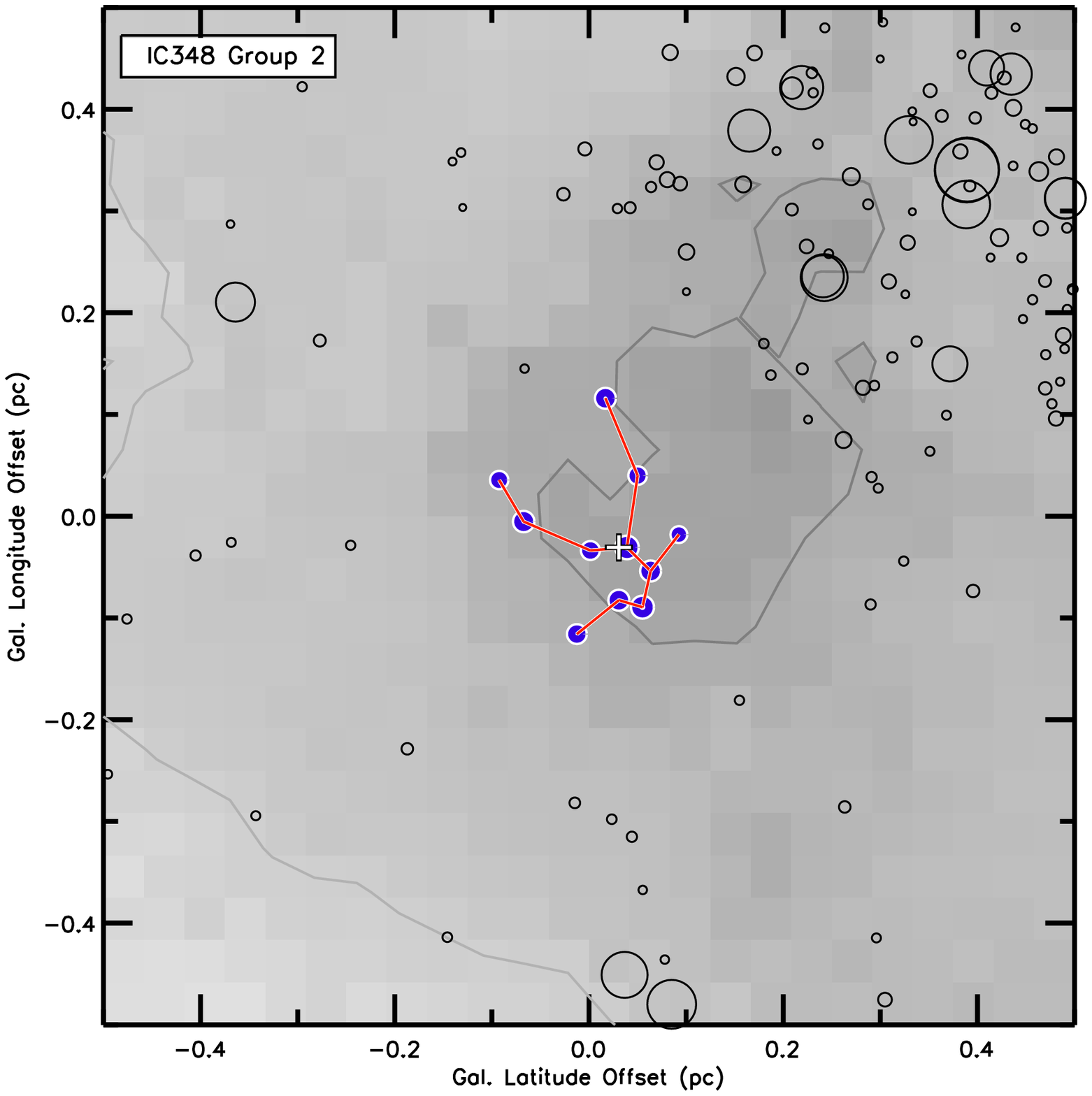} \\
\end{tabular}
\caption{Groups identified in IC348. 
	See Figure~\ref{fig_MST_gallery1} caption for more details.
	Note the linear scale is a factor of 3 smaller in these plots
	than the previous ones.  The direction of increasing RA and
	dec is shown in the first panel.}
\label{fig_MST_gallery4}
\end{figure}

As can be seen from the figures, some
of the groups are irregularly shaped, particularly when the
total number of members is small.  Some groups appear
filamentary, such as B213 in Taurus, unlike the conventional
picture of a group, which tend to be more circular. 
Despite these irregularities, we
find that the median member position (shown as white plus signs
in Figures~\ref{fig_MST_gallery1} through \ref{fig_MST_gallery4})\footnote{ The median position was
calculated from the median galactic latitude and longitude of the
positions, to allow for easier comparison with the large-scale
extinction maps of the regions in Figures~\ref{fig_MST_gallery1} to
\ref{fig_MST_gallery4}.} is usually a good 
representation of the apparent
centre of the group, although this becomes poorer
in the case of the most filamentary groups, as illustrated
in Taurus Groups 1 and 2 (Figure~\ref{fig_MST_gallery1}) versus
Taurus Group 6 (Figure~\ref{fig_MST_gallery2}), for example.
The mean member position is sometimes a worse descriptor of the 
group centre than the median, 
as it is more likely to be skewed by one or two
group members lying preferentially in one direction away from where
most of the group is concentrated.  The centre of mass tends to lie
between the median and mean positions. 
The centres of the groups as defined by the median positions of the
group members are also given in Table~\ref{tab_groups}.

\section{RESULTS}

\subsection{Mass Distribution}
 
\begin{figure}[hp]
\plotone{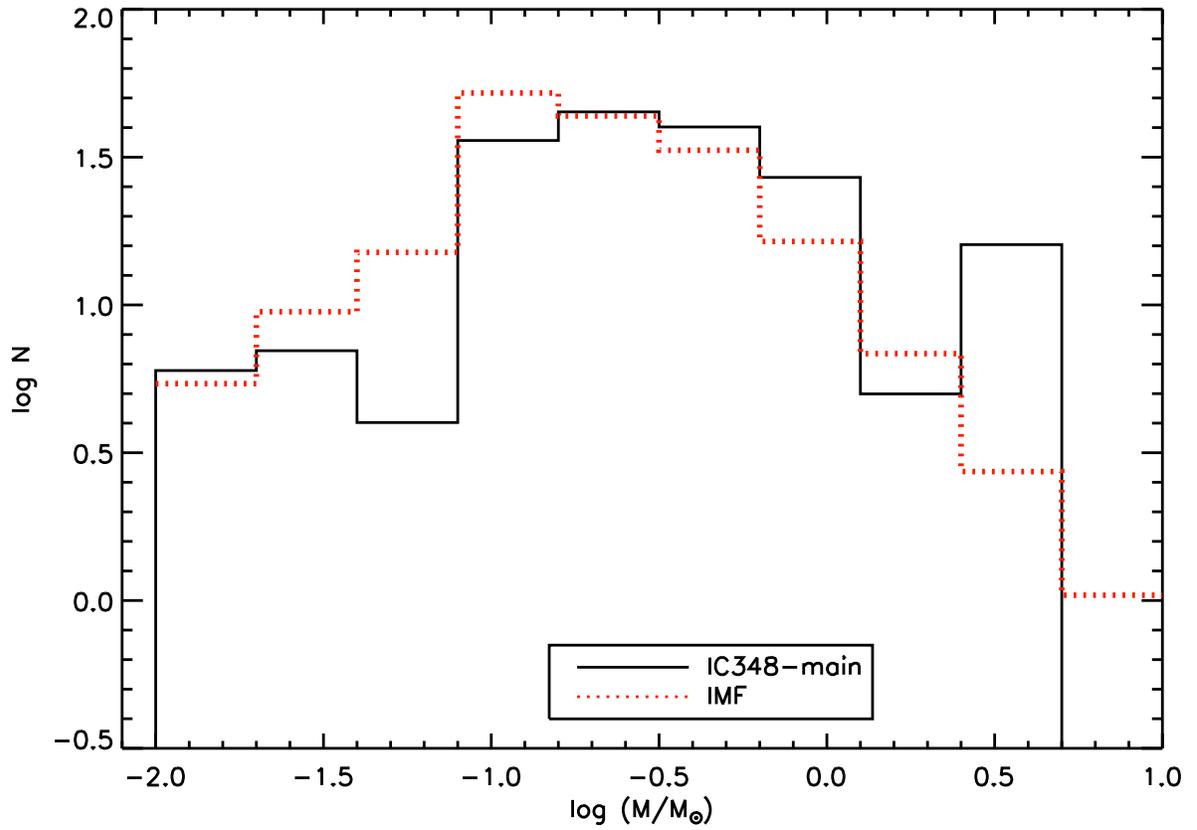}
\caption{The distribution of masses within the main IC348 group (solid line).
	The dotted line shows what would be expected from the IMF, using
	the formulation given in \citet{Weidner10}, with the minimum mass
	set equal to that in the observations and scaled to the total 
	number of YSOs in the main IC348 group.}
\label{fig_IC348_massdistrib}
\end{figure}

Using the masses estimated for each YSO, we examine the
distribution of masses within each group.  Figure~\ref{fig_IC348_massdistrib}
shows the distribution of masses within the main IC348 group.  A
prominent excess of sources with masses around 2-3~\Msol\ is clearly
evident; a similar trend is seen in many of the other groups, although
at a much lower level, since the groups are smaller.  This excess
is unlikely real, and is at least in part caused by our adoption
of a single age for all of the stars.  Assuming a single age of 2~Myr 
instead of 1~Myr, for example, reduces the size of the excess (for
a given spectral type, assuming an older age will reduce the mass
estimated); allowing for an age spread would eliminate it completely
\citep[see, e.g., the IC348 mass distribution in][]{Luhman03}.  
In order to keep our analysis simple, we maintain our
assumption of a single age, recognizing that the details of the mass
distributions in the groups are affected by this assumption.  For
our analysis, however, the detailed mass distribution is not
important; we are concerned primarily with the rank of the masses
(Sections 4.2-4.4), which is determined solely by the spectral type,
as well as other broad properties (this section).

\begin{figure}[h]
\plotone{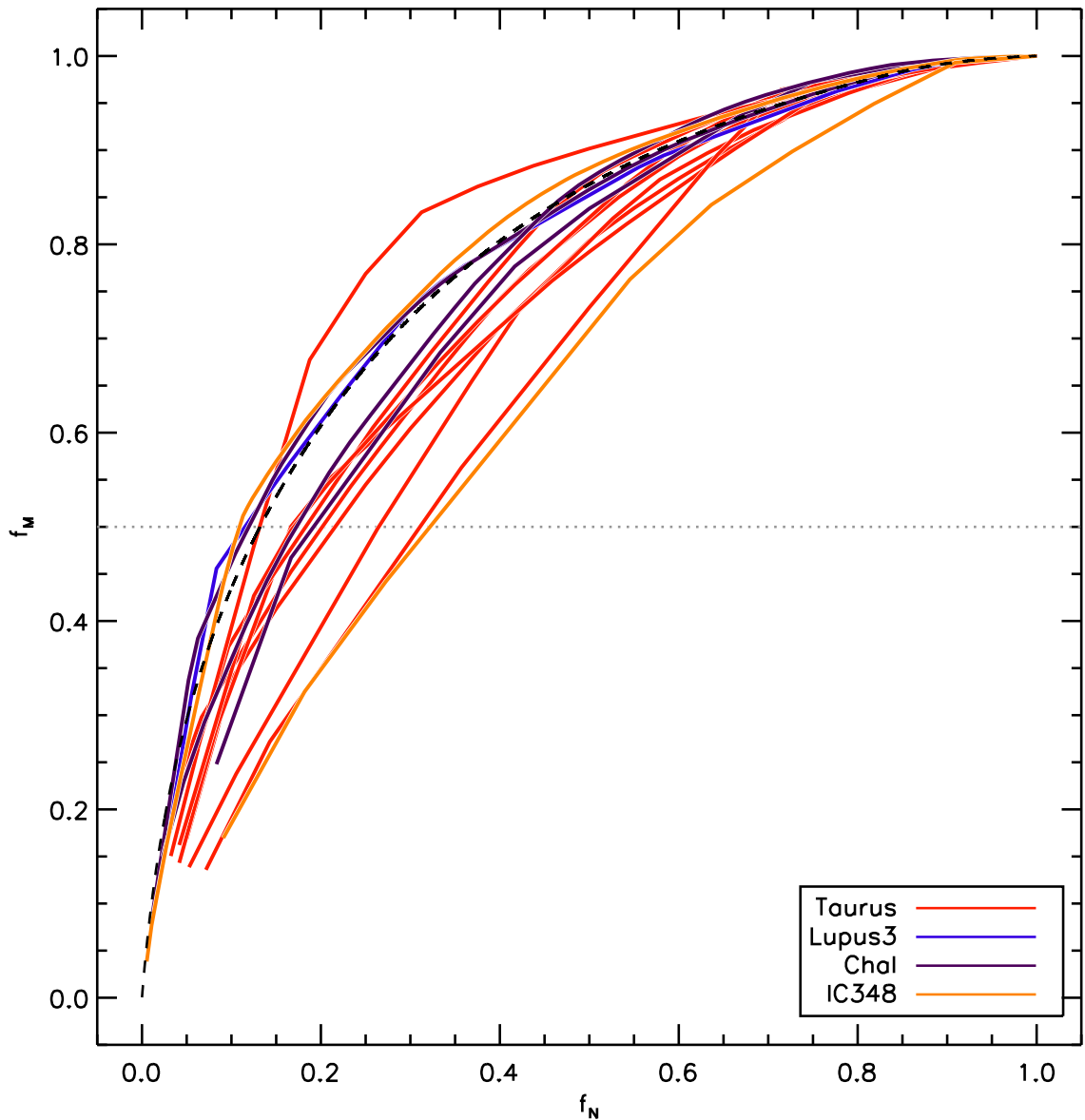}
\caption{The fraction, $f_M$, of the group mass in stars having mass
	greater than $M$ as a function of the fraction, $f_N$,
	of the number of stars having mass greater than $M$, as $M$
	decreases from the largest to smallest value for the group.  
	The coloured curves
	show the values for each observed group, while the dashed black
	line shows the values expected for the IMF.
	The grey dotted line indicates a fraction of 50\%.}
\label{fig_cuml_mass_num}
\end{figure}

Within each group, a significant fraction 
of the mass in concentrated in the
few most massive members.  This is illustrated in 
Figure~\ref{fig_cuml_mass_num} which shows the fraction of cumulative
mass, $f_M$, as a function of the fraction of cumulative number of
group members, $f_N$.  While there is variation between the
groups, some due to mass estimation uncertainties and some due to
small number statistics, the overall trend is clear -- half of the
mass of the group is found within the 10-30\% most massive members.
This property is similar to what would be expected for cluster whose
members follow the IMF.
The black dashed line in Figure~\ref{fig_cuml_mass_num}
shows the profile expected for the Kroupa IMF, using the formulation
given in \citet{Weidner10}, within the mass range spanned by our
groups.  For comparison, we also calculated the relationship 
expected for a symmetic, log-normal mass function, given, e.g.,
by \citet{Chabrier05}, and found a nearly identical relationship
to the one shown for the IMF.  Both describe the relationship
seen in the observed groups reasonably well.
It is notable that this property of a significant portion of the group
mass being found in a relatively small fraction of the group members
arises for groups whose most massive members have masses much greater
than the typical mass for the group, but much less than that of O stars
which dominate the largest clusters.

\begin{figure}
\plotone{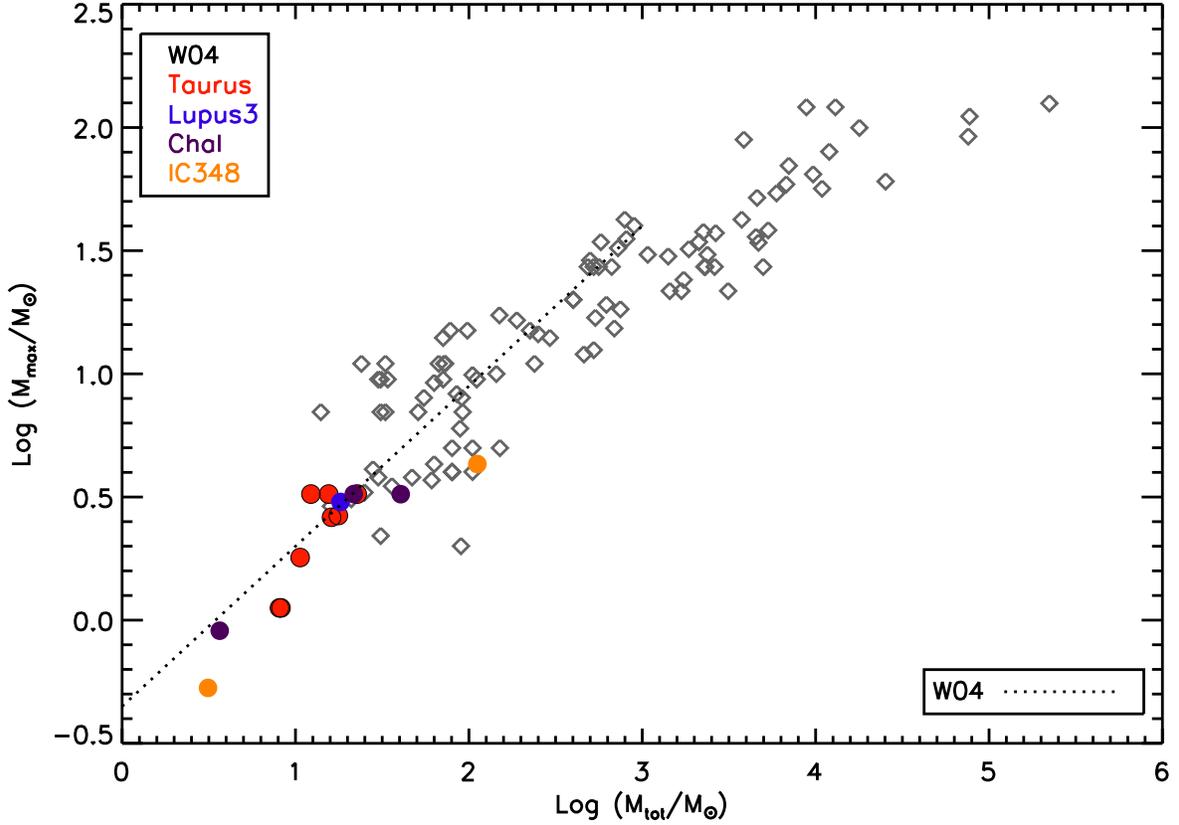}
\caption{The maximum mass member of each group versus the total mass in the
	group.  The black diamonds represent the data in \citet{Weidner10}
	(using the new dynamical mass estimates where appropriate), and the
	dotted line shows approximately the linear tail to the 
	\citet{Weidner10} relationship, {\hk assuming a Salpeter slope for
	the upper end of the IMF}.  Our groups fit the trend seen
	in higher mass clusters quite well.}
\label{fig_max_mass_vs_tot}
\end{figure}

Within clusters, the total cluster mass is correlated with either
the mass of the most massive member or the total number of members 
\citep{Weidner10}.
While the mass of the most
massive member of each group, and to a lesser
extent, the total mass in the group in our dataset, are uncertain, 
both should
be good to $\sim$50\% or better. 
We compare the mass of the most
massive member within each of our groups to the total group mass in 
Figure~\ref{fig_max_mass_vs_tot}.  In that figure, all of our groups
are plotted (note that in Taurus, several of the groups have nearly
identical values), along with the data in \citet{Weidner10}, using the
new dynamical masses for the most massive members where available.
\citet{Weidner10}
also derived an analytic expression for the most massive cluster member
expected, based on a random sampling of the IMF, and setting the maximum
mass possible for a star to be 150~\Msol; the dotted line in
Figure~\ref{fig_max_mass_vs_tot} shows the approximate linear slope 
found by \citet{Weidner10}, {\hk assuming a Salpeter slope for the
high mass tail of the IMF.  Regardless of the model fit to the
relationship between the most massive star and total cluster masses,
it is clear that our groups follow the same trend as the clusters do.}

%The consistency of the mass function of each group with the IMF resembles
%the finding of \citet{Massi06} for six clusters associated with IRAS 
%sources having luminosity of $> 10^3$~L$_{\odot}$ in the Vela D cloud.
%\citet{Massi06} use the K-band luminosity function and pre-main sequence tracks
%from \citet{PS99} to estimate masses.  When the mass functions of all groups
%in our sample are combined, we find a distinct deficit of massive
%stars compared to the IMF.  Sampling the IMF would predict several early B
%stars, whereas the most massive star in our sample is a late B star.
%\citet{Massi06} reached a similar conclusion for their clusters.
%%[HK thoughts -- This does appear to be roughly true for our sample; using
%%Weidner / Kroupa MF, expect 2 stars with masses $\ge$ 5.6~M$_{\odot}$, 
%%corresponding to a B3 star using Palla \& Stahler tracks.  This result does 
%%rely heavily on the PS models, however, and should be checked once 
%%D'Antona tracks are also tried.]

On the other hand, the correlation between the maximum stellar mass
and the total stellar mass for the groups studied here does not imply
that these groups have a particular upper limit on the stellar
mass for their model.  The groups do not extend to high enough masses to show a 
reduction in slope, unlike the largest of the \citet{Weidner10} clusters
shown in Figure~\ref{fig_max_mass_vs_tot} {\hk which suggested
the 150~M$_{\odot}$ maximum stellar mass.  At total cluster
masses below 100~M$_{\odot}$, \citet{Weidner10} found their
data to be consistent with a random sampling of the (Salpeter) IMF 
without the requirement of a maximum stellar mass.}

The foregoing properties of the mass distributions in
nearby groups suggest that similar physical processes are 
responsible for the mass distributions of stars in groups and in clusters, 
independent of their number of members.

\subsection{Location of the Most Massive Group Member}
\label{sec_mass_offset}
\begin{figure}[hp]
\plotone{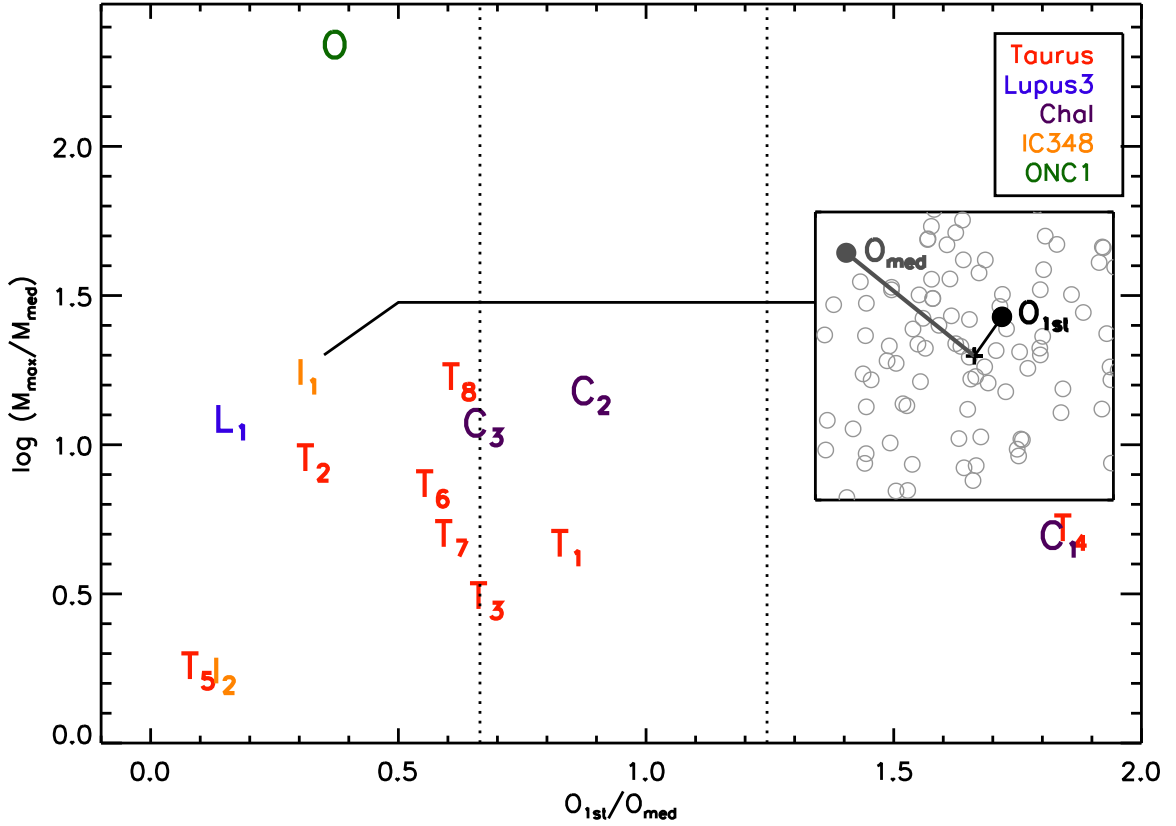}
\caption{Mass segregation observed in the groups.  The vertical axis shows
	the ratio in the mass of the most massive group member to the
	median group mass (an indication of how easily the most massive
	member is distinguishable), while the horizontal axis shows 
	the ratio in offsets from
	the cluster centre for the most massive member and the median
	value.  Coloured letters denote the various regions; the Trapezium
	cluster in Orion is also plotted for comparison.
	The vertical dotted lines indicate the 25th and 75th
	percentile values expected for a uniform random 
	sampling of group
	positions.  The inset shows the central part of the
	main IC348 group: the
	circles mark the positions of group members, while the
	plus indicates the group centre.  The offset of the most massive
	group member, $O_{1st}$, is shown in black, while the offset
	of the group member at the median separation, $O_{med}$ is
	shown in dark grey.  As shown in the main figure,
	$O_{1st} / O_{med} = 0.3$ in this group. 
	}
\label{fig_mass_posn}
\end{figure}

Within each group, the most massive group member tends
to lie close to the group centre.
We find that for the fourteen groups, the median offset 
(or separation from the group centre) of the most massive group member,
$O_{1st}$, 
is 0.6 times the median offset for all of the members of the same group,
$O_{med}$.
Figure~\ref{fig_mass_posn} shows the ratio of $O_{1st}$ to $O_{med}$
for all of the groups.  (Note that in the few groups where there were
two equal mass most massive members, the one closer to the centre is
used for the $O_{1st}$ calculation and the other is used for $O_{2nd}$
discussed below.)  The vertical
axis of Figure~\ref{fig_mass_posn} shows the ratio of the mass of the
most massive member, $M_{1st}$, and the median group member mass, 
$M_{med}$, indicating that
in most groups, regardless of the value of $M_{1st}/M_{med}$,
the most massive member of each group tends to
lie nearer to the group centre than typical.  For comparison, the 
ONC1 cluster is also shown
(see Section~\ref{sec_onc1}).  
These results are not sensitive to the method used to determine
the group centre -- we find $O_{1st}/O_{med}$ values which are
nearly always less than one using the centre
of mass as the group centre instead of the median position.

As might be naively expected, a random distribution of group
members tends to yield a ratio of offsets above and below one
with roughly equal probability.
We determine the ratio of offsets expected from a random uniform
distribution by running 200,000 simulations for a variety of numbers of YSOs
placed within a 2D circular or 3D spherical region.  We measured
the same ratio, $O_{1st}/O_{med}$, as described above.
The vertical dotted lines on Figure~\ref{fig_mass_posn}
show the 25th and 75th percentile value of the ratio found in the random
simulations for a uniform distribution of twenty five randomly distributed
YSOs within a 3D sphere.  These values do not change substantially
for either a 2D circular distribution or a different number of YSOs in the
sample.  We also ran similar tests on positions following a random {\hk 3D}
isothermal (probability $\propto 1/r^2$) profile and found similar
results.  The {\hk 3D isothermal}
distribution of $O_{1st}/O_{med}$ values tends to be more peaked
than the uniform distribution below $O_{1st}/O_{med} = 1$, and is shallower
with a long tail for values above 1 {\hk (the tail actually extends
beyond $O_{1st}/O_{med}$ = 100, much larger than the range shown in 
Figure~\ref{fig_mass_posn})}.
The 25th and 50th percentile values for $O_{1st}/O_{med}$ 
are nearly identical for both the isothermal and uniform distributions,
while the 75th percentile
value tends to be larger for the isothermal than the uniform 
distribution.  The 75th percentile value varies
with the level of truncatation the isothermal 
distribution ($O_{1st}/O_{med}$ is $\sim1.9$ in the absence of truncation
and decreases with the degree of truncation).
Regardless of the distribution adopted, 
the observed groups have a more centrally located
most massive member than would be expected from a random distribution
of positions.  

As a further test of whether these massive members are more centrally
located than would be expected in a random distribution, we
ran simulations where we kept the group members' positions
the same, but randomized which mass belonged to each member.
We did this for 10,000 trials for each group, and calculated
the fraction of simulations where the most massive member
had an offset ratio less than or equal to that which was
observed.  Where the mass segregation appears to extend
to the second (and third) group member(s) (as discussed
in the following subsection), we also computed
the joint likelihood of having the two (or three) members
with offset ratios smaller than or equal to those observed.
Barring L1551 and ChaI-Southwest, whose most massive members 
have large offset ratios, the probabilities found for the
observed group mass configurations were small -- at most, around
a percent, and often lower.  We also ran the same test on the
three {\it least massive} group members and nearly always
found substantially higher probabilities, since the
least massive members do not show a central concentration.

Regarding the two outlying groups, L1551 and ChaI-Southwest,
it is clear that neither has the appearance of typical groups.
ChaI-Southwest has a linear morphology and does not contain any
particularly massive stars (Figure~\ref{fig_MST_gallery3}), 
while L1551 (Figure~\ref{fig_MST_gallery2}) consists of a more 
typically-shaped group (bottom right in figure),
connected to the most massive group
members (top left in figure) through a series of relatively widely spaced
YSOs.  Had the critical MST branch length been 5\% smaller in L1551, the
most massive group members would not have been considered group members,
and the remaining group would show a much stronger central concentration
of the most massive members.  In most of the groups, the longest MST
branches only connect a few low mass YSOs to the rest of the
group, and the overall group structure is little affected with their
inclusion or exclusion, as discussed in more detail in 
Appendix~\ref{app_dendro}.  

While neither L1551 or ChaI-Southwest have
the appearance of traditional groups, this characteristic is
insufficient to explain the different behaviour between these two
groups and the others.  As can be seen in Figures~\ref{fig_MST_gallery1} 
through \ref{fig_MST_gallery4}, several other groups have a 
linear geometry (e.g., B213 / Taurus group 3) and loose spacings between
members (e.g., L1527 / Taurus group 7; see also Table~\ref{tab_groups} for
a comparison of median YSO spacings in each group)
while still possessing a centrally-located most massive member.

\subsection{Mass Segregation}
The central location of the most massive group member is
suggestive of a more general property of mass segregation, where
more massive members are progressively more centrally concentrated.
Within larger clusters, mass segregation is often observed
to varying extents.  For example, \citet{Stolte06} found
evidence of mass segregation at all masses in NGC~3603, with the degree of
segregation lessening in the lower mass bins, while \citet{Carpenter97}
found mass segregation only for the most massive members of the Monoceros
R2 cluster.
How do the groups which we analyze compare?

In clusters, mass segregation is measured in a variety of 
ways including the change in slope in the mass function (or luminosity
function) with radius, the ratio of the number of high- and low- mass
stars as a function of radius, and the mean radius of different masses
of stars \citep[see e.g.,][and references therein]{Bonnell98,Ascenso09}.
There are too few YSOs in each of the groups we identify to be able to
use any of the usual mass segregation measures.  We can, however,
extend the offset ratio measurement discussed in the previous 
section to look at the 
distribution of offsets from the group centre as a function
of mass rank. 

\begin{figure}[hp]
\plotone{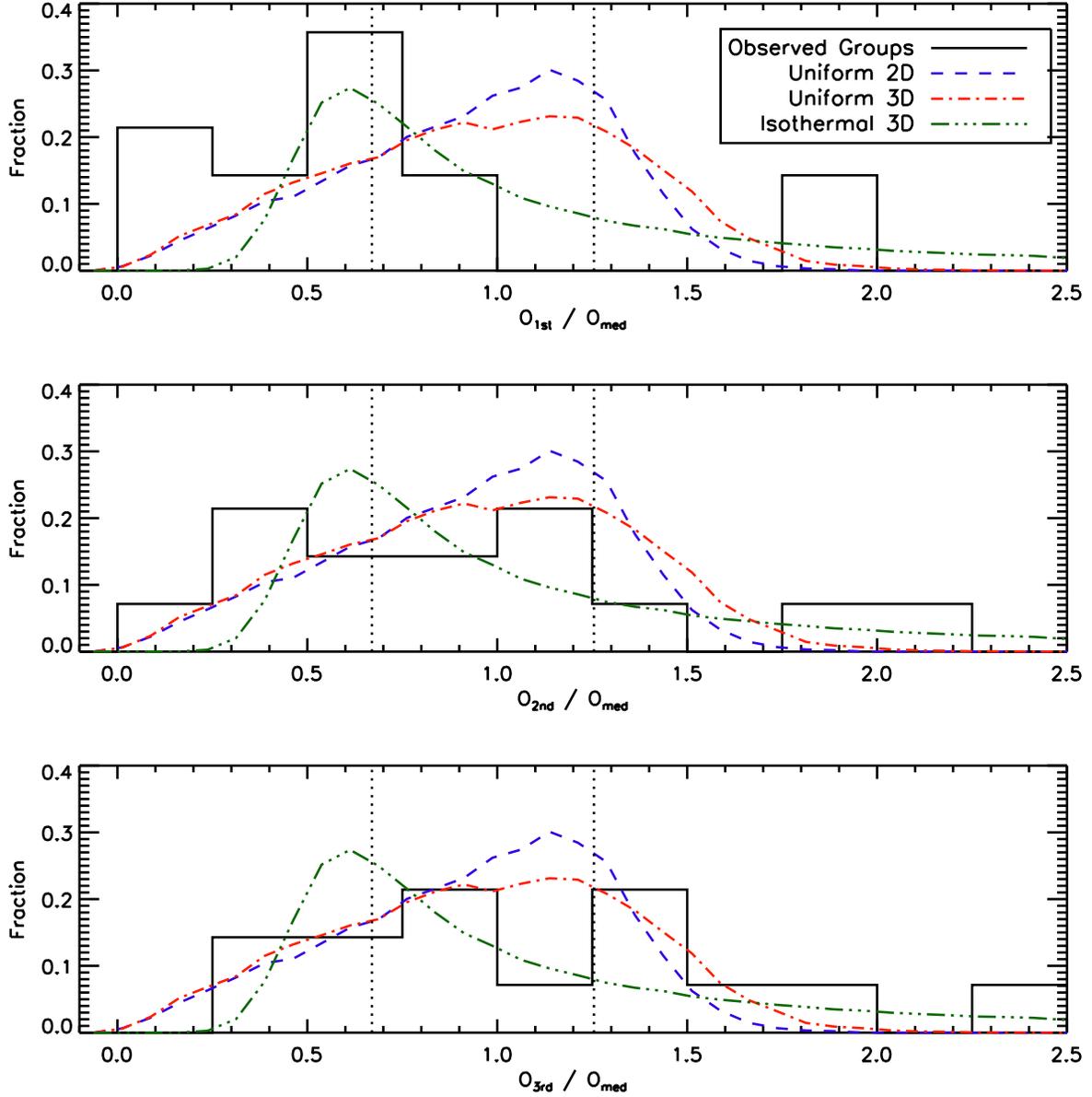}
%this is now stat_sample/output/compare_all_offsetratio.ps
\caption{The distribution of offset ratios for the most massive member 
	(top panel), second most massive (middle panel), and
	third most massive (bottom panel) member of each group.  The
	blue dashed line shows the values expected for group members randomly
	distributed uniformly over a 2D circular area for 
	25 group members, while
	the red dash-dotted line shows the same for 25 group members
	randomly distributed uniformly over a 3D spherical volume,
	{\hk and the green dash-triple-dotted line shows the same for
	25 group members in a random isothermal distribution over a
	3D spherical volume.}  
	The vertical
	dotted lines show the 25th and 75th percentile values for the 2D
	{\hk uniform} distribution, as shown in Figure~\ref{fig_mass_posn}.
	}
\label{fig_offset_hists}
\end{figure}

Figure~\ref{fig_offset_hists} shows the distribution of offset
ratios for each of the three most massive members in the groups
(top to bottom panels).  The overplotted dashed, dash-dotted,
{\hk and dash-triple-dotted}
lines show the distributions expected from our random simulations
discussed in the previous section.  
The most massive group member (top panel) shows a large excess 
of sources with
offset ratios smaller than expected from random sampling, but this
excess is diminished considerably for the second most massive
member (middle panel), and is gone completely for the third most massive 
member (bottom panel).

A simple comparison between the distribution of observed
offset ratios and those found in our random 3D simulations with a
two-sided KS test give probabilities of being drawn from the same
parent sample of less that 0.2\% for the most massive group member,
and 62\% and 50\% for the second and third most massive members 
respectively {\hk for the uniform distribution, and probabilities
of 0.1\%, 47\%, and 56\% for the non-truncated isothermal distribution,
with typically smaller probabilities for truncated isothermal distributions}.
Our largest source of error in the offset ratios is due to the
definition of the centre of the groups.  Using instead the centre
of mass of each group to compute the offset ratios, we again find
similar distributions.   {\hk The} two-sided KS test probabilities 
{\hk for the 3D uniform distribution} are
0.2\%, 22\% and 70\% for the
first, second, and third most massive members respectively,
{\hk and 0.2\%, 10\%, and 8\% for the 3D isothermal distribution.
The large change in the probabilities for the 
isothermal distribution is caused by a smaller minimum value of
$O_{xth}/O_{med}$ for the observed groups using the centre of mass (instead
of the median position).  $O_{xth}/O_{med}$ decreases sharply for the
3D isothermal distribution below 0.5, so the KS test is
highly sensitive to how low the observed $O_{xth}/O_{med}$ values
extend.}

Figures~\ref{fig_mass_offset_frac1} through \ref{fig_mass_offset_frac3}
{\it (available only online)} illustrate the trend towards less-central
locations with lower mass in an alternate manner, showing the mass of
each source versus the order of the offsets in the group,
visually confirming the statistic measures discussed above.  It is
interesting to note that the central location of the most massive
member(s) in each group is not a function of the mass of the most
massive member.  In the
ONC1 cluster, \citet{Hillenbrand98} found evidence for mass segregation
extending down to 5~\Msol, and possibly beyond.  Visually, a similar
divide is seen in Figure~\ref{fig_mass_offset_frac3} for ONC1. 
Quantitatively, the offset ratio $O_{1st}/O_{med}$
is less than 1.1 for all members with masses above 4.8~\Msol;
below this mass, the maximum of $O_{1st}/O_{med}$ rapidly increases.

\subsection{Massive YSOs and Clustering}
Another property which our groups share with larger clusters
is a central concentration.  Large clusters tend to have much smaller
separations between members towards the centre, coincident
with where the most massive members tend to be located.
While Figures~\ref{fig_MST_gallery1} through \ref{fig_MST_gallery4}
suggest that the most massive group members lie in or near zones with
higher than average degrees of clustering within the groups, 
we can quantify this.

\begin{figure}[hp]
\plotone{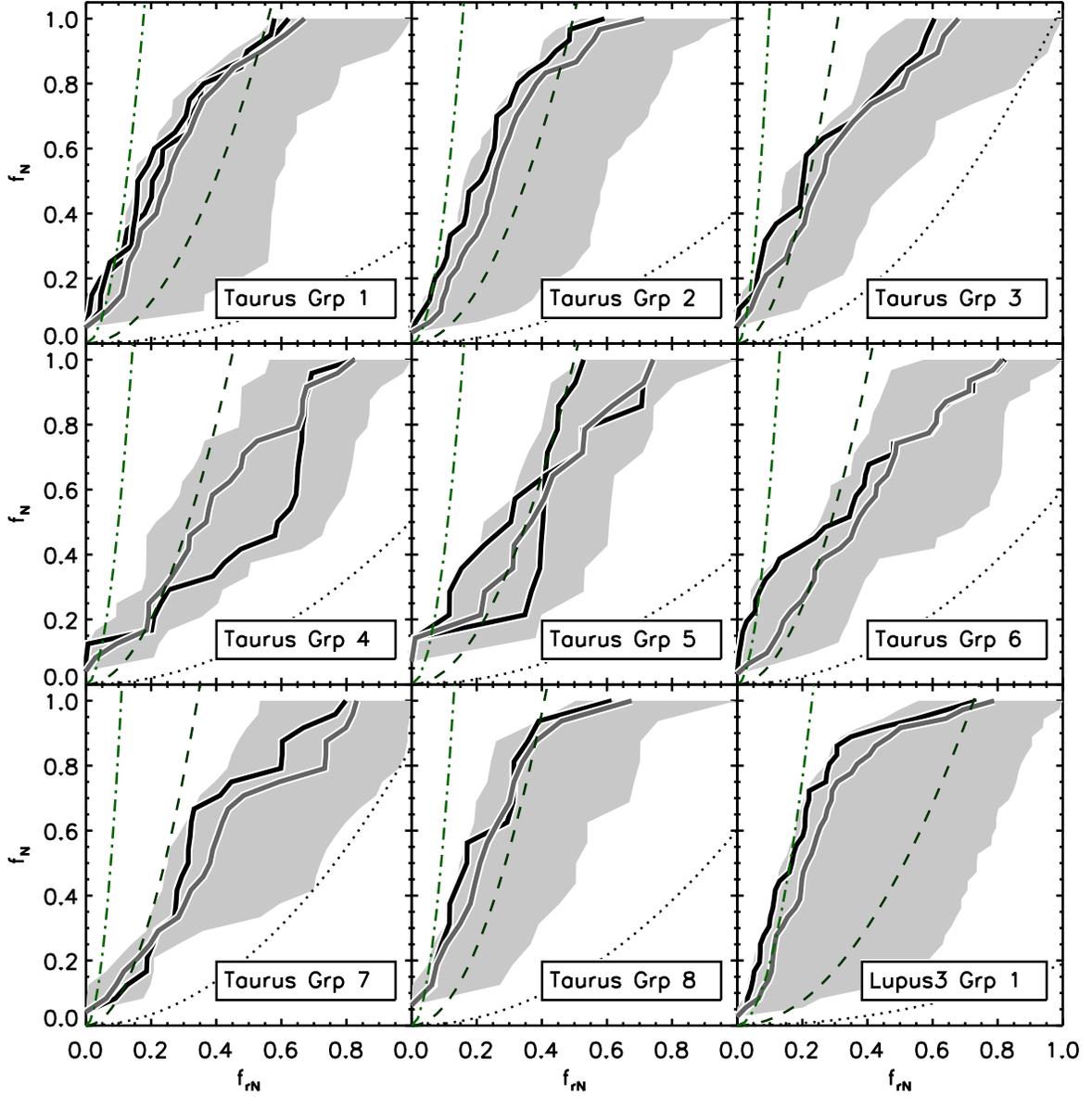}
\caption{The fraction of the total number of group members, $f_N$ versus
	$f_{rN}$, the radius enclosing the N nearest sources from each
	group member normalized by the maximum value of the 
	enclosing radius in the group. 
	The shaded grey region indicates the range of values
	spanned by all group members.  The black line shows the
	value for the most massive group member, while the 
	grey line shows the median value for all group members.
	Group members
	located in a more clustered environment will show a
	steeper rise in number at low separations compared to
	members located in more isolated parts of the group.
	Lines of constant surface density are shown in green,
	with values of (from dark to light): 1 (dotted), 10
	(dashed), and 100 (dash-dotted)~pc$^{-2}$.
	The groups in Taurus and Lupus3 are shown in this figure;
	the remaining groups are shown in the following figure.}
\label{fig_cuml_rad_sample}
\end{figure}

\begin{figure}[h]
\plotone{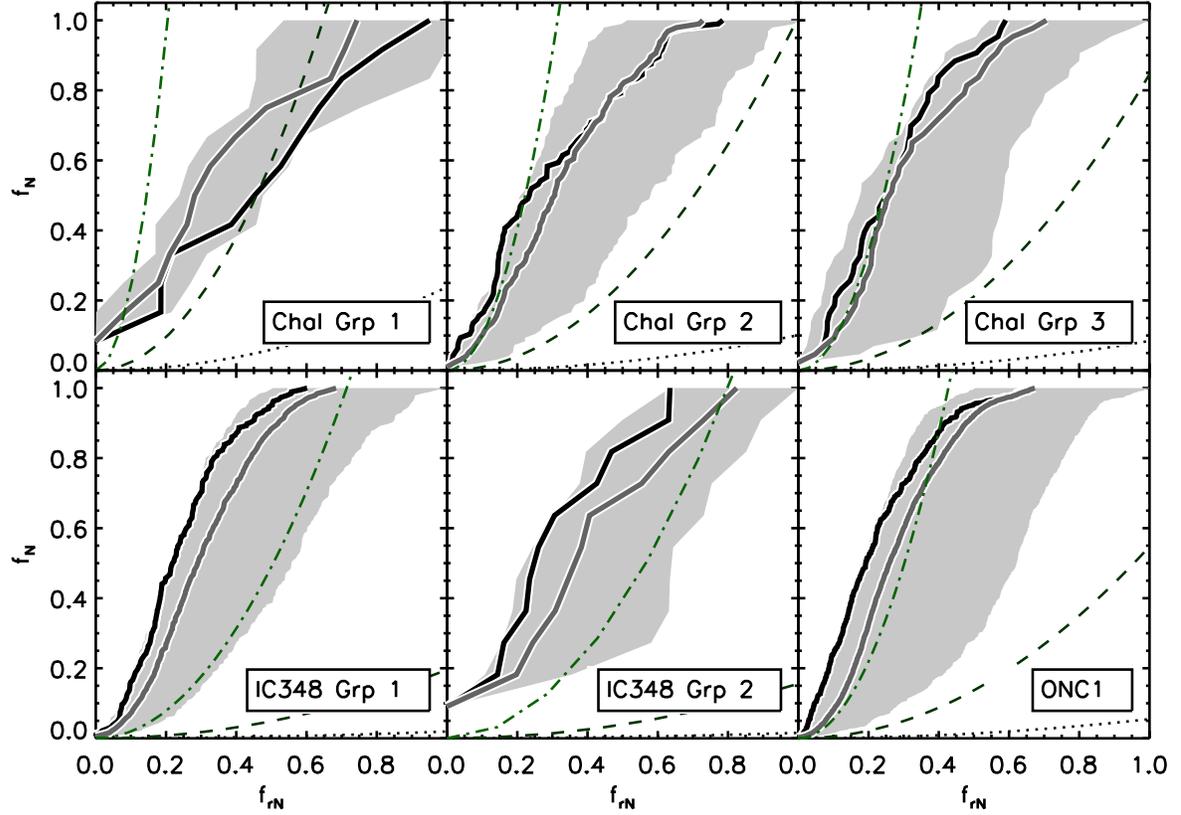}
\caption{The fraction of the total number of group members, $f_N$ versus
        $f_{rN}$, the radius enclosing the N nearest sources from each
        group member normalized by the maximum value of the
	enclosing radius in the group
	for groups in ChaI (top row), IC348 (bottom row, left and
	middle), and the ONC1 cluster (bottom right) for comparison.
	See Figure~\ref{fig_cuml_rad_sample} for the plotting conventions
	used.}
\label{fig_cuml_rad_sample2}
\end{figure}

The measurement of the surface density of stars often
provides a useful criterion for determining the clustering
properties.  Since our groups often have a very small number
of sources, these measures are more vulnerable to errors from
small number statistics, and so we follow a different approach;
surface density measures are discussed briefly in Section~5.1.
Here, we instead compute the radius, $r_N$, which
encloses the N nearest sources for each group member.  Group members
which are located near the most clustered part of the group
should show a sharper rise in a plot of $N$ versus $r_N$ 
than members located in a sparser part
of the group.  
Figures~\ref{fig_cuml_rad_sample} and \ref{fig_cuml_rad_sample2} 
show the fraction of total group members, $f_N$ versus 
$f_{rN}$, the radius $r_N$ normalized to its maximum value
in the group
for all members of each group.
For clarity, all group member profiles are not shown
in the figures.  Instead, the range in values 
spanned by all members is indicated
by the light grey shading, while the median profile is overlaid
as the dark grey line.  The most massive group member is shown 
in black.
These plots allow comparison of the local surface density for a 
given radius $r_N$, or for a given number of neighbours N, as a function
of member mass. 

Figures~\ref{fig_cuml_rad_sample} and \ref{fig_cuml_rad_sample2}
show that the most massive star is projected on
a region of relatively high local surface density for a large fraction
of the groups considered.  The black line lies to the left of the
grey line for the nine groups Taurus~1, 2, 3, and 6, Lupus3~1, ChaI~2 and 3,
and IC348~1 and 2, and also for ONC1, the prototype of a large cluster with
mass segregation.  In contrast, the black line lies to the right of the
grey line for two groups, Taurus~4 and ChaI~1.  In the three
groups Taurus~5, 7, and 8, the lines are too close to make a clear
discrimination.

In combination with the result from Section~4.2, this result indicates
that the most massive star in a group generally has a position which is 
central, and which is associated with a high surface density of lower-mass
stars.

%{\hk In Figures~\ref{fig_cuml_rad_sample} and \ref{fig_cuml_rad_sample2},
%the} most massive group member usually lies
%on or near the upper {\hk left} boundary of the range spanned in the
%group for moderate cumulative numbers, indicating that they reside 
%in or near the denser part of the group.  

\begin{figure}[t]
\plottwo{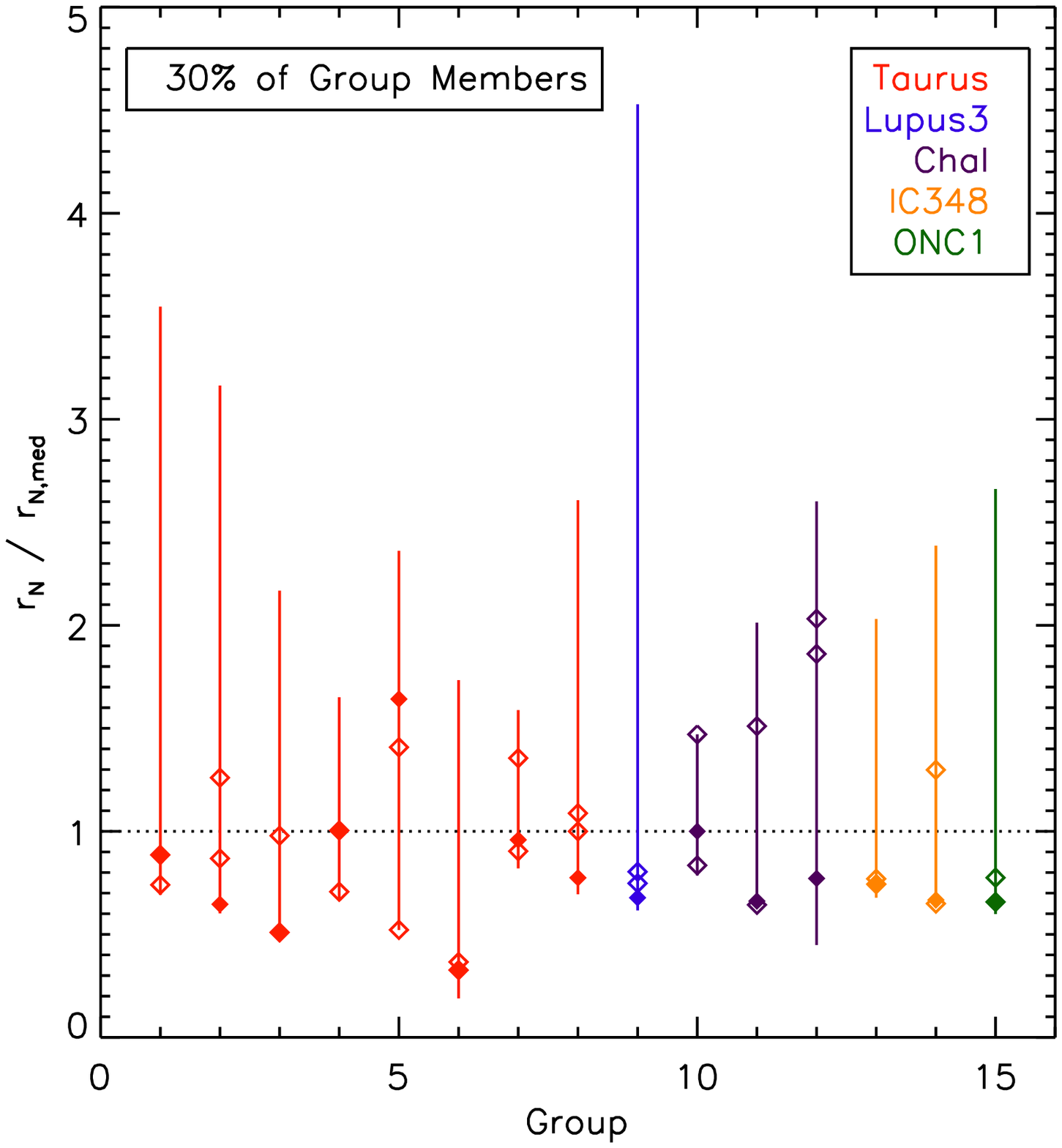}{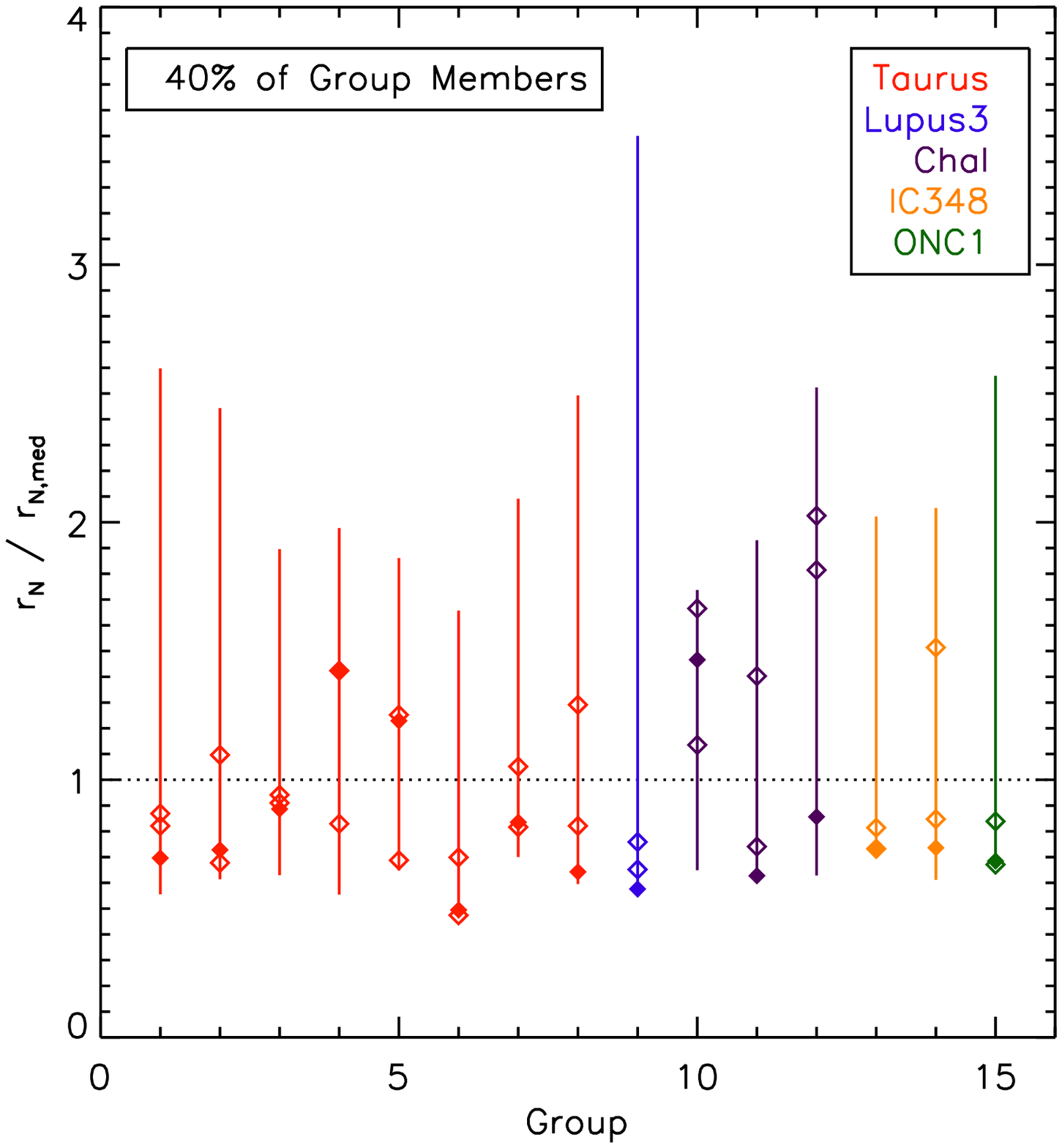}
\caption{
	A comparison of the local surface density of YSOs around
	the most massive group member relative to typical values in
	its group.  Each vertical line represents the data taken along
	a horizontal cut of Figures~\ref{fig_cuml_rad_sample} and
	\ref{fig_cuml_rad_sample2} for each group at $f_N = 30\%$
	(left panel) and 40\% (right panel).  Values in this figure
	are normalized to the median value of $r_N$ for each group
	at $f_N = 30\%$ and 40\%.  The vertical lines show the
	range of (normalized) $r_N$ 
	(light grey shading in the previous figures) for the 
	specified values of $f_N$,
	and the solid diamond shows the (normalized) $r_N$ for 
	the most massive group member (black line in the previous 
	figures).  The open diamonds show the (normalized) $r_N$ for 
	the second and third most massive group members.  Members with 
	(normalized) $r_N$ values below 1 lie in the most 
	clustered parts of their group.
}
%	The ratio of the radial separation required to 
%	enclose 30\% (left) and 40\% (right) of the total number of
%	group members for the most massive member ($r_{N,1st}$)
%	to the median value ($r_{N,med}$) are shown as filled diamonds.
%	The open diamonds show a similar ratio for the two next most
%	massive members.
%	The vertical lines show the total
%	spread in the values for the group, while the colours denote the
%	different star-forming regions.  The dotted horizontal line
%	indicates the median value for the group.}
\label{fig_allseps}
\end{figure}

The tendency for the most massive member of each group to lie
in a region of higher than average group member surface density
can be quantified further.  As already discussed, 
the most massive group member tends to have a profile which lies 
to the left of the median profile in the preceeding two figures.  
Put another way, for a fixed value of $f_N$,
either $f_{rN}$ or $r_N$ are smaller for the most massive member than the
median value.  For each group, we compare $r_N$ for the
most massive member and the median value at fixed levels of
$f_N$.  Figure~\ref{fig_allseps} shows $r_N$ for the
most massive member divided by the median value for $f_N = 30\%$
(left panel) and 40\% (right panel) for each of the fourteen
groups (solid diamonds).  For each group, the full range in range 
$r_N$ normalized by the median value at $f_N$ is indicated by the vertical
line.  Each vertical strip in Figure~\ref{fig_allseps} 
therefore summarizes a horizontal cut along $f_N = 30\%$ (left) and 
40\% (right) in Figures~\ref{fig_cuml_rad_sample} and 
\ref{fig_cuml_rad_sample2}, normalized
to the median $r_N$ at that $f_N$ (i.e., the dark grey line would
always lie at 1 in Figure~\ref{fig_allseps}). 

Plotted in this manner, it is easy to additionally compare the
behaviour of the most massive member of each group with the next
two most massive members, shown in Figure~\ref{fig_allseps} by the
open diamonds.
There is some scatter between the two panels, which is expected
since the most massive member and median profiles shown in 
Figures~\ref{fig_cuml_rad_sample} and \ref{fig_cuml_rad_sample2}
do not remain a fixed distance apart.  The two panels shown in
Figure~\ref{fig_allseps} are, however, representative of the
general trend -- we made similar figures for 
values of $f_N$ ranging from 25 to 50\% in intervals of 5\% and found
similar results to the cases shown.
Despite the scatter, it is easy to 
see from the plot that most of the most massive members (filled
diamonds) lie near or below a value of 1 in most cases, i.e., they lie in
locations which tend to be more clustered than typical for the group.
The second and third most massive members only shown in 
Figure~\ref{fig_allseps} follow this trend in only a limited number of
groups.

%We {\hk further} quantify the tendency of the most massive star to lie in the
%densest part of the group 
%through
%a comparison of the most massive member's profile with the
%median profile also shown in Figures~\ref{fig_cuml_rad_sample} and
%\ref{fig_cuml_rad_sample2}.  Between 25\% and 50\% of the total number
%of group members, $N_{tot}$, in intervals of 5\%, we compared $r_N$ for
%every group member, and found all values tend to increase at
%approximately the same rate. 
%Figure~\ref{fig_allseps} shows the comparison for 30\% (left panel) and 
%40\% (right panel) of 
%$N_{tot}$.  Plotted is the ratio of $r_N$ for the most massive
%group member to the median value in solid diamonds.  A similar ratio
%for the second and third most massive group members are indicated
%by the open diamonds, while the vertical lines show the range of
%values spanned by all group members.  Group members with a ratio
%of less than one tend to lie in a more clustered environment,
%while members with ratios greater than one tend to lie in a
%more isolated environment.  There is some scatter in the ratios
%at 30\% and 40\% of $N_{tot}$, but the main trend, also visible
%in Figures~\ref{fig_cuml_rad_sample} and \ref{fig_cuml_rad_sample2},
%is that in all groups except L1551 and ChaI-Southwest (where
%the most massive group member is also not centrally-located),
%the most massive group member lies in or near the most clustered
%part of the group.
%In this respect, the groups again have similar properties to 
%those observed in large clusters.

\section{DISCUSSION}
\subsection{Number and Surface Density}
The groups that we analyze tend to have properties scaled-down,
but similar to those in larger clusters.  In the literature, clusters
are often defined and described using criteria based on the
surface or volume density of sources.  \citet{Lada03} defined
clusters as having a minimum stellar density of 1~\Msol~pc$^{-3}$
and greater than 35 members, in order to be resistant to quick
dissolution.  \citet{Jorgensen08} found that protostars in the
nearby Perseus and Ophiuchus molecular clouds tended to have 
denser clustered substructures and used a threshold of 1~\Msol~pc$^{-3}$ to
define loose associations and 25~\Msol~pc$^{-3}$ to define
tight associations within these.  A minimum number of 35 members
was again used to define a cluster; associations with smaller
numbers were termed groups.  
\citet{Porras03} similarly used a maximum of 30 members to define
groups, with 31 to 100 members corresponding to small clusters and
over 100 members corresponding to large clusters.
\citet{Adams01} examined the
transition between groups and clusters in more detail, and
argued that groups containing roughly ten to one hundred members
evolve differently than large clusters -- the groups tend
to disperse quickly and are likely to be unaffected
by supernovae or strong UV radiation
\citep[note the number of members is higher than in]
[as the role of gas in the cluster dissipation was also considered]{Lada03}.

How do our groups compare?  In terms of numbers of members, all but
the main group in IC348 (with 186 members) fall easily within the
\citet{Adams01} definition of a group, and many 
also do so with the N=30-35 definition of \citet{Lada03}, \citet{Jorgensen08},
and \citet{Porras03}.
In terms of surface density, our groups tend to also lie below the
standard cluster values.

Figures~\ref{fig_cuml_rad_sample} and \ref{fig_cuml_rad_sample2}
show lines of constant surface density from 1 to 100~pc$^{-2}$
separated by factors of ten.  In the Taurus and Lupus3 groups,
values tend to range from a minimum of around 1~pc$^{-2}$ to a
few times 10~pc$^{-2}$.  The ChaI groups tend to have slightly
larger surface densities (particularly ChaI-Southwest / Group 1), while
the IC348 groups have surface densities nearly 100 times larger.
%We calculated surface density maps for all of our groups,
%using the  
%distance to the Nth nearest source (e.g., if the second nearest source
%is 1~pc away, the surface density is 1 per $\pi$~pc$^{-2}$).  
%Use of a higher number of neighbours increases
%the accuracy (the effect of chance alignments is diminished),
%with the fractional uncertainty being (N-2)$^{-0.5}$ \citep{Casertano85},
%but extending to a higher number of neighbours also decreases 
%the `resolution' of the resulting surface density map.
%\citet{Gutermuth09} found a reasonable balance between
%uncertainty and spatial smoothing using the 6th nearest neighbour.
%
%Using N=6, we find that most of our groups have peak surface
%densities of 10 to 50~pc$^{-2}$, falling off to around a few tenths to
%one~pc$^{-2}$ at the group edge; the groups in IC348 are a factor of
%roughly twenty higher.  Similar surface densities are found if
%the total number of group members are averaged over the area spanned
%by the group (taking into account the non-circular geometry in some
%groups).  
%HK note -- these values are in filen/analysis/MST/output/group_area.txt
% For Lupus3, look only at the circular area, as the pixel size for the
% ext map is so small to make the other measure not comparable with the
% other groups
The embedded clusters in the sample of 
\citet{Gutermuth09} tend to have much higher surface densities,
typically peaking around a few hundred per square parsec.
Assuming, as in \citet{Jorgensen08}, spherical symmetry and 
a typical YSO mass of 0.5~\Msol, then the volume density thresholds
of 1 and 25~\Msol~pc$^{-3}$ correspond to roughly twice those
values in number per square parsec for surface density.
The groups in this paper are denser than their surroundings, 
but are generally much
less dense than those considered in any of the above-cited works.

\subsection{Predictions of Mass Segregation}
In most of the groups, we find the single (or, for L1356, Lupus3-main,
and IC348-main, several) most massive
group member(s) are located near the group centre, and near a region
of higher than average surface density of sources.
Mass segregation is often observed in large clusters, with
some regions showing evidence of segregation at all masses,
with the degree of segregation lessening in the lower mass
bins \citep[e.g.][]{Stolte06}, while other regions appear to
show mass segregation only for the most massive members 
\citep[e.g.,][]{Carpenter97}.
One complication in mass segregation measurements in clusters,
particularly more distant clusters, is the observational
bias due to crowding.  \citet{Ascenso09} argue that this
bias, which leads to increasing levels of incompleteness in the
lower mass objects at smaller cluster radii, may be partially
or even fully responsible for the observed mass segregation
in clusters.  The central location of the most massive cluster
members would appear to be robust, however, since these stars
would be easily detectable in the outskirts of the clusters
as well. 

The young age of some of these
clusters, coupled with the degree to which the most massive
cluster members are concentrated in the centre has led
to the argument that at least some of the observed mass segregation is
primordial, rather than dynamical \citep[e.g.][]{Bonnell98}.
More recently, \citet{Moeckel09} have argued that the primordial
mass segregation must be confined to only the most massive
stars, as any initial amount of mass segregation in the lower-mass
population leads to an over-prediction of the mass segregation
that should be currently observable in those clusters.
Other work has questioned whether any primordial mass segregation
is required \citep{Allison09} to match present-day observations
of clusters. 

In the groups we study, the mass segregation observed for the
most massive members appears to require an early central 
concentration of these objects.  The groups have ages around
1~Myr, while the crossing times are typically between 1-3~Myr, assuming
a velocity dispersion of 1~km~s$^{-1}$ \citep[the approximate value
found in the proper motion groups in Taurus in][]{Luhman09}.  Since there
are too few group members for the scenario simulated in \citet{Allison09}
to be applicable, it seems that the most massive members of 
the groups must have formed near the group centres,
rather than migrating there later.   
For the majority of their lifetime, the YSOs have been
embedded in their natal gas, hence the timescale for dynamical 
evolution can be estimated as
\begin{equation}
t_{relax} = \frac{N\epsilon^{-2}}{10 ln(N/\epsilon)} t_{cross}
\end{equation}
where $N$ is the number of stars, $\epsilon$ is the star formation
efficiency of the group, and $t_{cross}$ is the crossing time
of the group \citep{Adams01}.  For a star formation efficiency value 
of 10\%, the relaxation timescale is nearly five times the crossing
time for the smallest groups, and becomes larger for both larger groups
and lower star formation efficienies; {\hk the relaxation timescale
well exceeds 5~Myr for the groups under any reasonable assumption
of the star formation efficiency}.  Given the ages estimated for the 
groups, the central locations of the most massive members cannot be
wholly attributed to dynamical relaxation.

Some degree of early mass segregation appears to be
consistent with models for massive star formation.
In the competitive accretion scenario, protostars forming in the
centre of the cluster potential inhabit a higher density environment
and hence accrete more mass than those formed at the cluster
periphery \citep[e.g.,][]{Bonnell01}.
In the monolithic collapse scenario \citep{McKee03}, stars form
out of quasi-equilibrium clumps.  Clusters require several
times the dynamical timescale to form, which enables mass
segregation to occur in a greater amount than would be
anticipated from a faster formation scenario \citep{Tan06}.
In the stationary accretion model of \citet{Myers09}, high mass stars
are only able form in the densest environments, where
the accretion rate is highest, while lower mass stars
are able to form in the surrounding higher and lower density environments.
It is unclear, however, whether these models can predict the 
clear transition between the location of the most massive
few group members and the other group members, or, in some
cases, operate at all for such small groups.

Association of the most massive star with a high surface
density of lower-mass stars tends to rule out a very simple 
cluster formation scenario, where the massive star accretes all the mass
within a certain radius, causing a local minimum in the density of
lower-mass stars.  Instead, it may be more realistic that a spherical
zone around an accreting massive star feeds the massive star as well
as numerous lower-mass stars \citep{Smith09,Wang10}.

In their Herbig Ae/Be survey, \citet{Testi99} found instances
where the massive star appeared to be relatively isolated,
as did \citet{Massi03} in their
survey of luminous IRAS sources in the Vela C and D molecular clouds.
We also find a few instances of isolated massive stars 
in our catalogs --
one A2 star in Taurus, one B6.5 and one F0 star in ChaI, 
one B4 star in Lupus3, and several A stars in IC348 which
fall near but outside of the main group.  These sources 
represent a much smaller fraction of isolated early-type 
sources than were found in \citet{Testi99}'s work, 
suggesting that isolated A and B stars are uncommon
in young star-forming regions.  Future observations may
lower this fraction even further, either with evidence
that these apparently isolated stars are interlopers or the discovery of
more lower mass members nearby.  The isolated early-type star in Lupus3,
for example, lies well outside the main group, where there
have likely been fewer surveys for low mass members
in the region.

\section{CONCLUSION}
We present a study of groups of young stars within four nearby
star-forming regions -- Taurus, Lupus3, ChaI, and IC348.  The census
of stars within each of these regions is complete down to very low
masses, typically late-M, corresponding to $\sim$0.02~\Msol.  
YSO masses are estimated from spectral
types following \citet{Luhman03}.  Using a minimal spanning tree
algorithm and following the procedure of \citet{Gutermuth09}, we
identify fourteen groups of YSOs within the regions, with the total
number of members ranging from 11 to 186, with most in the 
range of 20 to 40.  The total number and surface density of
group members tends to be smaller than in clusters by a factor of five to 
ten, or more.  The groups are sufficiently young
that their configurations should be similar to their primordial
configuration.

Within these groups, we find the following:
\begin{enumerate}

\item The groups have a wide range in masses; the maximum mass member
is typically more than five times the median member mass.
\item The maximum member mass and total group mass are correlated
and follow a similar relationship to that seen in clusters.  
%{\hk The total group masses are, however, not high enough to probe
%the regime where \citet{Weidner10} argue observations show a maximum
%stellar mass is required.}
\item Most of the mass in each group is found in a small fraction
of the group members.
\item In most groups, the most massive star tends to be 
centrally located.  In a few groups, this property extends to the
second- or third- most massive star.
\item In most groups, the most massive star is associated with a
relatively high surface density of lower-mass stars.
\item The central concentration of massive stars is much more than
expected for a random distribution YSOs.
\item The central concentration of massive stars occurs even
if the most massive star is only 1~\Msol.
\end{enumerate}
%I didn't think it looks like it depends on the relative mass
%ratio either...}

%Our analyses indicate that the few most massive members within
%each group are distinct from the rest of the group -- they tend
%to be located close to the group's centre and in or near the
%most clustered part of the group.
%This is the case even when the absolute mass off these most 
%massive group members is not large and does not appear to be a function
%of absolute mass.  The regions tend to
%have ages similar to a single crossing time, suggesting
%that these group properties are at least in part primordial.
Due to the proximity and sensitivity of the coverage of these 
star-forming regions, our
analysis does not suffer from the problems of crowding and variable
completeness which may affect more distant clusters \citep[e.g.,][]{Ascenso09}.
The similarity in the properties of these small groupings of stars 
therefore offers a complementary avenue to explore the some of 
the processes which influence massive cluster-forming regions
which are more distant.

\appendix

\section{ADOPTED SOURCE CATALOGS}
\label{app_catalogs}

\subsection{Taurus}
We analyze the 352 Taurus members discussed in \citet{Luhman09} and 
given in Table~7 of \citet[][hereafter L10]{Luhman10}.
For the binary pair HD 28867A+C and B, we adopt the positions 
given in \citet{Walter03}; the \citetalias{Luhman10} catalog names
correspond to identical positions for both sources.
Where the data exists, members were confirmed using proper
motion data, as discussed in the appendix of
\citet{Luhman09}.  Otherwise, membership was based on a variety
of observations including Ca II emission, H~$\alpha$ emission, 
X-ray data, spectral energy distributions, and / or spectra
from optical through far-IR observations \citep[e.g.,][]{Kenyon08}.
Table~\ref{tab_Taurus} {\it (available online)} summarizes the data --
the position, \citetalias{Luhman10} name(s), spectral type,
estimated mass (Section~\ref{sec_mass_est} and Appendix C), 
and the group (Section~\ref{sec_mst}).
Some of the spectral types are highly uncertain, and were estimated
based on the bolometric luminosity of the source, as
indicated by footnotes in Table~\ref{tab_Taurus}.  In these cases, 
the spectral type and mass should be treated as being in the 
range given in the footnote.

The {\it Spitzer} Taurus team has also published a full catalog
of YSOs in Taurus, including both {\it Spitzer} photometry of previously
known members and new and candidate members based on {\it Spitzer}
photometry, and in many cases, additional follow-up spectroscopy
\citep[][hereafter R10]{Rebull10}.  For the analysis discussed in this paper, we
use the \citetalias{Luhman10} catalog because it spans a larger area
of the cloud -- the \citetalias{Rebull10} catalog covers only 
the region mapped by {\it Spitzer}.  Our results are largely unchanged when
using the \citetalias{Rebull10} catalog instead, as described in more
detail in Appendix~\ref{app_rebull}.

We adopt a distance of 140~pc to the Taurus cloud, following
\citet{Torres07}.

\subsection{ChaI}
In ChaI, we analyze 237 sources whose properties are summarized in
Table~\ref{tab_ChaI} (available online; the same columns are used as in
Table~\ref{tab_Taurus}).  This list
includes the 226 known members of ChaI discussed in \citet{Luhman07};
215 of these sources are given in Table~6 of \citet{Luhman07}, 
while the remaining 11 were
excluded because they lacked accurate spectral types.  These 11
sources are J11094192-7634584 and J11095505-7632409 from Table~5 of
\citet{Luhman07}, J11011926-7732383B from Table~1 of \citet{Luhman04b}, 
and Ced110-IRS4, ISO86, Ced110-IRS6, ISO97, B35, IRN, ISO192, and ISO209
from Table~5 of \citet{Luhman04a}.  We also add the 8 new members 
identified in Table~1 of \citet{LuhmanMuench08} and 4 new members
identified in Table~4 of \citet{Luhman08}.  Following the discussion in
\citet{Luhman08}, ISO130 was excluded, as it is likely a galaxy,
and sources J11183572-7935548, J11334926-7618399, J11404967-7459394,
and J11432669-7804454 were removed as their proper motions indicate
they are more likely to be members of $\eta$ Cha than ChaI.
Additionally, four new members were added : RXJ1129.2-7546, RXJ1108.8-7519a,
RXJ1108.8-7519b, and Cha-MMS1, for which proper motion measurements
indicate that they are likely ChaI members.
As with the Taurus catalog, proper motion data was used where available to
verify membership, otherwise, a variety of indicators of youth were used.

Following the discussion in \citet{Luhman_handbook}, we adopt a distance
of 160~pc to the region.

\subsection{Lupus3}
In Lupus3, we analyze 70 YSOs (Table~\ref{tab_Lupus3}, available online)
from the compilation of \citet{Comeron08}.
We include all of the sources in Comer\'on's Table~11 
\citep[well-known classical T-Tauri 
stars from][]{The62,Krautter97,Hughes94}, Table~14 \citep[additional
low-mass members from][]{Comeron03} and Table~16 \citep[possible low mass
members of Lupus3 from][]{Lopez05}.  The source list in Table~16 
in \citet{Comeron08} gives less accurate positions than in \citet{Lopez05}, 
so we use the original data.

We do not include the sources given in Comer\'on's Table~13
\citep[suspected Lupus3 members from][]{Nakajima00}, as the survey
these data originate from only spans an RA of roughly 16:09:44 to
16:08:44 (in J2000), which is much smaller than the region spanned
by the Lupus3 group.  Inclusion of these sources could bias the
clustering statistics due to their limited areal range (which is 
not centred on the apparent group centre), and furthermore the
spectral types of all of these sources are unknown.
We also exclude the list of weak T-Tauri stars in Lupus3, as 
these stars are thought to be older and not associated with the 
current groups \citep{Comeron08}.

We adopt a distance to Lupus3 of 200~pc, as recommended by
\citet{Comeron08}.

\subsection{IC348}
In IC348, we analyze a total of 363 sources whose properties are given in
Table~\ref{tab_IC348} (available online).
This includes the 307 sources listed in \citet[][Table~2]{Lada06} and
the 41 sources identified in \citet[][Table~1]{Muench07}.
We supplement this list with all other likely members with
known spectral types: seredipitously discovered members 273 and 401 
discussed in Appendix~C of \citet{Muench07}, as well as
source 30074, the companion of 166 
\citep[][and listed as 166B there]{Luhman05}, 
and 8078, the companion of 9078 
\citep[][listed as 78B and 78A respectively there]{Luhman03}.
We use updated spectral classifications for 7 of the sources in the
\citet{Lada06} catalog (141, 174, 294, 334, 366, 1050, and 2103),
and also add 11 sources with spectra (249, 250, 307, 313, 340, 1686, 1779, 
1840, 6005, 10074, and 10095) recently obtained by K. Luhman (private
communication).
Note that in the \citet{Muench07} catalog, where there are multiple spectral
classifications, we use the optical classification.

We adopt a distance of 300~pc, following the discussion in \citet{Herbst08}.

\subsection{Completeness}
The YSO catalogs of all four regions have good completeness.  In Taurus,
a comparison of X-ray and optical/IR survey data shows the catalog
should be complete to $\sim$0.02~\Msol\ for class II and III stars
and brown dwarfs.
The completeness is good for class~I stars, but is difficult to determine
for class~I brown dwarfs (later than $\sim$M6) due to confusion in 
{\it Spitzer} bands with faint red galaxies 
\citep{Luhman09}.  Well-known young protostars such as L1527-IRS1,
L1521F, and IRAM~04191+1522 are included in the catalog. 
Other candidate class 0/I protostars with very weak {\it Spitzer}
fluxes may be missing, such as J041757.75+274105.5 \citep{Barrado09}. 
Using the overly conservative estimate that all class~I brown dwarfs 
and class~0 sources are missing from the catalog suggests only 6\%
of the total number of sources are missing from the catalog, using the
ratio of classes of objects given in \citet{Evans09}.

\citet{Lopez05} estimate they are complete in Lupus3
down to an R-band magnitude of 20 and and I-band magnidue of 19,
which corresponds to $\sim$0.02~\Msol\ at 1~Myr and 0.03~\Msol\ at
5~Myr in the \citet{Chabrier00} models, at a distance of 200~pc.
The completeness
level is not explicitly given in the \citet{Hughes94} sample
of T-Tauri stars.  In the brown dwarf mass regime, the
Lupus3 sample may be less complete than in Taurus; \citet{Comeron08}
lists several additional studies with lists of candidate
members that have not yet been spectroscopically-confirmed.
In particular, sources detected only in {\it Spitzer} have not
yet had spectroscopic follow-up, so it is likely that most, if not all, 
class~I sources are not included in our analysis.  
Class~I sources consistute only $\sim 4 - 5\%$
of the population in Lupus3 \citep{Merin08}, however.

The ChaI catalog described in \citet{Luhman07} was found to be complete
for masses above 0.01~\Msol\ in regions where A$_J \le 1.4$~mag.
{\it Spitzer} data was not used for that catalog, hence class~I
sources are likely missing.  Subsequent spectral 
surveys \citep{Luhman08,LuhmanMuench08}
based on {\it Spitzer} data do include a limited number of class~I
sources (four or five).  If the class~I population in ChaI is similar
to that in ChaII \citep{Evans09}, then roughly one quarter of the
class~I's are currently included in the catalog, implying only
$\sim$7.5\% of the total sources are missing from the catalog.

In IC348, the central region covered in the \citet{Luhman03} catalog
was found to be complete to $\sim$0.03~\Msol\ for $A_V < 4$~mag.
The {\it Spitzer} data included in \citet{Muench07} identified very
few new members within the \citet{Luhman03} survey bounds, and
instead extended the catalog to larger distances from the centre,
with an estimated completeness of $>$80\% for YSOs with H-band
magnitudes of 16; this corresponds to a mass of $\sim$0.015~\Msol\ 
at 1~Myr and 0.02~\Msol\ at 5~Myr in the \citet{Chabrier00} models,
at a distance of 300~pc.  In the \citet{Muench07} dataset, a total of
20 class~0/I sources were identified, corresponding to 
$\sim$6\% of the total.  Other
work \citep{Jorgensen08} suggests that the fraction of class~0/I
sources is $\sim9\%$ of the total IC348 population, which implies that 
only roughly 3\% of the total sources are missing from the catalog.

\section{ALTERNATE CATALOG OF TAURUS YSOS}
\label{app_rebull}
Two groups have independently released Taurus YSO catalogs
recently -- \citetalias{Luhman10} and \citetalias{Rebull10}.
We adopted the \citetalias{Luhman10} catalog in our analyses because
of the wider spatial coverage.  While not identical, the bulk of the
catalogs agree within the area covered by \citetalias{Rebull10}
(i.e., the extent of the {\it Spitzer} coverage).  
The \citetalias{Rebull10} catalog contains several types of listings --
definite members (previously identified and newly confirmed members),
as well as candidate YSOs described both by a likelihood of
membership (probable or possible member, needing additional 
spectroscopic follow-up and pending spectroscopic follow-up), in
addition to a rank (likelihood of membership given the available
data, ranging from A+ to C-).  
%A comparison between the
%various \citetalias{Rebull10} members and candidate members with
%the \citetalias{Luhman10} sources shows that some
%with lower masses are only found in one of the two catalogs, but at
%the higher masses, as expected, the sources are listed in both.
%Moving to decreasing levels of likelihood of membership in the
%\citetalias{Rebull10} candidate member list increases the number
%of sources without a match in the \citetalias{Luhman10} list, while
%adding few sources (only 4 over the entire range of candidates)
%that do have a match in the \citetalias{Luhman10} catalog.
Within a central part of the {\it Spitzer} coverage
(4:15:45 to 4:35:45 in RA and 23:15:00 to 27:00:00 in dec), we 
found 79 sources common to both catalogs,
2 additional un-matched sources in \citetalias{Rebull10} and 
7 additional un-matched sources in \citetalias{Luhman10}, of
which 6 were unresolved secondaries in the \citetalias{Rebull10} catalog.
%the above is identical for definite or probable sources
Where the sources are listed in both, the agreement in spectral
classification is generally quite good -- 77\% agree within the
expected error of 1 spectral sub-class, and an additional 11\%
have no spectral classifications in \citetalias{Rebull10}.  Only
5\% of the sources have classifications that differ by more
than two sub-classes, and some of those are listed as being
uncertain classifications in each paper. [These numbers
are for the definite members in \citetalias{Rebull10} catalog
only; the fractions are nearly identical when including the
candidate members, since there are few (or no) additional matches
for each broader \citetalias{Rebull10} category.]

%from Rebull_data/compare_reb_luh.pro
%239/249 Rebull sources have assoc with Luhman source
%10/239 Rebull sources have no spectral types (-> cannot match)
%152/239 have identical spec types
%32/239  have diff of 1 sub-class
%15/239 have diff of 2 sub-classes
%13 have diff of > 2 subclasses:
	%3 has Luhman bolom range
	%(diff) 1 has Rebull range
	% HOWEVER, the most discrepant one F1/G8 star (R/L) not a range
	% also A6/B9 star (R/L) and some more discrep at lower masses (less imp)

We ran all of our analysis on the \citetalias{Rebull10} catalog and found
similar results to those using the \citetalias{Luhman10} catalog.
Two of the Taurus groups we identified in \citetalias{Luhman10} 
(L1551 and L1517) fall outside the
spatial range covered by the \citetalias{Rebull10} catalog.  All 
remaining groups were recovered, although one group (L1536) was
split into two groups.  The critical branch length fit for the
\citetalias{Rebull10} stars was nearly 10\% smaller than the value
found in the \citetalias{Luhman10} catalog, causing the linkage between 
the most clustered part of the group to become
separated from the more filamentary part of the group.
%Table~\ref{tab_luh_vs_reb} compares the total number of sources
%per group using
%both catalogs.  In L1536, both \citetalias{Rebull10} groups were summed
%together for the table; the total number of sources
%are similar using the two catalogs.  Furthermore,
As with the \citetalias{Luhman10} Taurus groups, 
the \citetalias{Rebull10} Taurus groups nearly always had the most massive
group member(s) located near the centre of the group, and in or
near the region of highest surface density of sources.
Our conclusions are therefore unaffected by the choice in source
list for Taurus.

\section{MASS ESTIMATION}
\label{app_mass_est}
We adopt the conversion between spectral type and effective
temperature used by \citet{Luhman04a} with the addition from 
\citet{Luhman08} of a temperature of 2200~K for L0 stars.  The
full set of effective temperatures we adopt is included in 
Table~\ref{tab_spec_mass}.  
%This conversion uses the \citet{SchmidtKaler82}
%main sequence temperature scale 
%%{\hk (HK note -- question for Kevin here)}
%%(HK nitpick note: temperature scale same as 
%%Table 3 of text for **MS** stars, but Kevin's list is more finely divided;
%%no dwarf table in the book)
%for spectral types of M0 and earlier, and for later spectral types,
%uses a temperature scale determined
%by \citet{Luhman03} to keep the YSOs in each of Taurus and IC348 
%along a single isochrone for the stellar evolution models adopted.
We then follow the procedure of Luhman \citep[e.g., see discussion in Appendix~B
of][]{Luhman03}, using several different stellar evolution models 
to estimate the mass based on the effective temperature, which are outlined
in more detail below.  The mass we estimate for each spectral type is 
also given in Table~\ref{tab_spec_mass}.

Above 1~\Msol, we use the \citet{PS99} models at an age of 1~Myr.
All of their models adopt a ratio of mixing length to local pressure
scale height of 1.5, and a helium fraction, Y, of 0.28.

Between 0.6 and 1~\Msol, we use the \citet{Baraffe98} models with
a mixing length of 1.9 (and Y of 0.282).  The youngest ages available
in these models are 2~Myr, which we adopt.  Despite having twice the
age of the \citet{PS99} models, there is good agreement between the
two models at 1~\Msol.

Between 0.1 and 0.6~\Msol, we use a different set of the \citet{Baraffe98}
models -- those with a mixing length of 1 (and Y of 0.275), and again
the youngest available age of 2~Myr.  These models were run on a much
finer grid than the mixing length of 1.9 Baraffe models, particularly
at the lower mass regime, and hence provide a much more precise 
estimate of the mass based on effective temperature.  The mixing
length 1 models are not in agreement with the \citet{PS99} models
at 1~\Msol, hence cannot be used for the entire range.  At 0.6~\Msol,
they are consistent with the mixing length 1.9 \citet{Baraffe98} models,
and hence this is a reasonable mass at which to switch the model used.

Below 0.1~\Msol, the \citet{Chabrier00} models at 1~Myr are used.  This 
has good agreement with the \citet{Baraffe98} models with
mixing length of 1 at 0.1~\Msol\footnote{Note that while the 1~Myr models
are not directly given in \citet{Chabrier00}, they can be downloaded by
anonymous ftp from the authors.  See \citet{Baraffe02} for details 
on how to download the models.}.
None of the other models extend to such low masses, and so cannot
be used in this regime.

\begin{figure}[h!]
\plotone{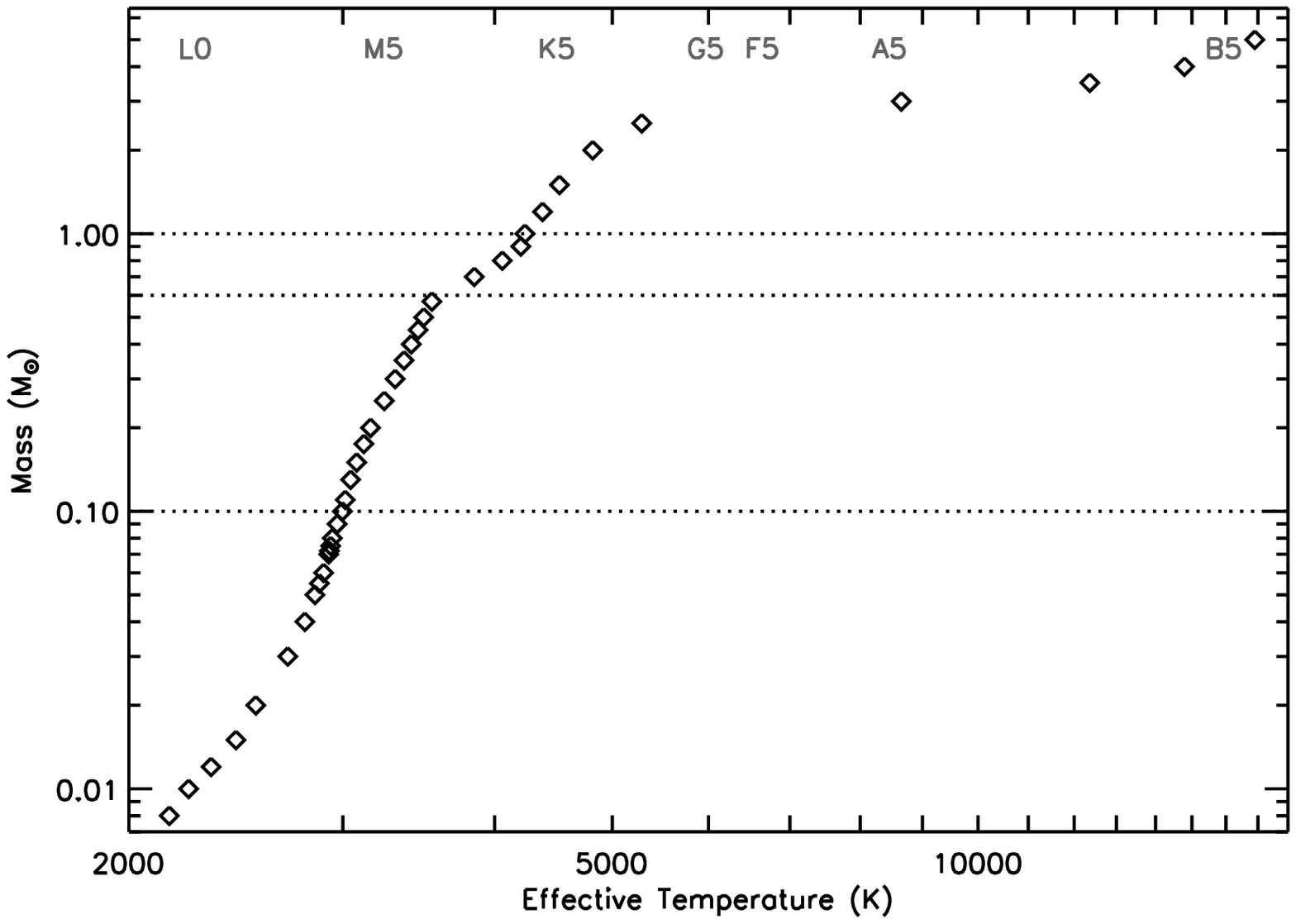}
\caption{The mass estimate based on effective temperature 
	given by the stellar evolution models adopted (diamonds).
	The horizontal dotted lines indicate the transition
	between the various stellar evolution models used: \citet{PS99} 
	above 1~\Msol, \citet{Baraffe98} between 0.6 and 1~\Msol\
	and 0.1 and 0.6~\Msol, and \citet{Chabrier00}
	between 0.01 and 0.1~\Msol.  The effective temperature
	of selected spectral types are indicated along the top
	of the plot.
	See Appendix C for more details.
	}
\label{fig_mass_spectype}
\end{figure}

Figure~\ref{fig_mass_spectype} shows the mass versus the effective
temperature given in the above models.  The transitions between
the various models are indicated by the horizontal dotted lines.
The transition between models is relatively continuous,
thus the combination of models is reasonable.
Comparison with \citet{Dantona94} model 1, over its full range
in masses (0.02 and 2.5~\Msol), and again adopting the Luhman temperature
scale, we find the \citet{Dantona94} models predict that the stars
are $\sim$30\% less massive.  
%This may be an overprediction of the
%differences between the models, since the temperature scale we adopt
%was fine-tuned to fit the combination of models we adopt, and may be
%less appropriate for the \citet{Dantona94} models.
This is slightly smaller than the uncertainty due to assuming
a single constant age, as discussed in Section~\ref{sec_mass_est}.

%Our main results concerning the mass segregation within the
%groups is unaffected by the uncertainty in the masses -- the only
%quantity of importance in the mass {\it rank} of the group members,
%which is a function only of the spectral type.  Assuming a 30\% smaller
%mass at all spectral types, the deficit of early
%B-type stars within the sum of the groups discussed in Section~4.1
%would actually become worse.  For low total group masses, the maximum
%mass expected from random sampling of an IMF varies as the total mass 
%to a power less than one \citep[see, e.g.,][]{Weidner10}.  
%\citep[While the][model strictly applies to randomly sampling an
%IMF which has a maximum mass of 150~\Msol, this has a negligible
%effect on the low-mass tail of the relationship.]{Weidner10}

\subsubsection{A Note on Uncertainties in Spectral Types}
\label{sec_mass_uncert}
The spectral types are typically uncertain to 
one subclass (K. Luhman, private communication).  
As can be seen in 
Tables~\ref{tab_Taurus} through \ref{tab_IC348}, 
however,
there are some instances of uncertain or unknown spectral types
in most of the regions, which are discussed below. 

Sources with limits on their spectral types (e.g., less than K5)
are assigned a spectral type equal to the limit for the purpose
of estimating the mass (this affects three, four, and two sources in 
Lupus3, ChaI, and IC348 respectively).  Sources with a range in spectral types
given are assigned to the
midpoint spectral type for the mass calculation (two K7-M0 sources in
Lupus3).  
Finally, sources with completely unknown spectral types 
are assigned to have the median mass of YSOs in their region,
in order to avoid bias in our later analysis
(five, eleven, and seven soures in Lupus3, ChaI, and IC348 
respectively).

Only some of the sources with uncertain spectral types fall
within the groups we analyze.  In Taurus, 16 of the 31 sources with
spectral types estimated from bolometric luminosities
(Appendix~\ref{app_catalogs})  are group members; none are the most
massive few members, hence the uncertain
spectral type has minimal impact on our analysis.  All of the sources
with uncertain spectral types in ChaI fall within ChaI-South (12 of
96 group members) and ChaI-North (3 of 43 group members), 
and again do not have mass rankings 
within the top few group members.  In Lupus3, none of the sources
with uncertain spectral types fall within the group, and in IC348,
only one falls within IC348-North.

\section{GROUP IDENTIFICATION}
\subsection{MST Critical Branch Length}
\label{app_dendro}
\begin{figure}[h]
\plotone{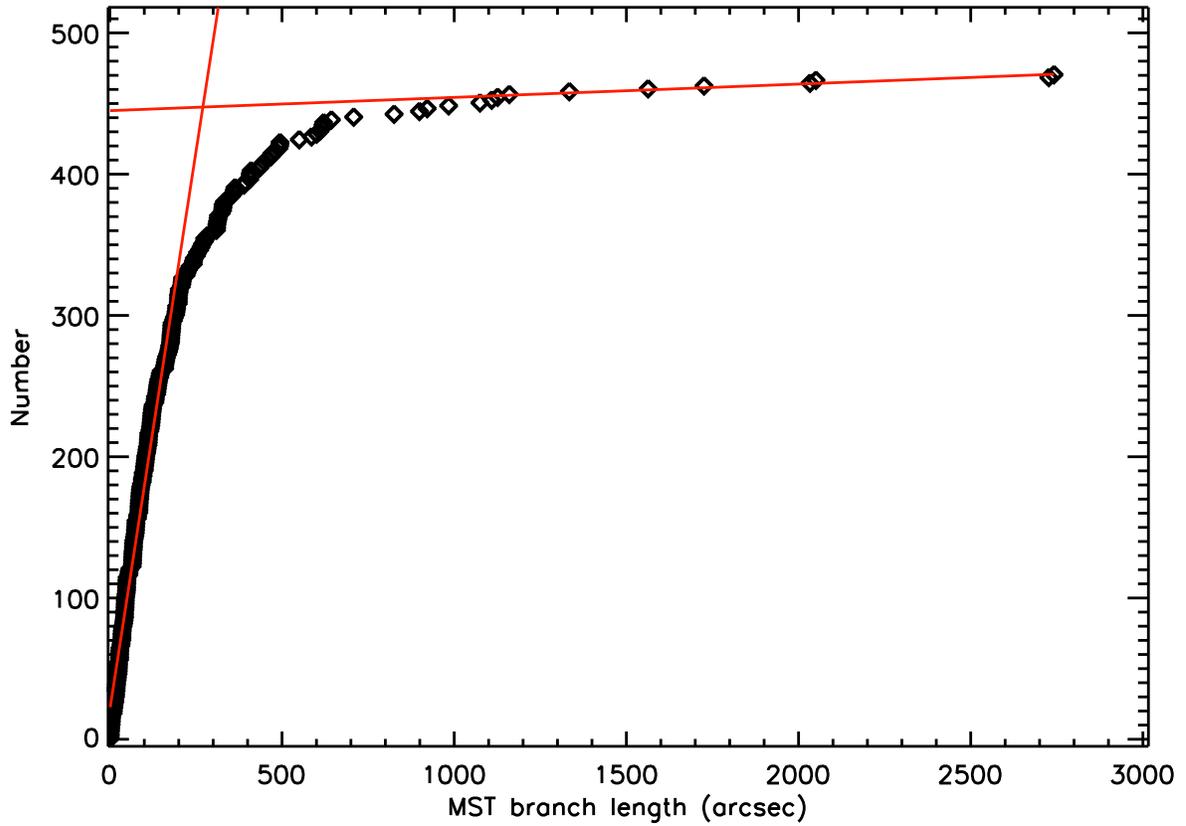}
\caption{The determination of the critical branch length for a region.
	Plotted is the cumulative number of branches
	with length x or smaller in the ChaI region.  
	The two ends of the distribution are
	well fit by straight lines.  The critical branch length is
	defined as the intersection between these two lines.  See
	Appendix~D.1 for more details.}
\label{fig_MST_cutoff}
\end{figure}

The definition of the groups we identified relies on the value
used for the critical MST branch length; larger values tend
to increase the number of group members,
while smaller values tend to decrease the number of members.
Figure~\ref{fig_MST_cutoff} shows our method for determining 
the critical branch length -- the cumulative distribution of
MST branch lengths is well-described by a steep linear rise at
small branch lengths, followed by a turn-over and a shallow linear 
rise at large branch lengths.  Figure~\ref{fig_MST_cutoff} shows the
data for ChaI; the other three regions show a similar trend.
Following \citet{Gutermuth09}, we make linear fits to the two ends of the
distribution, and define the critical branch length as the
length at which the two best-fit lines intersect.  For each of the
four regions, we
tested the range of possible critical values that could be fit, 
and found that the variation was
less than 10\% for all regions, and less than 5\% in IC348.

\begin{figure}
\plotone{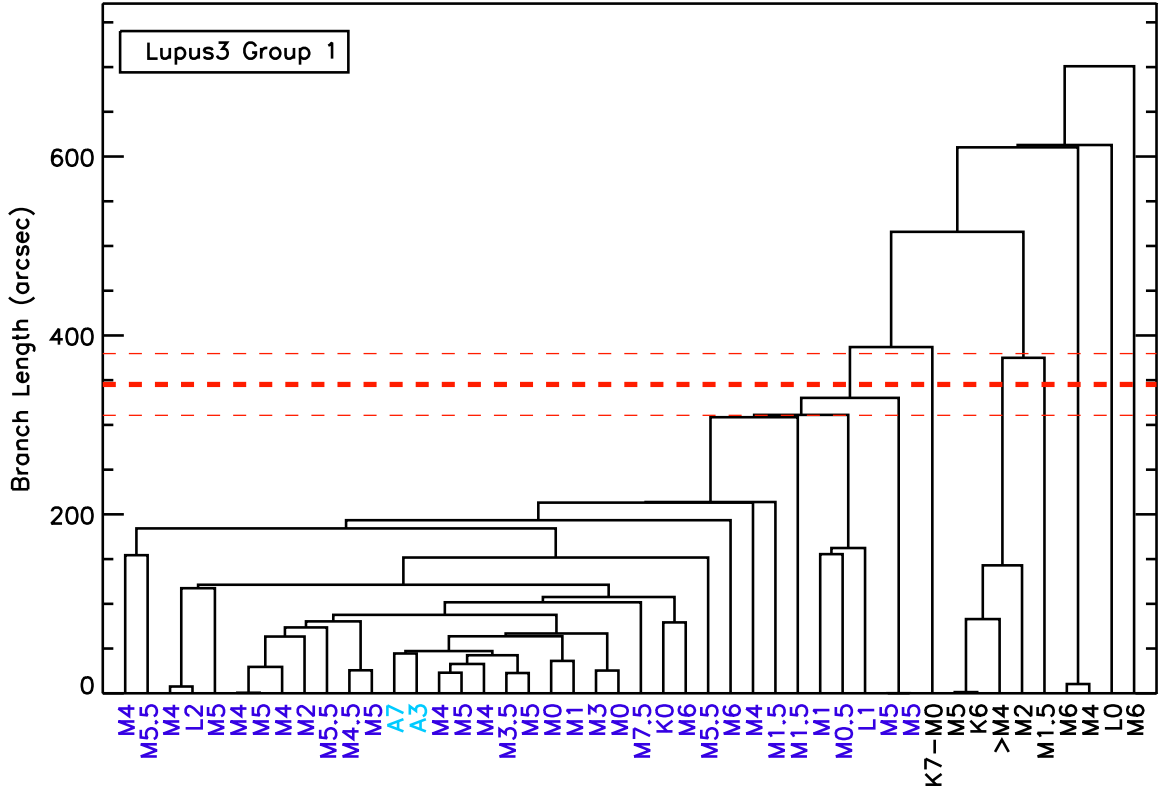}
\caption{The dendrogram structure for the main group and nearby
	YSOs.  The spectral type of each Lupus3 group member (blue), 
	and the ten nearest sources beyond the group (black) are shown
	along the horizontal axis.  The most massive members in Lupus3
	are shown in pale blue.  The thick horizontal dashed line
	indicates the critical branch length measured in Lupus3, while
	the thin dashed lines show $\pm 10\%$ of this value.
	}
\label{fig_Lupus3_dendro}
\end{figure}

\begin{figure}[p]
\begin{tabular}{cc}
\includegraphics[height=6cm]{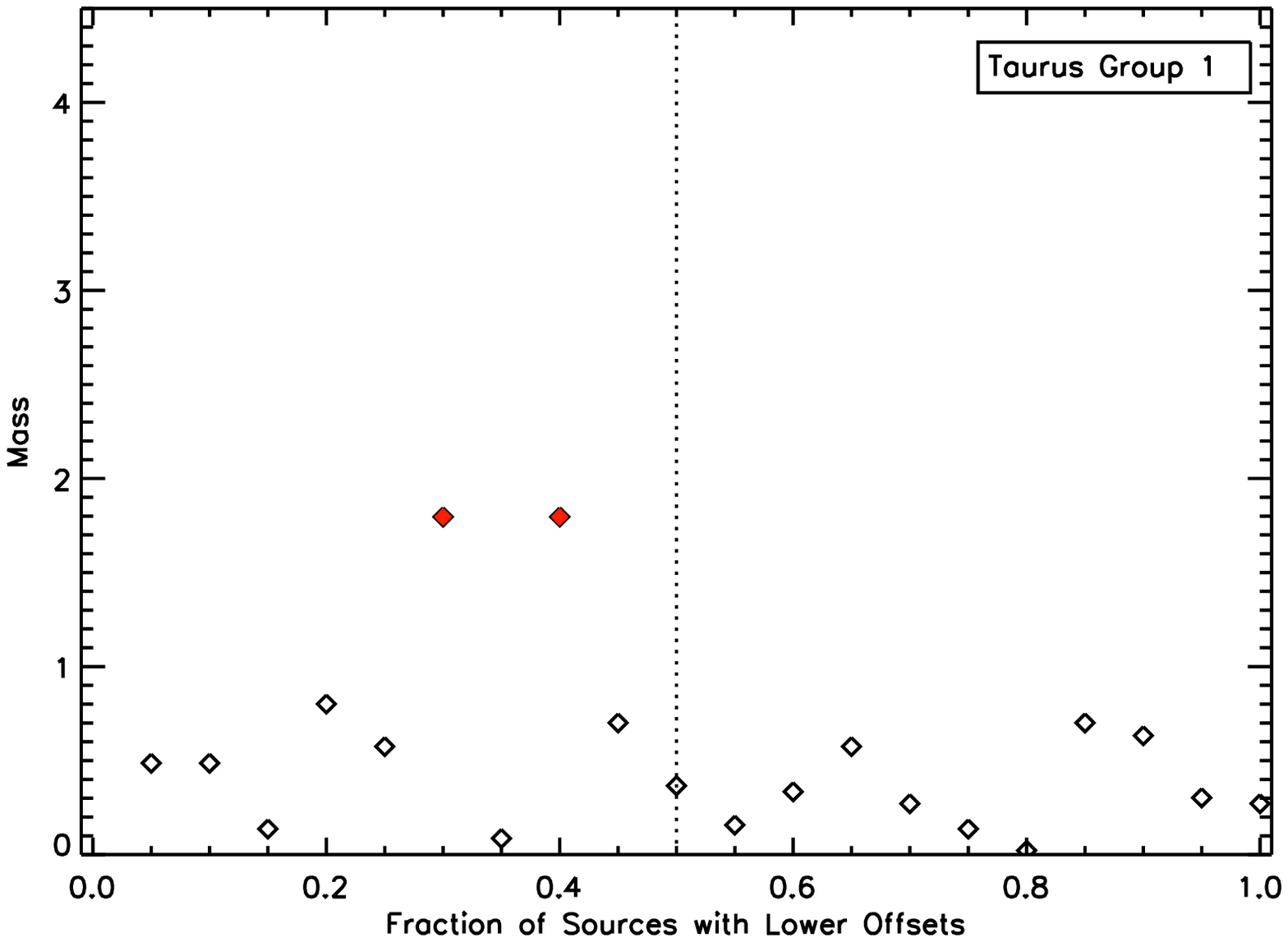} &
\includegraphics[height=6cm]{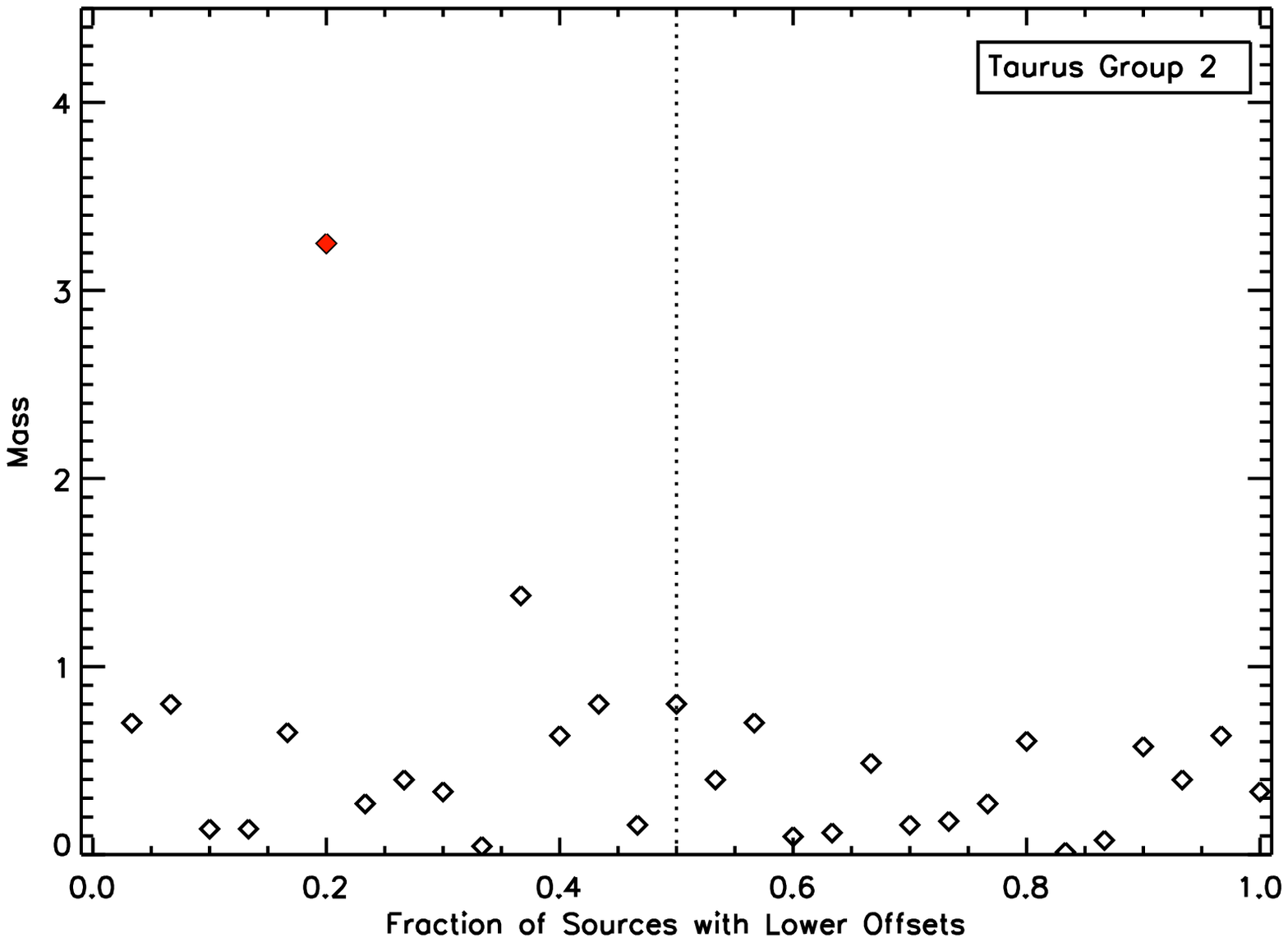} \\
\includegraphics[height=6cm]{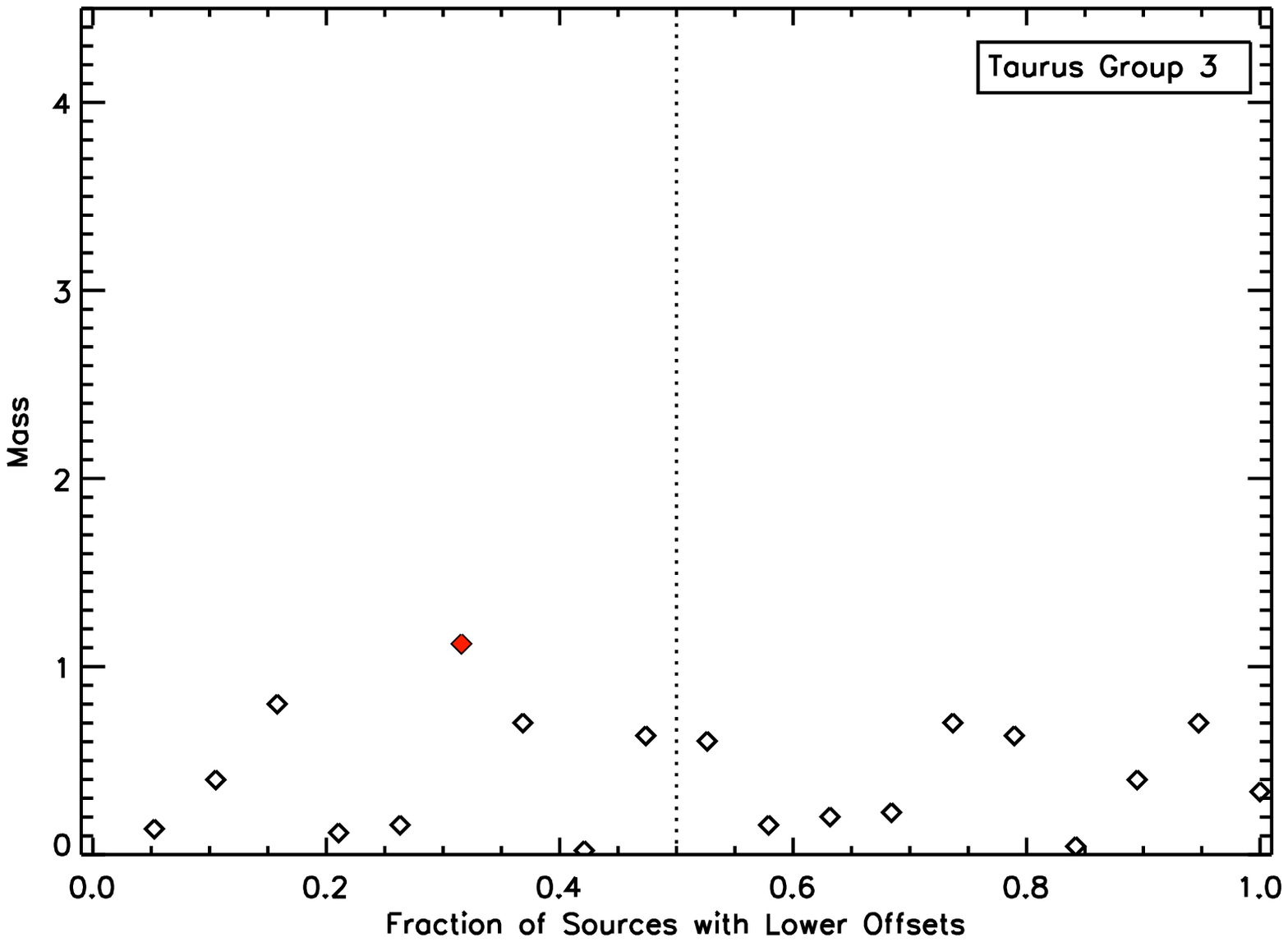} &
\includegraphics[height=6cm]{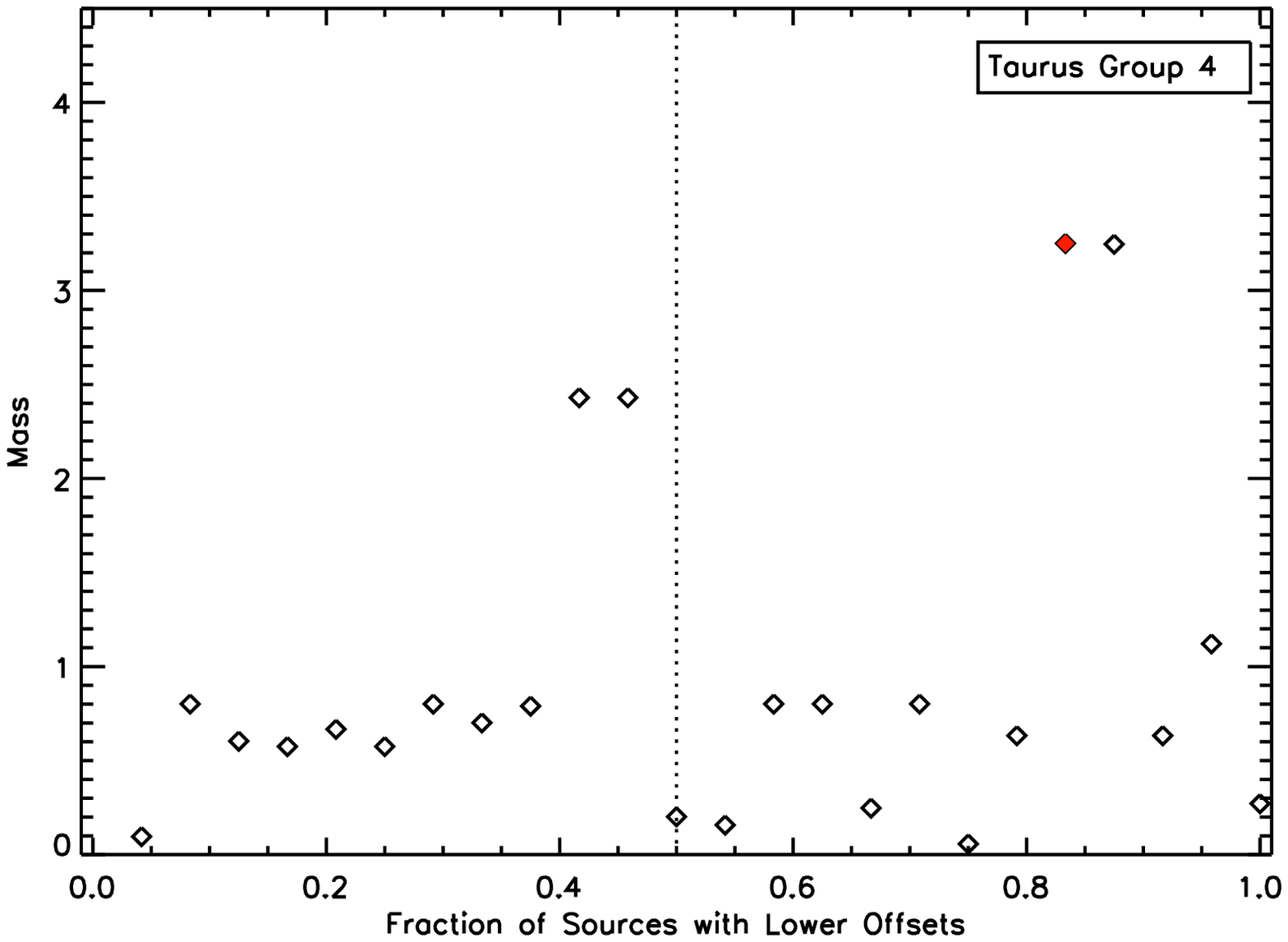} \\
\includegraphics[height=6cm]{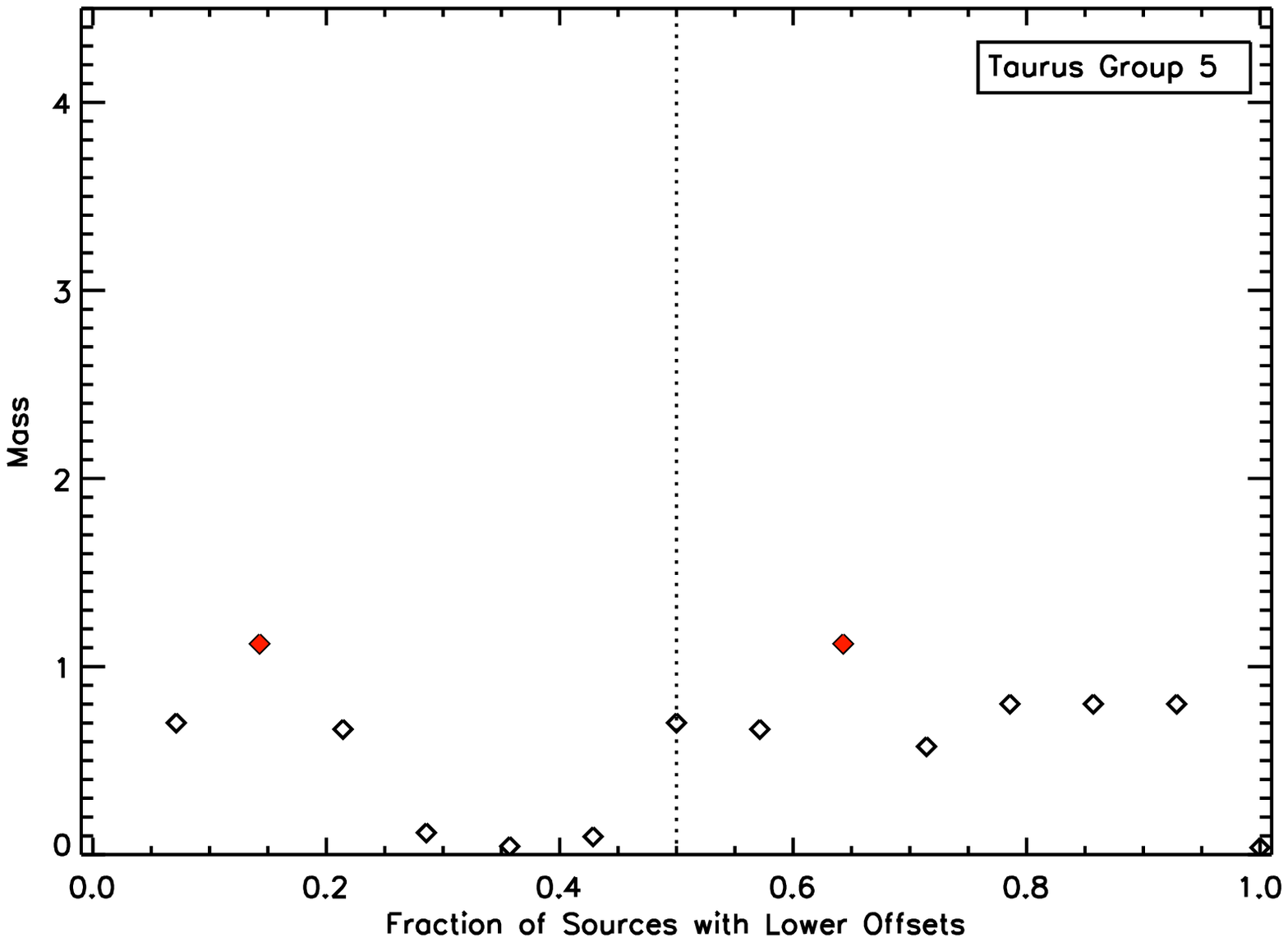} &
\includegraphics[height=6cm]{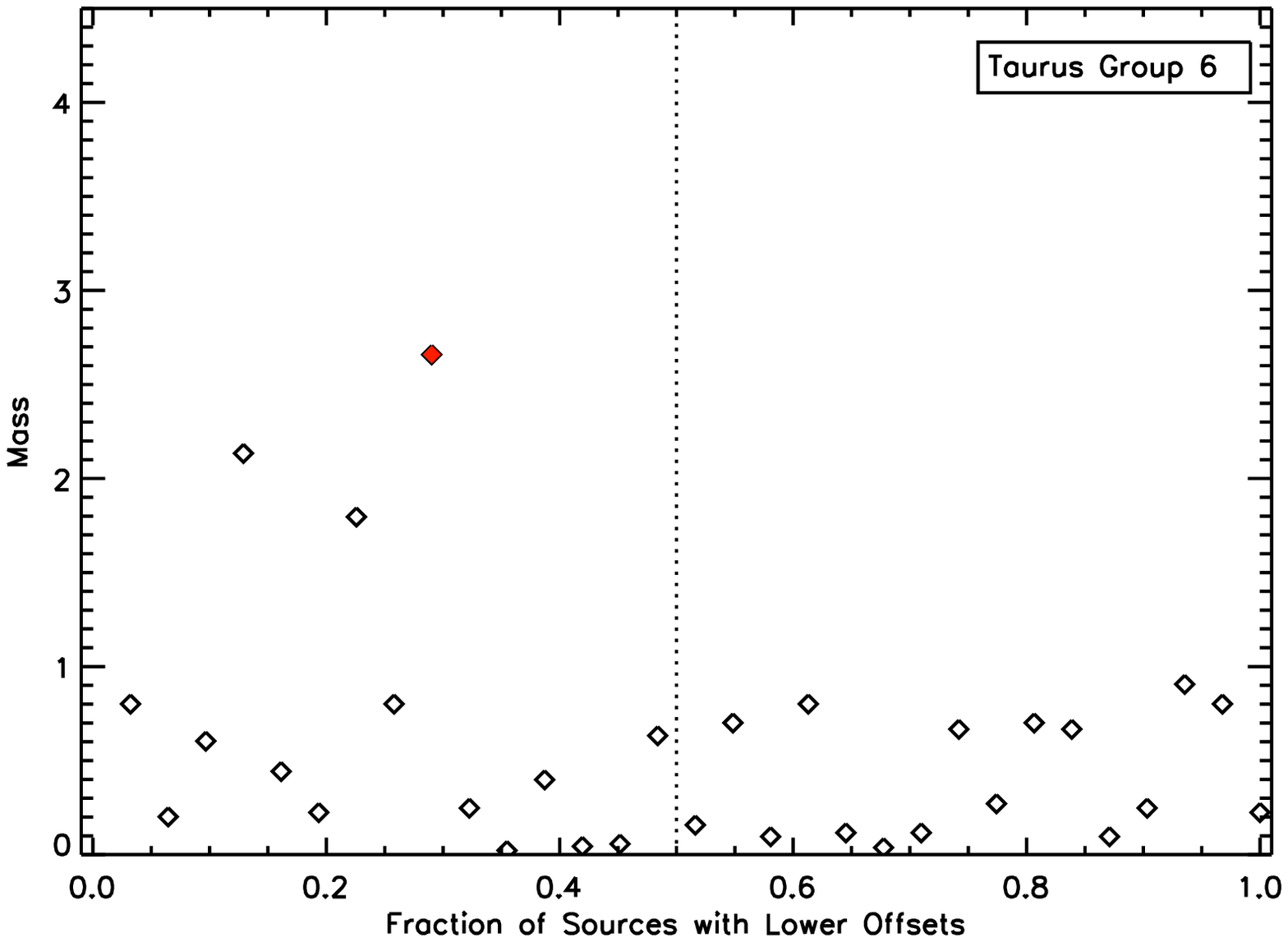} \\
\end{tabular}
\caption{ {\it (This figure for the online version only)}
	The mass of each YSO in the group versus the fraction of group members
	which have separations equal to or smaller than the offset of that
	group member.  The solid red diamonds show the most massive
	member(s) of each group, and the vertical dotted line indiates
	a fraction of 50\% for each group.  {\hk Groups with mass
	segregation should have their most massive members at 
	low fractions.  Groups with complete mass segregation 
	would follow a trend from the top left to the bottom right
	of the plot.  Instead, we find YSOs at lower masses to be
	roughly evenly distributed across low and high fractions, which
	indicates that they are not segregated.}
	The first six groups in Taurus are shown in this figure.
	}
\label{fig_mass_offset_frac1}
\end{figure}

\begin{figure}[p]
\begin{tabular}{cc}
\includegraphics[height=6cm]{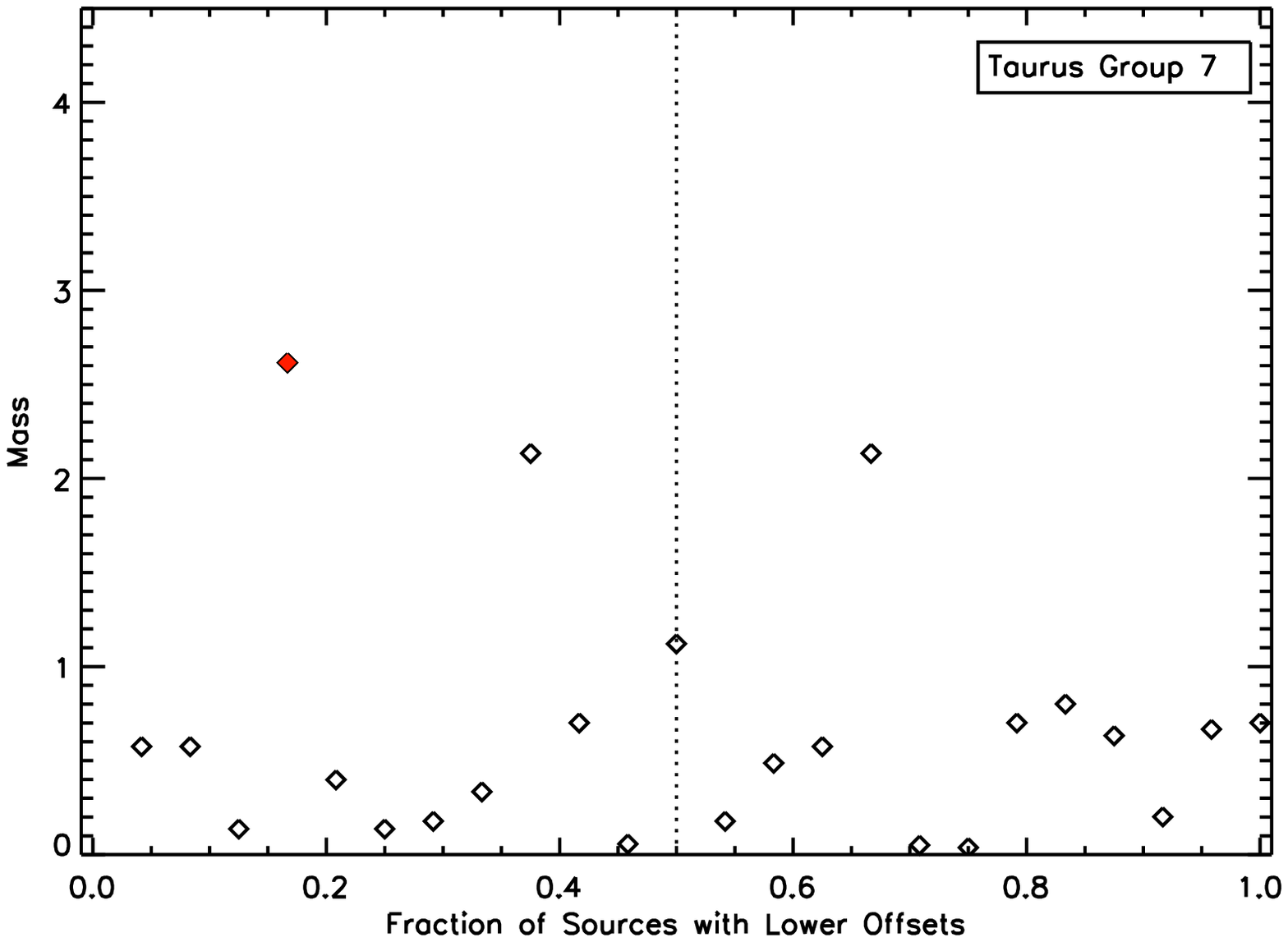} &
\includegraphics[height=6cm]{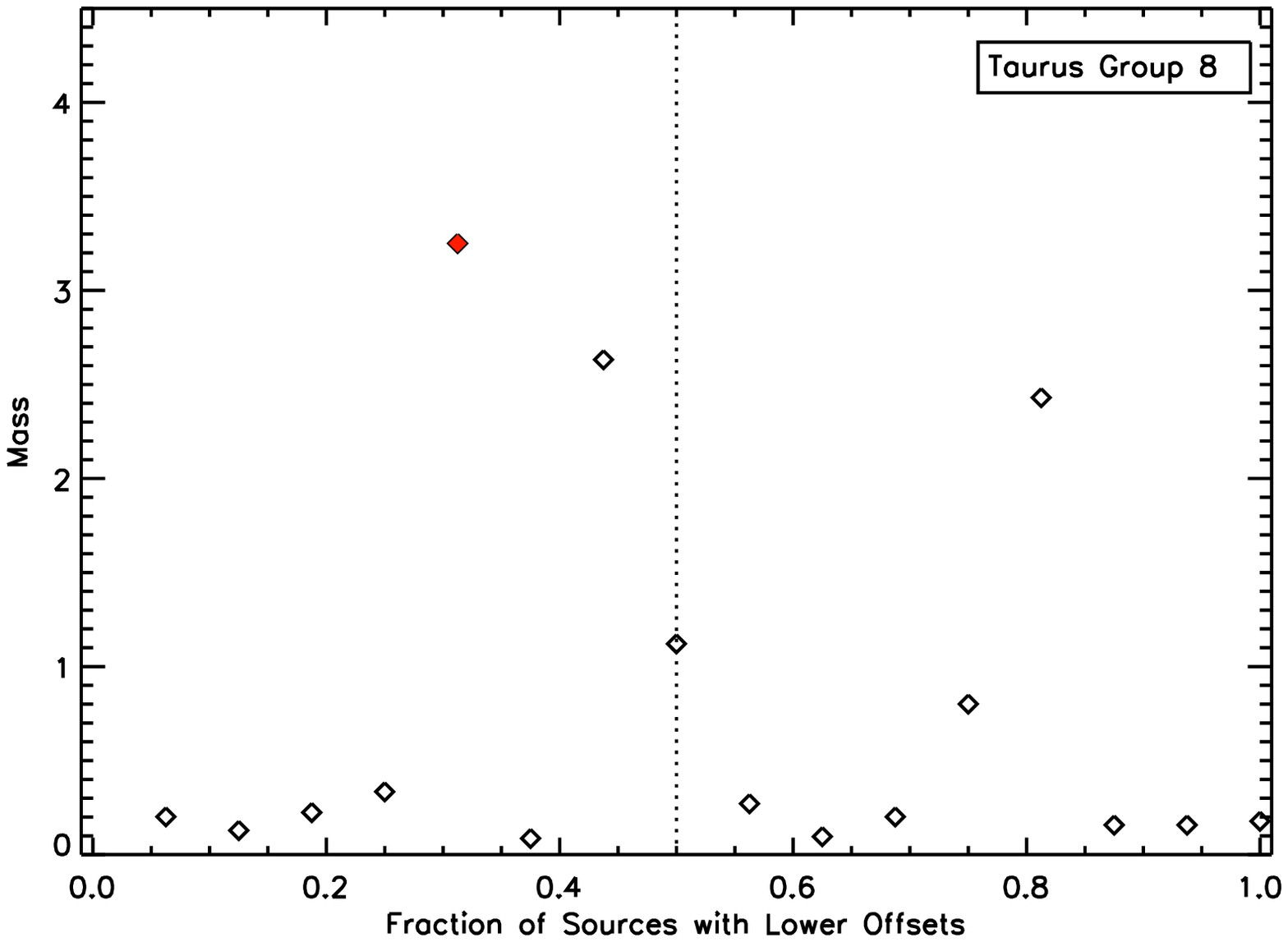} \\
\includegraphics[height=6cm]{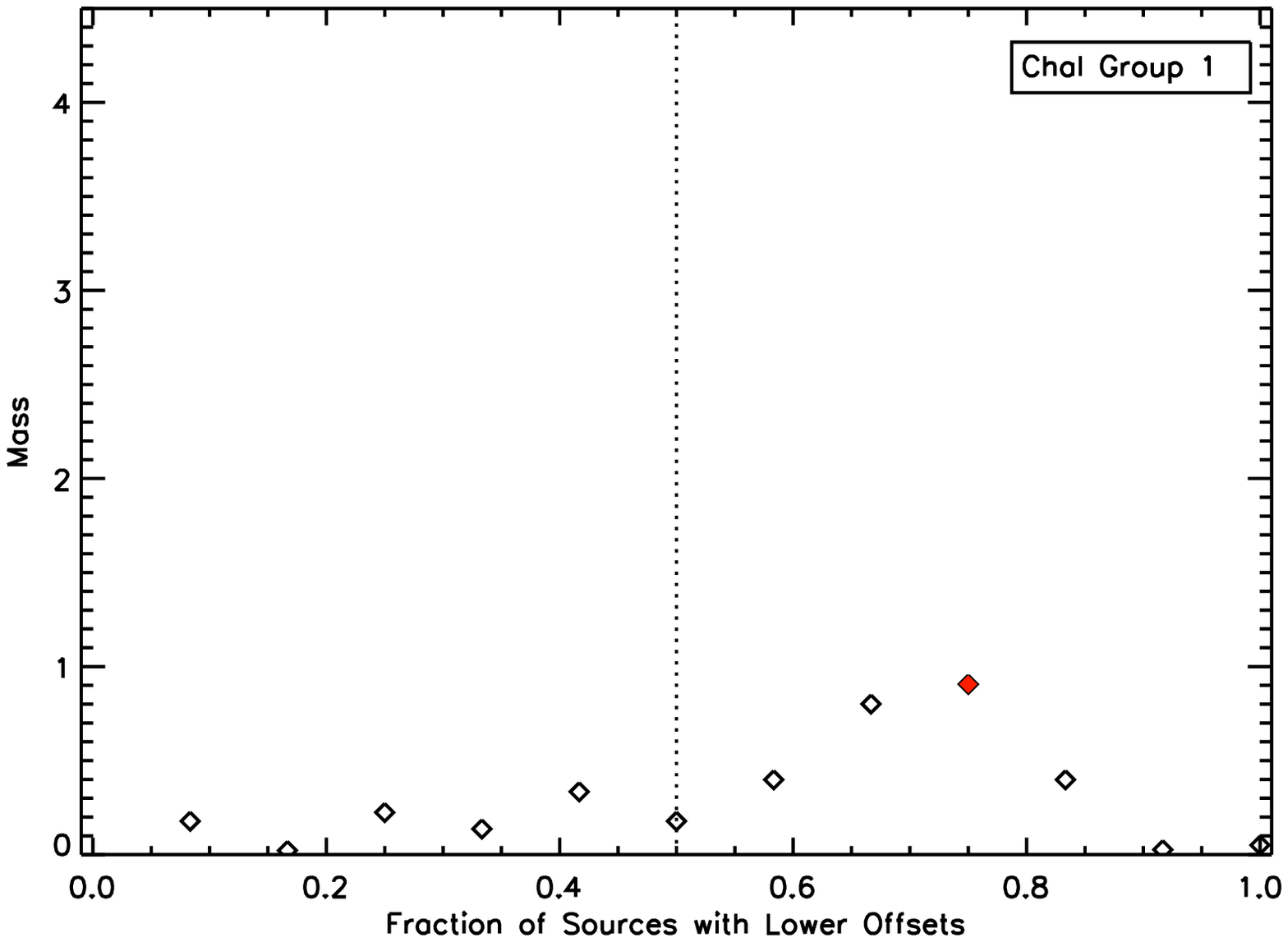} &
\includegraphics[height=6cm]{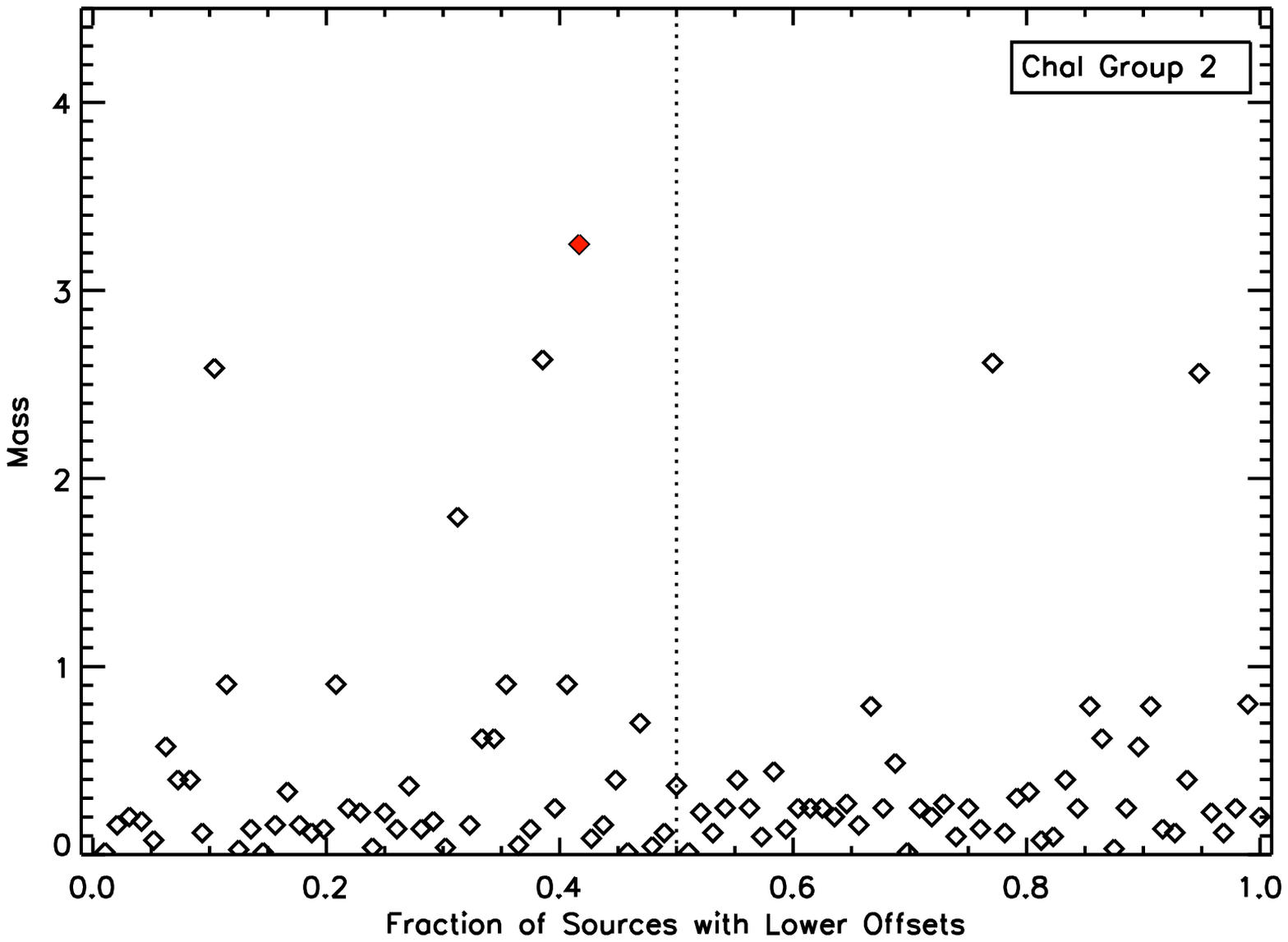} \\
\includegraphics[height=6cm]{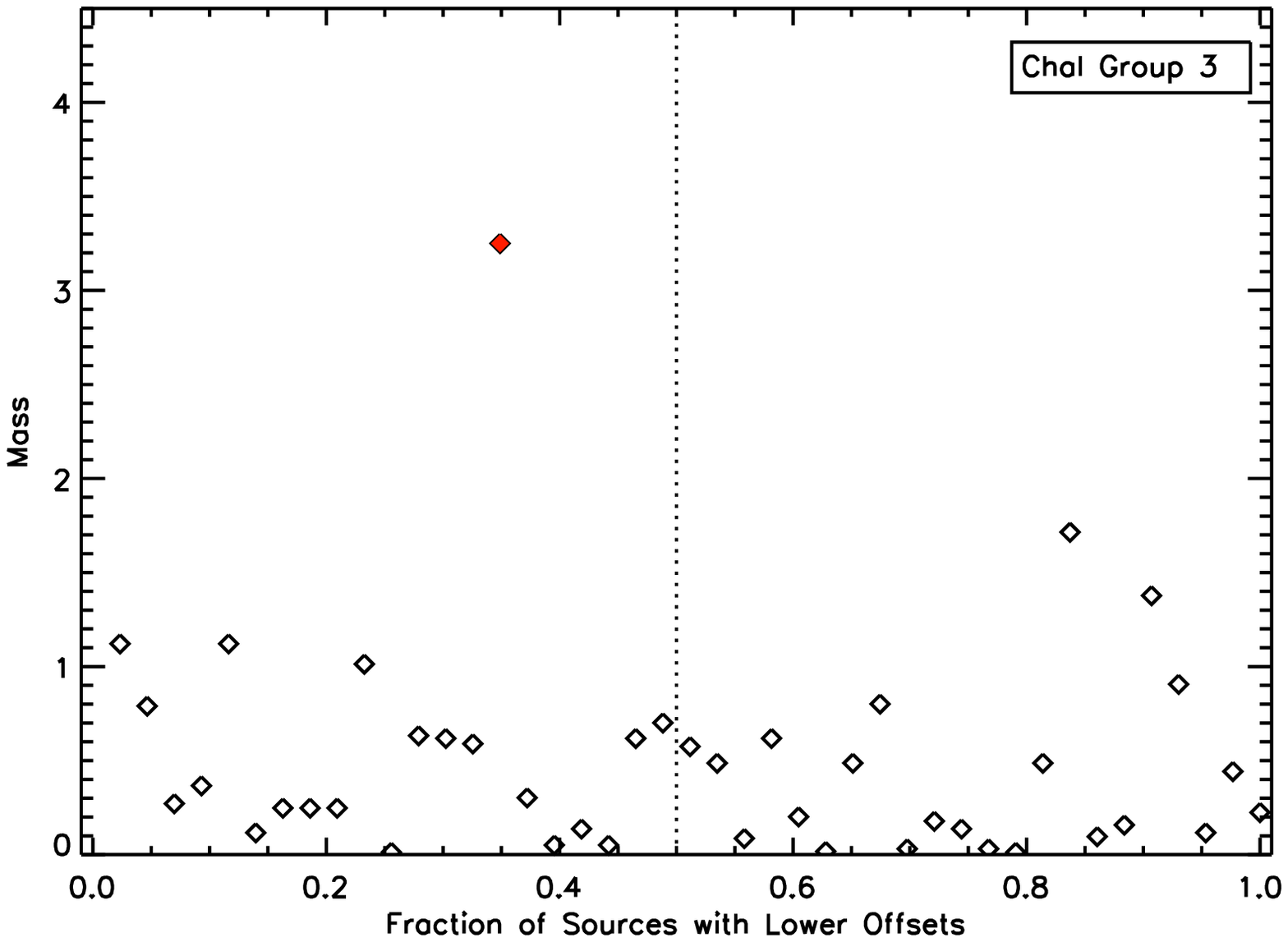} &
\includegraphics[height=6cm]{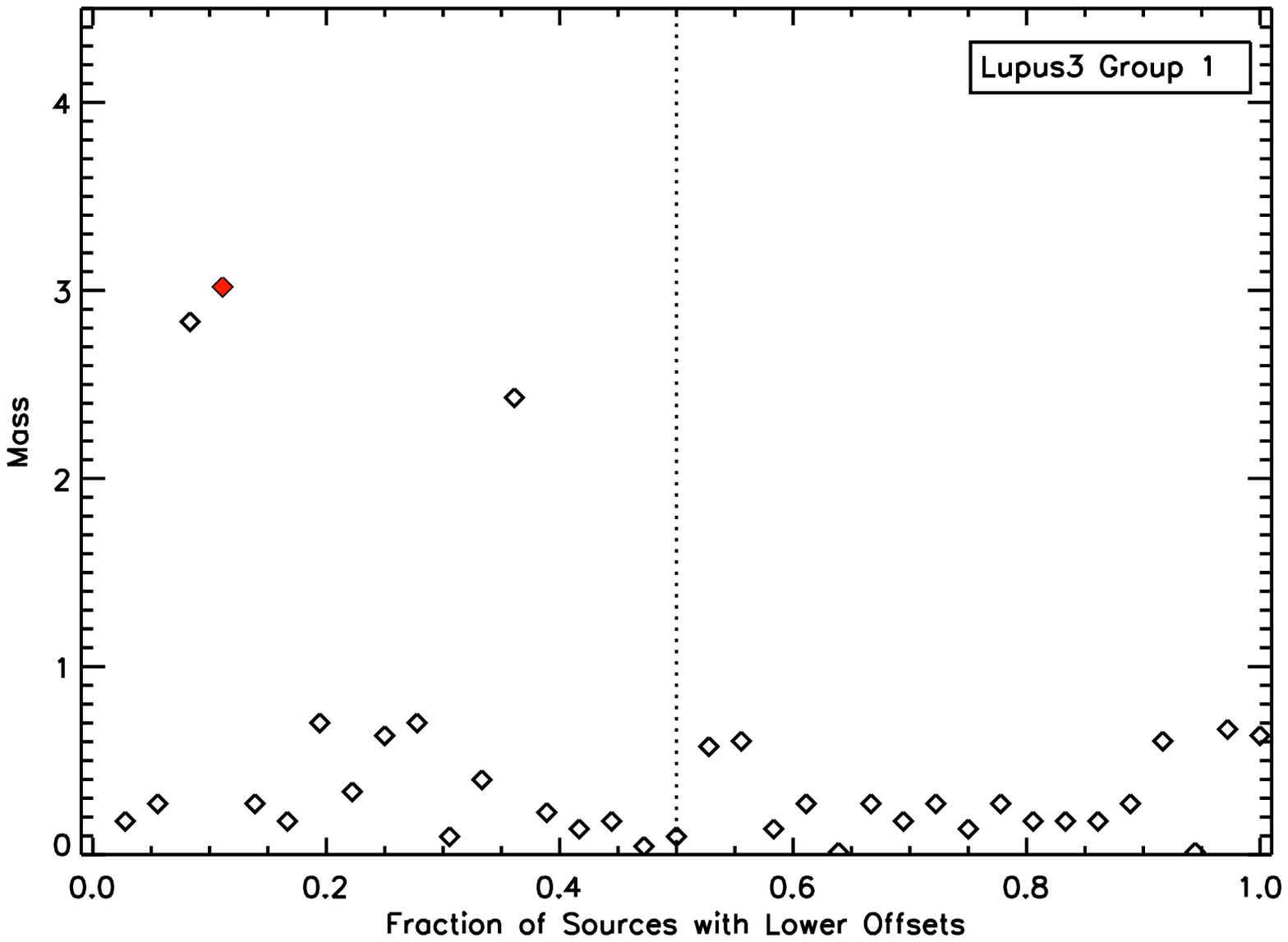} \\
\end{tabular}
\caption{
	A continuation of Figure~\ref{fig_mass_offset_frac1} showing the
        final two groups in Taurus and the groups in ChaI
	and Lupus3.}
\label{fig_mass_offset_frac2}
\end{figure}

\begin{figure}[ht]
\begin{tabular}{cc}
\includegraphics[height=6cm]{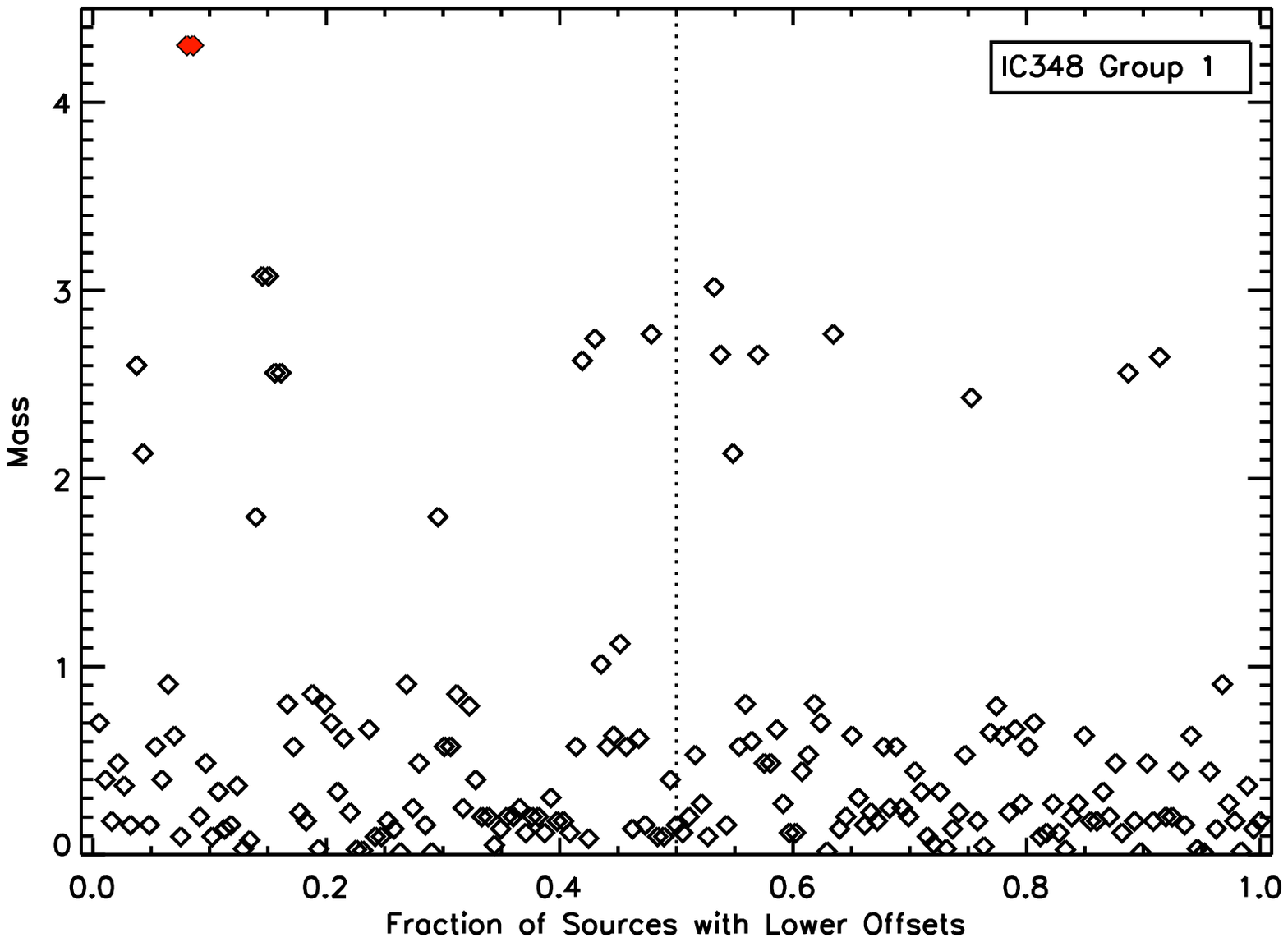} &
\includegraphics[height=6cm]{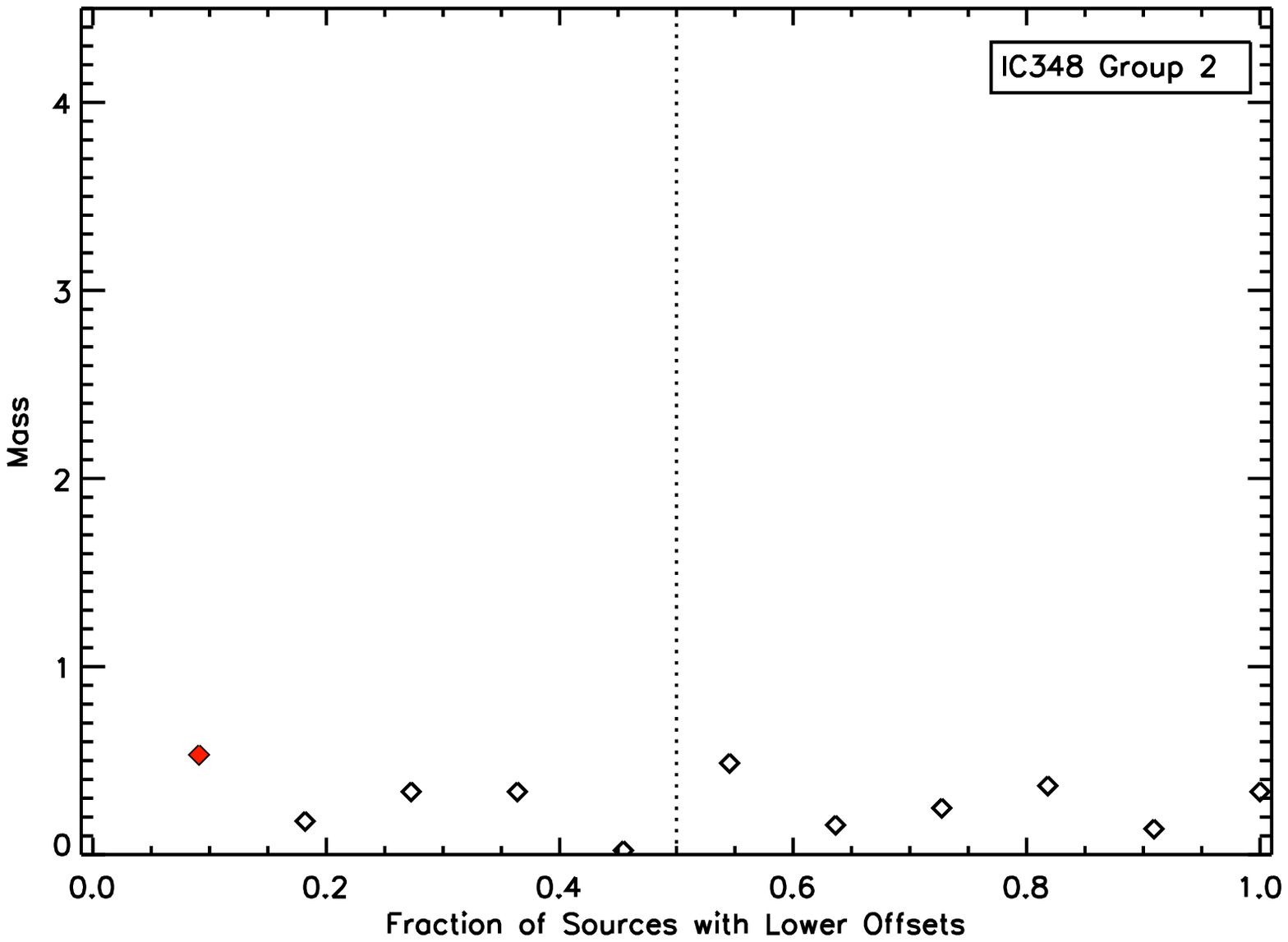} \\
\includegraphics[height=6cm]{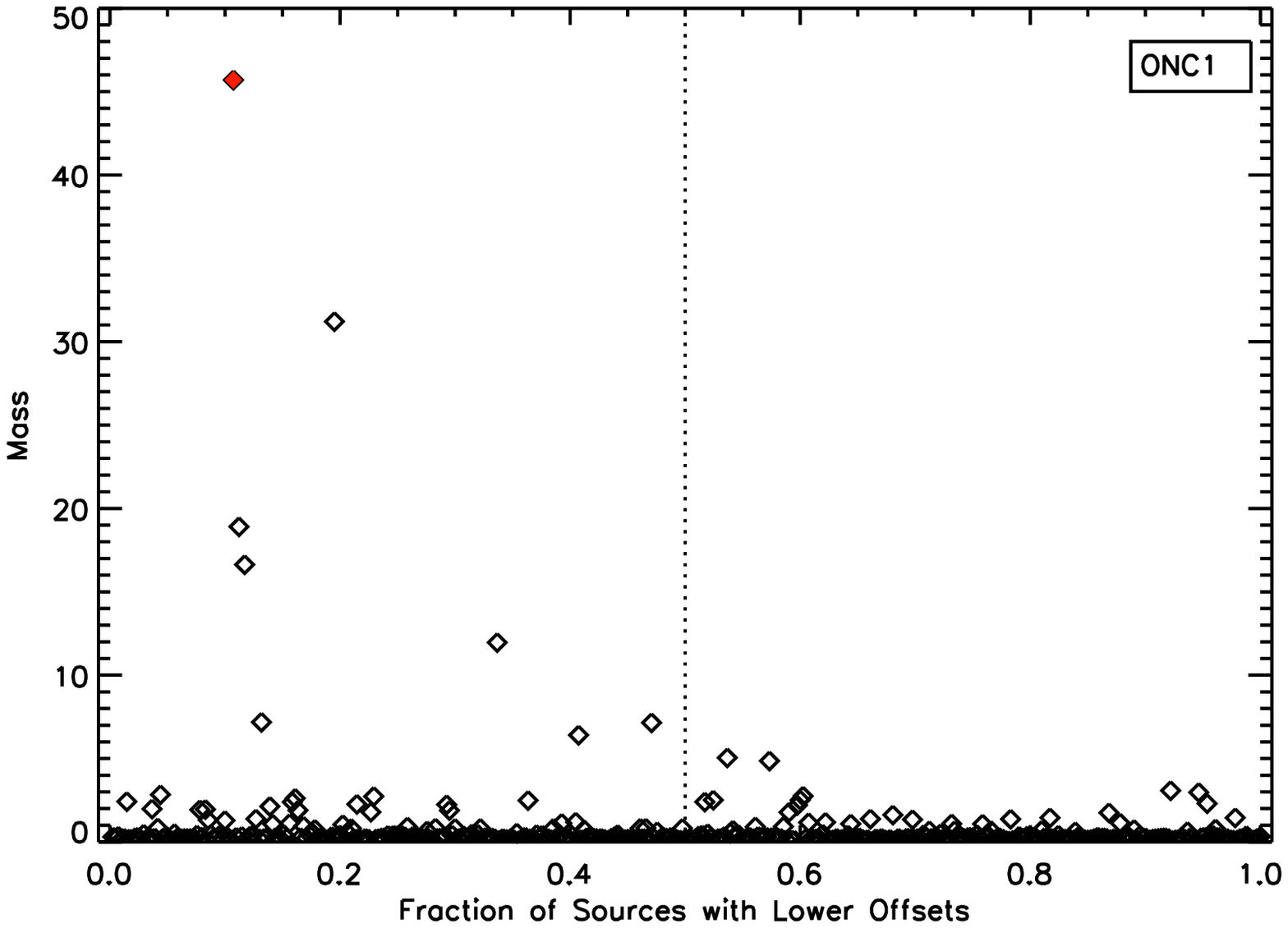} &
\\
\end{tabular}
\caption{
	A continuation of Figure~\ref{fig_mass_offset_frac1} showing the
	two groups in IC348.
	For comparison, the Trapezium cluster in Orion is also shown
	(bottom panel).}
\label{fig_mass_offset_frac3}
\end{figure}

We examined the effect on the group definitions of a 10\% larger or
smaller critical branch length.  One way
to examine this is through a dendrogram \citep[used recently
for analyzing structure in 3D data cubes in][for example]{Rosolowsky08}.   
%Dendrograms have
%recently been used in identifying self-gravitating parcels of
%gas in $^{13}$CO spectral cubes \citep{Rosolowsky08,Goodman09};
%we can make use of a much simpler implementation, however, as we
%are only interested in the separations between discrete sources.
Figure~\ref{fig_Lupus3_dendro} shows the dendrogram 
of the main group in Lupus3.  The MST branch length
connecting two sources is shown on the vertical axis at the
connection of each pair of sources or previously connected nodes.
The thick dashed
horizontal line shows the critical branch length measured in
Lupus3, while the two thin dashed lines indicate a range of 
$\pm10\%$ around the critical branch length.

In the Lupus3 group, clearly most of the members are tightly clustered
with separations well below the critical branch length.  An
increase in the critical branch length of 10\% 
would increase membership by only one late-type source
and a decrease of 10\% would result in the group decreasing by
two or up to six late-type YSOs (90\% of the critical branch
length is very nearly equal to the branch length required to 
joining four of the outlying group members to the rest of the group).

Most of the other groups show a similar behaviour -- 
an increase or decrease of the
critical branch length by 10\% changes the group membership 
by at most a handful of members (most often one or two).  
In the main group in IC348, the number of group members
included / excluded by a variation in the critical branch length
is larger, but still a small fraction ($< 5\%$) of the total number
of members.
The other exceptional group is L1536, also discussed in Appendix~B -- 
the critical branch length is only marginally larger than the
branch length connecting two sub-groupings of YSOs,
and would be split into two groups with
a slightly smaller critical branch length, as occurred with
our analysis using the \citetalias{Rebull10} catalog.  L1527 also
shows a similar behaviour although to a lesser extent
-- two smaller sub-groups would be lost if the critical branch
were reduced by 5 to 10\%, but the structure as a whole is more robust
to smaller perturbations in the critical branch length.

In terms of our analysis, the variation in the number of group members
is less important than the impact 
on the relationship between the most massive group
members and the rest of the group.  In the Lupus3 group, as seen in
Figure~\ref{fig_Lupus3_dendro}, the most massive group members lie
in a much more highly clustered part of the group (as found in Section~4.4), 
and their relationship with the bulk of the group is little affected
by the loss or gain of YSOs at the group outskirts.  A similar
result is found upon examination of dendrograms of most of the other
groups and their nearest neighbours.  As expected, the two groups
whose most massive member is not centrally-located (ChaI-Southwest
and L1551) are more liable to be excluded
from the group structure if the critical branch length is 
decreased sufficiently.  In L1536, if the group is split in
half, each piece has a centrally-located most massive member.

\subsection{Other Cluster-Identification Algorithms}
The MST algorithm identifies groups by linking members together through
their closest neighbour, referred to as a `single linkage' technique
for cluster- (or group-) identification.  
In fields outside of astronomy, the MST
technique is often less popular than other linkage techniques
which are better-suited for identifying round clusters 
(private communication, E. Feigelson).  There are two other main classes
of routines -- `average linkage', which use the distance of an object to the
cluster centre, and `complete linkage', which
use the distance of an object to the furthest cluster member
(Feigelson \& Babu, in prep);
the latter is useful for identifying very concentrated clusters.

We experimented with both techniques to identify groups in our 
dataset, using IDL's {\texttt cluster\_tree} function.  
As with the MST or `single-linkage' technique, the maximum linkage 
length to define a group must still be chosen.  We 
use the same method as we adopted for the MST, i.e., the critical
length, measured using the
cumulative distribution of lengths (discussed in more detail in 
Appendix~D.1).  Since the distance to a group's centre or furthest
member is larger than to the nearest neighbour, the critical lengths
fit for the average and complete linkage techniques tend to be significantly
larger than the value we found for the MST analysis.  
A (small) range of critical lengths can
provide a good fit to the cumulative distribution; for reasons
that will become apparent below, we use the largest critical
length possible.

Using the most conservative complete linkage technique, only 
small groupings are identified in our dataset.  IC348, for example, 
is split into four groups, each with only a handful of members, 
which do not appear as visually distinct groupings.
In our MST analysis, 182 members were found in the main group of IC348.
In Taurus, only three groupings are identified; other visually striking
groupings identified using independent methods (Section~3.2) are missed.  
We therefore conclude
that these young nearby stellar groups are too sparse to be effectively
identified using the complete linkage technique.

Using the less conservative average linkage technique, we have mixed
results.  In Lupus3 and Taurus, most visually-striking groupings
are identified.  There is a good correspondence between these groups 
and the MST groups, although the average linkage groups 
tend to have fewer members.  In Taurus, two of the eight groups 
identified with the MST (B213 and L1527) are no longer large enough 
to meet our minimum membership criterion ($>10$ members), and one 
MST-identified group
becomes split in half (L1536, discussed in Appendix~D.1 
for a similar reason).  
In ChaI and IC348, however, we encounter a similar
problem as found with the complete linkage technique --
the groups identified are overly-subdivided and do not
appear visually distinct.  IC348, for example, is split into 8 groups, 
most of which border directly on several 
other groups and do not appear to be separate.
The ChaI-south group identified using MSTs is similarly split into 
four groups with the average linkage technique.
This fragmentation occurs despite the fact that we pushed the linkage 
length to the upper end of the range possible.
In our datasets, it therefore appears that these other
linkage techniques are overly sensitive to small-scale substructure
perturbations, and are not optimal for identifying groups.  

In Taurus and Lupus3, where the average linkage technique works, the
most massive group members have a small offset from the centre.
The $O_{1st}/O_{med}$ values found are slightly smaller than those
measured with the MST groups.

\acknowledgements{The authors would like to thank Kevin Luhman for
providing us with his YSO catalogs, as well as fielding numerous
questions throughout this work, and Luisa
Rebull for sending us a copy of her Taurus catalog and paper
before publication and answering several queries.  We also thank
Francesco Palla and Steve Stahler for providing their stellar
isochrone data, Eric Feigelson for alerting us to alternative
cluster-identification techniques, and providing us with a copy
of the chapter from his upcoming astro statistics text on this topic,
and Paula Teixeira for sending us her Lupus3 extinction data. 
We are grateful for illuminating discussions with many people including
Alyssa Goodman, Rob Gutermuth, Eric Mamajek, Tom Megeath, Manon Michel, 
Gus Muench, and Tom Robitaille, and thank the referee for a thoughtful
report which improved this paper.

HK is supported by a Natural Sciences and Engineering Research
Council of Canada Postdoctoral Fellowship, with additional support
from the Smithsonian Astrophysical Observatory. 
}
%Acknowledgements:
% Luisa & Kevin for data
% Palla & Stahler for models
% Eric M for proper motion
% Gus Muench for ks discussion
% COMPLETE for suggestions (dendrograms)
% Eric Feigelson for cluster id techniques (other than MST)

%Tables for print edition
\include{tab1}
\include{tab2}

\clearpage

%Central positions of most massive member

\clearpage

%ratios of the above measure for all groups at 30 & 40% levels

\clearpage
\section{Online Material}
%%Tables for online-only
\include{tab3}
\include{tab4}
\include{tab5}
\include{tab6}

%%Figures for online-only


\end{document}

%% file: tab1.tex
\begin{deluxetable}{ccccccccccccc}
\rotate
\tabletypesize{\scriptsize}
\tablecolumns{13}
\tablecaption{Properties of groups identified \label{tab_groups}}
\tablehead{
\colhead{Region\tablenotemark{a}} &
\colhead{\#\tablenotemark{a}} &
\colhead{centre RA\tablenotemark{a}} &
\colhead{centre Dec\tablenotemark{a}} &
\colhead{N\tablenotemark{a}} &
\colhead{M$_{med}$\tablenotemark{b}} &
\colhead{M$_{max}$\tablenotemark{b}} &
\colhead{O$_{med}$\tablenotemark{b}} &
\colhead{O$_{1st}$\tablenotemark{b}} &
\colhead{L$_{crit}$\tablenotemark{c}} &
\colhead{L$_{max}$\tablenotemark{c}} &
\colhead{L$_{med}$\tablenotemark{c}} &
\colhead{Description\tablenotemark{a}} \\
\colhead{} &
\colhead{} &
\colhead{(J2000.0)} &
\colhead{(J2000.0)} &
\colhead{} &
\colhead{(\Msol)} &
\colhead{(\Msol)} &
\colhead{(pc)} &
\colhead{(pc)} &
\colhead{(pc)} &
\colhead{(pc)} &
\colhead{(pc)} &
\colhead{} 
}
\startdata
    Taurus &   1 &   4:14:25.07 &  28:10:13.75 &   20 & 0.426 & 1.796 & 0.282 & 0.228 & 0.520 & 0.516 & 0.127 &            B209 \\
    Taurus &   2 &   4:18:51.68 &  28:23:46.33 &   30 & 0.398 & 3.250 & 0.400 & 0.118 & 0.520 & 0.448 & 0.135 &          L1495E \\
    Taurus &   3 &   4:21:28.19 &  27:02:57.18 &   19 & 0.398 & 1.121 & 0.512 & 0.330 & 0.520 & 0.481 & 0.168 &            B213 \\
    Taurus &   4 &   4:32:15.48 &  18:17:42.35 &   24 & 0.684 & 3.250 & 0.551 & 1.005 & 0.520 & 0.498 & 0.182 &           L1551 \\
    Taurus &   5 &   4:32:34.07 &  24:22:18.58 &   14 & 0.684 & 1.121 & 0.452 & 0.028 & 0.520 & 0.508 & 0.193 &           L1529 \\
    Taurus &   6 &   4:35:17.03 &  22:56:09.92 &   31 & 0.398 & 2.659 & 0.685 & 0.367 & 0.520 & 0.511 & 0.221 &           L1536 \\
    Taurus &   7 &   4:40:36.82 &  25:52:09.08 &   24 & 0.575 & 2.616 & 0.707 & 0.406 & 0.520 & 0.506 & 0.231 &           L1527 \\
    Taurus &   8 &   4:55:49.24 &  30:30:57.12 &   16 & 0.213 & 3.250 & 0.343 & 0.202 & 0.520 & 0.475 & 0.125 &           L1517 \\
    Lupus3 &   1 &  16:08:29.53 & -39:05:54.72 &   36 & 0.271 & 3.019 & 0.271 & 0.035 & 0.335 & 0.320 & 0.077 &     Lupus3-main \\
      ChaI &   1 &  11:02:56.17 & -77:21:40.87 &   12 & 0.201 & 0.906 & 0.250 & 0.449 & 0.210 & 0.209 & 0.135 &  ChaI-Southwest \\
      ChaI &   2 &  11:08:26.16 & -77:28:52.22 &   96 & 0.236 & 3.246 & 0.445 & 0.378 & 0.210 & 0.190 & 0.066 &      ChaI-South \\
      ChaI &   3 &  11:09:40.23 & -76:31:20.94 &   43 & 0.303 & 3.250 & 0.212 & 0.134 & 0.210 & 0.188 & 0.066 &      ChaI-North \\
     IC348 &   1 &   3:44:33.96 &  32:08:16.75 &  186 & 0.271 & 4.303 & 0.244 & 0.072 & 0.083 & 0.082 & 0.034 &      IC348-main \\
     IC348 &   2 &   3:44:33.10 &  32:14:51.26 &   11 & 0.335 & 0.531 & 0.064 & 0.008 & 0.083 & 0.083 & 0.048 &     IC348-North \\
\enddata
\tablenotetext{a}{Groups identified, position of centre, number 
of members, and descriptive names as discussed in 
Section~\ref{sec_mst} and following.  In Taurus, the descriptive 
names correspond to the \citet{Luhman09} proper motion groups 
where appropriate.}
\tablenotetext{b}{Median and maximum mass of members in the group, 
and the median offset and offset of the maximum mass group member 
as discussed in Section~\ref{sec_mass_offset}.}
\tablenotetext{c}{Critical branch length for the region and 
the maximum and median branch length found within each group, 
as discussed in Section~\ref{sec_mst}.}
\end{deluxetable}

%% file: tab2.tex
\begin{deluxetable}{ccc||ccc}
\tabletypesize{\small}
%(this version edited by hand to make shorter -- two sets of data columns
\tablecolumns{6}
\tablecaption{Adopted Mass Estimates \label{tab_spec_mass}}
\tablehead{
\colhead{Spectral Type} &
\colhead{T$_{eff}$ (K) \tablenotemark{a}} &
%\colhead{Mass (M$_{\odot}$) \tablenotemark{b}} &
\multicolumn{1}{c||}{Mass (M$_{\odot}$) \tablenotemark{b}} &
\colhead{Spectral Type} &
\colhead{T$_{eff}$ (K) \tablenotemark{a}} &
\colhead{Mass (M$_{\odot}$) \tablenotemark{b}}
}
\startdata
B5 &  15400 & 4.303  & G2 &   5860 & 2.632  \\
B6 &  14000 & 3.725  & G3 &   5830 & 2.627  \\
B7 &  13000 & 3.652  & G4 &   5800 & 2.621  \\
B8 &  11900 & 3.397  & G5 &   5770 & 2.616  \\
B9 &  10500 & 3.250  & G6 &   5700 & 2.602  \\
A0 &   9520 & 3.165  & G7 &   5630 & 2.587  \\
A1 &   9230 & 3.124  & G8 &   5520 & 2.562  \\
A2 &   8970 & 3.076  & G9 &   5410 & 2.535  \\
A3 &   8720 & 3.019  & K0 &   5250 & 2.430  \\
A4 &   8460 & 2.949  & K1 &   5080 & 2.265  \\
A5 &   8200 & 2.891  & K2 &   4900 & 2.134  \\
A6 &   8050 & 2.864  & K3 &   4730 & 1.796  \\
A7 &   7850 & 2.834  & K4 &   4590 & 1.634  \\
A8 &   7580 & 2.802  & K5 &   4350 & 1.121  \\
A9 &   7390 & 2.783  & K6 &   4205 & 0.906  \\
F0 &   7200 & 2.768  & K7 &   4060 & 0.801  \\
F1 &   7050 & 2.756  & M0 &   3850 & 0.701  \\
F2 &   6890 & 2.744  & M1 &   3705 & 0.633  \\
F3 &   6740 & 2.733  & M2 &   3560 & 0.575  \\
F4 &   6590 & 2.721  & M3 &   3415 & 0.398  \\
F5 &   6440 & 2.707  & M4 &   3270 & 0.271  \\
F6 &   6360 & 2.699  & M5 &   3125 & 0.178  \\
F7 &   6280 & 2.690  & M6 &   2990 & 0.096  \\
F8 &   6200 & 2.681  & M7 &   2880 & 0.057  \\
F9 &   6115 & 2.670  & M8 &   2710 & 0.031  \\
G0 &   6030 & 2.659  & M9 &   2400 & 0.013  \\
G1 &   5945 & 2.646  & L0 &   2200 & 0.009  \\
\enddata
\tablenotetext{a}{Effective temperatures from \citet{Luhman03}.}
\tablenotetext{b}{Mass estimates based on models of 
\citet{PS99}, \citet{Baraffe98}, and \citet{Chabrier00}.
See text for details.}
\end{deluxetable}

%% file: tab3.tex
\begin{deluxetable}{ccccccc}
\tabletypesize{\scriptsize}
\tablecolumns{7}
\tablecaption{YSO catalog for Taurus \label{tab_Taurus}}
\tablehead{
\colhead{RA} &
\colhead{Dec} &
\colhead{Name} & 
\colhead{Other} &
\colhead{Spectral} &
\colhead{Est. Mass \tablenotemark{a} } &
\colhead{Group \tablenotemark{b} } \\
\colhead{(J2000.0)} &
\colhead{(J2000.0)} &
\colhead{} &
\colhead{Designation} &
\colhead{Type} &
\colhead{(\Msol)} &
\colhead{} 
}
\startdata
   4:03:49.31 &  26:10:52.03 &   J04034930+2610520 &         HBC358A+B+C &                M3.5 &  0.335 & -- \\
   4:03:49.98 &  26:20:38.23 &   J04034997+2620382 &          XEST06-006 &               M5.25 &  0.157 & -- \\
   4:03:50.84 &  26:10:53.20 &   J04035084+2610531 &              HBC359 &                  M2 &  0.575 & -- \\
   4:04:39.37 &  21:58:18.65 &   J04043936+2158186 &              HBC360 &                M3.5 &  0.335 & -- \\
   4:04:39.85 &  21:58:21.53 &   J04043984+2158215 &              HBC361 &                  M3 &  0.398 & -- \\
   4:04:43.07 &  26:18:56.39 &   J04044307+2618563 &      IRAS04016+2610 &                  K3 &  1.796 & -- \\
   4:05:30.88 &  21:51:10.68 &   J04053087+2151106 &              HBC362 &                  M2 &  0.575 & -- \\
   4:08:07.82 &  28:07:28.05 &   J04080782+2807280 &              \ldots &               M3.75 &  0.303 & -- \\
\enddata
\tablenotetext{a}{Mass estimated using the procedure discussed in Appendix~C.}
\tablenotetext{b}{MST group the YSO belongs to.  See Section~3 and Appendix~D.1 for details.}
\tablenotetext{c}{Spectral type and corresponding mass based on estimate from bolometric luminosity measured.  Values should only be used within the following ranges: $<$K6, K6-M3.5, M3.5-M6, and M6-M8 corresponding to mass ranges of: $>$0.906, 0.906-0.3345, 0.3345 - 0.096, and 0.096 - 0.031 \Msol respectively.}
\end{deluxetable}

%% file: tab4.tex
\begin{deluxetable}{ccccccc}
\tabletypesize{\scriptsize}
\tablecolumns{7}
\tablecaption{YSO catalog for ChaI \label{tab_ChaI}}
\tablehead{
\colhead{RA} &
\colhead{Dec} &
\colhead{Name} & 
\colhead{Other} &
\colhead{Spectral} &
\colhead{Est. Mass \tablenotemark{a} } &
\colhead{Group \tablenotemark{b} } \\
\colhead{(J2000.0)} &
\colhead{(J2000.0)} &
\colhead{} &
\colhead{Designation} &
\colhead{Type} &
\colhead{(\Msol)} &
\colhead{} 
}
\startdata
  11:04:42.56 & -77:41:57.21 &   J11044258-7741571 &               ISO52 &                  M4 &  0.271 & -- \\
  11:05:14.72 & -77:11:29.35 &   J11051467-7711290 &                 Hn4 &               M3.25 &  0.366 & -- \\
  11:02:53.73 & -77:22:56.27 &   J11025374-7722561 &              \ldots &                M8.5 &  0.022 &  1 \\
  11:02:41.82 & -77:24:24.58 &   J11024183-7724245 &              \ldots &                  M5 &  0.178 &  1 \\
  10:58:16.72 & -77:17:17.11 &   J10581677-7717170 &                  T6 &                  K0 &  2.430 & -- \\
  11:04:22.73 & -77:18:08.13 &   J11042275-7718080 &                T14A &                  K7 &  0.801 &  1 \\
  11:02:06.06 & -77:18:07.96 &   J11020610-7718079 &              \ldots &                  M8 &  0.031 & -- \\
  11:01:13.69 & -77:22:38.76 &   J11011370-7722387 &              \ldots &               M5.25 &  0.157 & -- \\
\enddata
\tablenotetext{a}{Mass estimated using the procedure discussed in Appendix~C.}
\tablenotetext{b}{MST group the YSO belongs to.  See Section~3 and Appendix~D.1 for details.}
\end{deluxetable}

%% file: tab5.tex
\begin{deluxetable}{ccccccc}
\tabletypesize{\scriptsize}
\tablecolumns{7}
\tablecaption{YSO catalog for Lupus3 \label{tab_Lupus3}}
\tablehead{
\colhead{RA} &
\colhead{Dec} &
\colhead{Name} & 
\colhead{Other} &
\colhead{Spectral} &
\colhead{Est. Mass \tablenotemark{a} } &
\colhead{Group \tablenotemark{b} } \\
\colhead{(J2000.0)} &
\colhead{(J2000.0)} &
\colhead{} &
\colhead{Designation} &
\colhead{Type} &
\colhead{(\Msol)} &
\colhead{} 
}
\startdata
  15:53:41.20 & -39:00:38.00 &                Sz78 &              \ldots &                   ? &  0.271 & -- \\
  15:59:28.40 & -40:21:51.00 &               RYLup &              \ldots &                  K4 &  1.634 & -- \\
  16:03:05.50 & -40:18:26.00 &               EXLup &              \ldots &                  M0 &  0.701 & -- \\
  16:06:44.30 & -39:14:11.00 &                Sz86 &              \ldots &                   ? &  0.271 & -- \\
  16:07:00.60 & -39:02:19.00 &                Sz88 &              \ldots &                  M1 &  0.633 &  1 \\
  16:07:10.10 & -39:11:03.00 &                Sz90 &              \ldots &               K7-M0 &  0.786 & -- \\
  16:07:11.60 & -39:03:47.00 &                Sz91 &              \ldots &                M0.5 &  0.667 &  1 \\
  16:07:15.20 & -40:03:42.00 &                Sz92 &              \ldots &                   ? &  0.271 & -- \\
\enddata
\tablenotetext{a}{Mass estimated using the procedure discussed in Appendix~C.}
\tablenotetext{b}{MST group the YSO belongs to.  See Section~3 and Appendix~D.1 for details.}
\end{deluxetable}

%% file: tab6.tex
\begin{deluxetable}{ccccccc}
\tabletypesize{\scriptsize}
\tablecolumns{7}
\tablecaption{YSO catalog for IC348 \label{tab_IC348}}
\tablehead{
\colhead{RA} &
\colhead{Dec} &
\colhead{Name} & 
\colhead{Other} &
\colhead{Spectral} &
\colhead{Est. Mass \tablenotemark{a} } &
\colhead{Group \tablenotemark{b} } \\
\colhead{(J2000.0)} &
\colhead{(J2000.0)} &
\colhead{} &
\colhead{Designation} &
\colhead{Type} &
\colhead{(\Msol)} &
\colhead{} 
}
\startdata
   3:44:34.21 &  32:09:46.70 &                   1 &              \ldots &                  B5 &  4.303 &  1 \\
   3:44:35.36 &  32:10:04.60 &                   2 &              \ldots &                  A2 &  3.076 &  1 \\
   3:44:50.65 &  32:19:06.80 &                   3 &              \ldots &                  A0 &  3.165 & -- \\
   3:44:31.19 &  32:06:22.10 &                   4 &              \ldots &                  F0 &  2.768 &  1 \\
   3:44:26.03 &  32:04:30.40 &                   5 &              \ldots &                  G8 &  2.562 &  1 \\
   3:44:36.94 &  32:06:45.40 &                   6 &              \ldots &                  G3 &  2.627 &  1 \\
   3:44:08.48 &  32:07:16.50 &                   7 &              \ldots &                  A0 &  3.165 & -- \\
   3:44:09.15 &  32:07:09.30 &                   8 &              \ldots &                  A2 &  3.076 & -- \\
\enddata
\tablenotetext{a}{Mass estimated using the procedure discussed in Appendix~C.}
\tablenotetext{b}{MST group the YSO belongs to.  See Section~3 and Appendix~D.1 for details.}
\end{deluxetable}